\pdfoutput=1
\documentclass[usenatbib]{mnras}
\usepackage[utf8]{inputenc}
\usepackage{graphicx}
\usepackage{amssymb}
\usepackage{mathtools}
\usepackage{amsmath}
\usepackage{color}
\usepackage{rotating}
\usepackage{pdflscape}
\usepackage{array}
\usepackage{multirow}
\usepackage{tabularx}
\usepackage{booktabs}
\usepackage{cite}
\usepackage{threeparttable}
\usepackage{caption}
\usepackage{subcaption} 
\title[Dust emission from evolved stars]{Extended Dust Emission from Nearby Evolved Stars\thanks{The scripts and data required to reproduce the analysis, figures and tables presented in this paper can be downloaded from {\tt https://github.com/Thavisha/EvolvedStars\_DustEmission}}}
\author[T. E. Dharmawardena et al.]{Thavisha E. Dharmawardena$^{1,2}$\thanks{tdharmawardena@asiaa.sinica.edu.tw}, Francisca Kemper$^{1}$, Peter Scicluna$^{1}$, 
\newauthor
Jan G. A. Wouterloot$^{3}$, Alfonso Trejo$^{1}$, Sundar Srinivasan$^{1}$, Jan Cami$^{4,5}$, 
\newauthor
Albert Zijlstra$^{6,7}$, Jonathan P. Marshall$^{1}$\\
$^{1}$Academia Sinica Institute of Astronomy and Astrophysics, 11F of AS/NTU Astronomy-Mathematics Building, \\No.1, Sect. 4, Roosevelt Rd, Taipei 10617, Taiwan, R.O.C.\\
$^{2}$ Graduate Institute of Astronomy, National Central University, 300 Zhongda Road, Zhongli 32001, Taoyuan, Taiwan, R.O.C.\\
$^{3}$East Asian Observatory, 660 N A'ohoku Place, Hilo, Hawaii 96720, USA\\
$^{4}$Department of Physics and Astronomy and Centre for Planetary Science and Exploration (CPSX), \\The University of Western Ontario, London, ON N6A 3K7, Canada\\
$^{5}$SETI Institute, 189 Bernardo Ave, Suite 100, Mountain View, CA 94043, USA\\
$^{6}$Jodrell Bank Centre for Astrophysics, School of Physics and Astronomy, University of Manchester, Oxford Road, Manchester M13 9PL, UK\\
$^7$Laboratory for Space Research, University of Hong Kong, Pokfulam Road, Hong Kong
}

\begin{document}

\maketitle

\begin{abstract}

We present  JCMT SCUBA-2 $450\micron$ and $850\micron$ observations of 14 Asymptotic Giant Branch (AGB) stars (9 O--rich, 4 C-rich and 1 S--type) and one Red Supergiant (RSG) in the Solar Neighbourhood. We combine these observations with \emph{Herschel}/PACS observations at $70\micron$ and $160\micron$ and obtain azimuthally-averaged surface-brightness profiles and their PSF subtracted residuals. The extent of the SCUBA-2 850 $\micron$ emission ranges from 0.01 to 0.16 pc with an average of $\sim40\%$ of the total flux being emitted from the extended component. By fitting a modified black-body to the four-point SED at each point along the radial profile we derive the temperature ($T$), spectral index of dust emissivity ($\beta$) and dust column density ($\Sigma$) as a function of radius. For all the sources, the density profile deviates significantly from what is expected for a constant mass-loss rate, showing that all the sources have undergone variations in mass-loss during this evolutionary phase. In combination with results from CO line emission, we determined the dust-to-gas mass ratio for all the sources in our sample. We find that, when sources are grouped according to their chemistry, the resulting average dust-to-gas ratios are consistent with the respective canonical values. However we see a range of values with significant scatter which indicate the importance of including spatial information when deriving these numbers.  

\end{abstract}


\section{Introduction}

The origin of interstellar dust in galaxies remains only partially understood. A major problem faced by researchers currently in this area is the so-called \emph{dust budget crisis} \citep[e.g.~][]{Morgan2003,Rowlands2014}. The Asymptotic Giant Branch (AGB) stars in high-redshift galaxies have not yet had enough time to evolve into their mass-losing phase. Therefore assuming these stars are the primary contributors of dust and heavy elements of these galaxies we are unable to explain their interstellar dust masses observed, thus giving rise to the \emph{\rm dust-budget crisis}. Even in nearby galaxies, where sufficient time has elapsed for the formation of AGB stars, the dust-production rates (DPR) by AGB stars seem to be insufficient to explain the interstellar dust reservoir. For instance, at the current DPR by AGB stars and Red Supergiants (RSGs) for the Small Magellanic Cloud (SMC), it would take 92 Gyr, significantly longer than a Hubble time, to produce the existing interstellar dust reservoir \citep{Gordon2014, Boyer2012}. For the LMC similar conclusions can be drawn, as a replenishment time scale of 35 Gyr results from the DPR derived by \citet{Riebel2012}. While an ongoing effort to address the dust budget crisis is focused on the exploration of additional sources of dust that contribute to the interstellar dust reservoir, i.e., dust creation during supernovae explosions \citep[e.g.~][]{Bianchi2007,Gall_11_Production}, dust production by active galactic nuclei \citep{Elvis_02_Smoking}, and dust formation in the interstellar medium (ISM) \citep[e.g.~][]{Jones_01_Interstellar, Fulvio2017:C_inthe_ISM}; it remains crucial to fully understand the production of dust by evolved stars, particular those on the AGB. 

AGB stars have main-sequence masses in the range 1--8 $M_\odot$. They have evolved off the Main Sequence and are currently undergoing He- and H-shell burning, while a strong stellar wind develops, in which dust formation occurs \citep[see][and references therein]{Habing_03_Asymptotic}. Although individual, more massive stars (M $>$ 8 M$_{\odot}$) expel a larger dust mass into the ISM during both their supergiant and supernova stages, AGB stars are so numerous that as a group they contribute a significant amount of dust to the ISM. This makes them extremely effective in replenishing the interstellar dust reservoir, as well as contributing the products of nucleosynthesis to the galactic chemical evolution. 

To establish the importance of AGB stars in the life cycle of dust in galaxies, the DPR has been determined for a number of nearby galaxies, including the LMC \citep{Riebel_12_Mass}, the SMC \citep{Srinivasan2016}, in the central kpc$^2$ of M33 \citep{Javadi_13_UK}, and M32 \citep{Jones_15_Spitzer}. The advantage of studying the AGB population in external galaxies is that the distances -- and thus the luminosities -- of the stars are known, which is not the case in the Milky Way. 
Extinction in the Galactic Plane limits the determination of a Galactic
DPR in the Solar Neighborhood \citep[][Trejo et al. \emph{in prep.}]{Jura_89_Dust}, with significant uncertainties due to the difficulties
in determining the luminosities. All these determinations of the integrated DPR rely on fitting of the spectral energy distribution (SED) of point sources, dominated by mid-infrared emission, using for instance the GRAMS model grid \citep{Srinivasan2011,Sargent2011}, and assuming a constant mass-loss rate (MLR). 

Here, we investigate whether this constant MLR assumption is reasonable, by studying extended thermal dust emission at sub-millimetre (sub-mm) wavelengths from nearby AGB stars. We compare these resolved measurements to the expected results assuming a constant MLR. Past enhancement in MLRs will thus be revealed and we will be able to evaluate the reliability of DPRs determined by mid-infrared SED fitting. We are also able to then determine whether a cold dust reservoir exists that is not represented in these fits.

Two recent studies \citet{Ladjal2010} and \citet{Cox2012} have attempted to spatially resolve the thermal emission in the circumstellar environments of evolved stars. \citet{Ladjal2010} carried out aperture photometry of a sample of nine evolved stars observed at 870 \micron\ with the LABoCa instrument on APEX, showing that four of the nine stars were extended at 870 \micron. They also derived the total dust mass and the dust MLR for these objects. \citet{Cox2012} generated an image atlas of nearby AGB stars at 70 and $160 \micron$ using observations from the \emph{Herschel} Space Observatory's `Photometric Array Camera and Spectrograph' instrument \citep[PACS;][]{Poglitsch2010}. This revealed 78 sources with extended emission.

In order to expand the statistics provided by the sample studied by \citet{Ladjal2010}, we have selected a sample of 15 sources from the \emph{Herschel} atlas published by \citet{Cox2012} showing extended emission more than 1$^\prime$ away from the central position for followup with the Sub-millimetre Common-User Bolometer Array 2 (SCUBA-2) instrument on the James Clerk Maxwell Telescope (JCMT). Compared to LABoCa, SCUBA-2 has better sensitivity and a larger field of view of $\sim 45$ arcmin$^{2}$ \citep{Holland2013}. Combining these SCUBA-2 continuum observations at $450 \micron$ and $850 \micron$ with the \emph{Herschel}/PACS observations at $70\micron$ and $160\micron$ allows us to study the extended thermal emission due to a cold thermal dust component. Thus we are able to trace the dust mass-loss history in better detail than what may be achieved with \emph{Herschel} data alone. 

This study is a pilot for the Nearby Evolved Star Survey (NESS\footnote{\url{http://www.eaobservatory.org/jcmt/science/large-programs/ness/}}) (Scicluna et al., \emph{in prep}) which targets a volume-limited sample of Galactic AGB stars in both continuum and CO line emission. The NESS observations are being carried out using multiple sub-mm telescopes, including the JCMT, and the resulting data will be analysed to derive robust statistics of stellar dust and gas return to the Galactic ISM. 

This paper is organised as follows: In Sect. \ref{sec:obs} we present the sample and observation strategy along with the data reduction methods used. We then derive and analyse the radial surface-brightness profile and its PSF-subtracted equivalent (i.e: the residual profile; Sect. \ref{sec:emission}). Subsequently we derive and discuss radial profiles of the temperature ($T$), spectral index of dust emissivity ($\beta$), and dust column density ($\Sigma$) (Sect. \ref{sec:dust-props}). Section \ref{sec:dust-to-gas_ratios} present the total dust masses, dust mass-loss rates and dust-to-gas ratio measurements and their implications. Finally, our findings are summarised in Sect. \ref{sec:summary}.

\section{Observations and Data Reduction}
\label{sec:obs}

\subsection{Source selection and SCUBA-2 observations}
\label{sec:obs-selection}

The observations presented in this paper were carried out as part of programs M15AI65 and M15BI047 on the JCMT, which aimed to study both the dust and the gas mass-loss rates and histories of nearby AGB stars. Therefore these programs also included both SCUBA-2 continuum and HARP CO(3-2) line observations. 

The sources in both the M15AI65 and M15BI047 samples were selected from the \emph{Herschel} MESS (\textit{mass-loss from Evolved StarS}) survey sample \citep{Groenewegen2011}, showing extended emission out to a radius $\geq 1^\prime$ at PACS $70\micron$ and $160\micron$ \citep{Cox2012}. With CO(3-2) observations in mind, we also stipulated the presence of strong CO(3-2) line emission with a peak $T_\mathrm{MB} \gtrsim 2$ K. Observing time allocated and telescope scheduling constraints then further restricted us to SCUBA-2 observations for 14 objects, with a fifteenth object (IRC +10216) being a calibration source and thus having sufficient archival observations available to be included in the analysis of the sample. Furthermore, our observations of $o$ Cet, the archetypal Mira, another JCMT pointing calibration source, have been supplemented with archival data. 

The sample is summarised in Tab.~\ref{table:Targets}, which shows the source IRAS name, Identifier, RA and Dec. The table also presents the source type and chemistry (C-rich AGB star (C-AGB), O-rich AGB star (O-AGB), or Red Supergiant (RSG)). The distance and terminal velocities obtained from literature. Further we also present the observation program ID and total observing times for the JCMT SCUBA-2 observations.

We used the SCUBA-2 receiver \citep{Holland2013, Chapin2013} on the JCMT to simultaneously generate continuum maps at both $450 \micron$ and $850 \micron$ of the 13 AGB stars and one RSG. The targets were observed using the SCUBA-2 {\sc daisy} scan pattern observing mode with a scan velocity of $155\arcsec$s$^{-1}$ along the scan axis.  

We also obtained JCMT HARP CO 3-2 observations for our sample via the  M15AI65 and M15BI047 programs in order to study the gas mass emission in our sample. However this is beyond the scope of this paper and these observations will be combined with the NESS sample and analysed for a later publication.  

\begin{table*}
  \centering
  \caption{Source Information and Observation Details}
  \begin{threeparttable}
    \begin{tabular}{lllllllllll}
    \hline
    \hline %
           \multicolumn{1}{c}{\multirow{2}[0]{*}{IRAS}} & \multicolumn{1}{c}{\multirow{2}[0]{*}{Identifier}} & RA (J2000) & Dec(J2000) & \multicolumn{1}{c}{\multirow{2}[0]{*}{Type}} & \multicolumn{1}{c}{\multirow{2}[0]{*}{Chemistry}} & Distance & $v_{\infty}$\tnote{(e)} & \multicolumn{1}{c}{\multirow{2}[0]{*}{Program ID}} & Observation \\
          
           & & (hh:mm:ss) & (dd:mm:ss) & & & (pc) & (km s$^{-1}$) & & Time (s) & \\
     \hline
     \\
    10131+3049 & CIT 6 & 10:16:02.28 & +30:34:19.0 & AGB & C-rich & 440 $\pm$ 132\tnote{(a)} & 20.8 & M15BI047 & 7591 \\
    & (RW LMi) & & & & & & & &  \\
    
    21439-0226 & EP Aqr & 21:46:31.85 & $-$02:12:45.9 & AGB & O-rich & 113.6 $\pm$ 8.1\tnote{(b)} & 11.5 & M15BI047 & 5606 \\
    
     03507+1115 & IK Tau & 03:53:28.87 & +11:24:21.7 & AGB & O-rich & 260.0 $\pm$ 10.0\tnote{(b)} & 18.5 &  M15BI047 & 7583 \\

     01037+1219 & IRC+10011 & 01:06:25.98 & +12:35:53.1 & AGB & O-rich & 740.0 $\pm$ 222 \tnote{(a)} & 19.8 & M15BI047 & 7448\\
    & (WX Psc) & & & & & & & &  \\
    
     09452+1330 & IRC+10216 & 09:47:57.41 & +13:16:43.6 & AGB & C-rich & 130.0 $\pm$ 13.0\tnote{(b)} & 14.5 &  archival & $\sim$156000\\
    & (CW Leo) & & & & & & & &  \\
    
    23320+4316 & LP And & 23:34:27.53 & +43:33:01.2 & AGB & C-rich & 630.0 $\pm$ 189\tnote{(a)} & 14.0 & M15BI047 & 7475\\
    
    \ldots & NML Cyg & 20:46:25.54 & +40:06:59.4 & RSG & \ldots & 1610.0 $\pm$ 120.0\tnote{(c)} & 33.0 & M15AI65 & 7116 \\ 
    
     02168-0312 & \textit{o} Ceti & 02:19:20.79 & $-$02:58:39.5 & AGB & O-rich & 91.7 $\pm$ 10.2\tnote{(b)} & 8.1 &  M15BI047 & $\sim$179000\\
    & (Mira) & & & & & & & + archival &  \\
    
   23558+5106 & R Cas & 23:58:24.87 & +51:23:19.7 & AGB & O-rich & 125.8 $\pm$ 16.4\tnote{(b)} & 13.5 & M15BI047 & 7541\\
    
    09448+1139 & R Leo & 09:47:33.49 & +11:25:43.7 & AGB & O-rich & 71. 3$\pm$ 13.5\tnote{(b)} & 9.0 & M15BI047 & 7473\\
    
   14219+2555 & RX Boo & 14:24:11.63 & +25:42:13.4 & AGB & O-rich & 190.8 $\pm$ 22.9\tnote{(b)} & 9.0 &  M15BI047 & 9448\\
    
    04566+5606 & TX Cam & 05:00:51.22 & +56:10:54.2 & AGB & O-rich & 380.0 $\pm$ 114\tnote{(a)} & 21.2 & M15BI047 & 7649 \\
    
     10350-1307 & U Hya & 10:37:33.27 & $-$13:23:04.4 & AGB & C-rich & 208.3 $\pm$ 10.0\tnote{(b)} & 8.5 & M15BI047 & 7481 \\
    
    19126-0708 & W Aql & 19:15:23.35 & $-$07:02:50.4 & AGB & S-type & 340.0 $\pm$ 102\tnote{(d)} & 20.0 & M15AI65 & 7275\\
    
  13462-2807 & W Hya & 13:49:02.00 & $-$28:22:03.5 & AGB & O-rich & 104.3 $\pm$ 12.2\tnote{(c)} & 8.5 & M15AI65 & 5618 \\

    \hline
    \hline
    \end{tabular}%
    
         \begin{tablenotes}
    \item The distances were obtained from the following publications: $^{\textrm{(a)}}$ \citet{DeBeck2010}, $^{\textrm{(b)}}$ \citet{McDonald2017}, $^{\textrm{(c)}}$ \citet{Zhang2012NMLCyg}, $^{\textrm{(d)}}$ \citet{Guandalini2008}.
    \item (e): The terminal velocities were obtained from \citet{DeBeck2010}. For sources with distances from \citet{DeBeck2010} and \citet{Guandalini2008} no published uncertainties on the distances were available and therefore we assume an uncertainty of $\pm 30\%$.  
    \end{tablenotes} 
    
    \end{threeparttable}
  \label{table:Targets}%
\end{table*}%

\subsection{SCUBA-2 data reduction}
\label{sec:obs-SCUBA-2_reduction}


The SCUBA-2 data was downloaded from the Canadian Astronomy Data Centre (CADC)\footnote{\url{http://www.cadc-ccda.hia-iha.nrc-cnrc.gc.ca/en/jcmt/}} digital archives. The reduction of this data was done using the Starlink ORAC-DR pipeline \citep{Currie2014,Jenness_15_ORAC}. Specific reduction recipes are available for extended emission ({\sc reduce scan extended source}) and point sources ({\sc reduce scan isolated source}; \citet{Chapin2013}). Although we are looking for extended emission, the extent of this emission is relatively small and weak compared to the bright central point source, causing the radial profile to deviate only slightly from the point spread function (PSF). Thus, we elected to use the {\sc reduce scan isolated source} recipe with some edited parameters, as it uses the position of the peak emission by fitting a PSF to each observation to calibrate the pointing of each observation. 

By default, the {\sc reduce scan isolated source} recipe assumes the emission to be zero beyond a radius of $1^\prime$ from the central pixel\footnote{http://starlink.eao.hawaii.edu/devdocs/sc21.htx/sc21.html} for the first few iterations. As we are interested in detecting emission extending beyond $1^\prime$, we adjusted the recipe and the constraint radius was set to $1.5^\prime$. Extending the radius beyond this value would result in an artificial effect known as \textit{blooming} where the source emission is blended with and hence confused with the background emission. A smaller radius than the chosen one would result in an artefact known as \textit{negative bowling} where the pipeline enhances the central point source significantly while resulting in an artificial negative region beyond the central point source. The chosen radius results in a compromise between the two effects to enhance as much extended emission, while minimising the said artefacts. 

However, this also means we may be ignoring possible extended emission beyond $1.5^\prime$. The detected extended emission limits can therefore be interpreted as the lower limits to the possible extended emission if this were the case. However we demonstrate in section \ref{sec:extended-emission} that this is very unlikely to be an issue and that the extended emission determined is reliable.   

\subsection{IRC+10216 (CW Leo) and \textit{o} Cet (Mira) SCUBA-2 data reduction}
\label{sec:CWLeo+Mira_Reduction}

IRC+10216 and \textit{o} Cet have been used by the JCMT as SCUBA-2 calibration sources since 2009 and 2011 respectively. Therefore in addition to the few scan science observations, a large number of short pointing observations ($\sim 1100$) are available via the CADC archive. We reduced all publicly available data to obtain two sets of deep maps at $450 \micron$ and $850 \micron$ suitable for detecting extended circumstellar shell emission. The total observation times for the two sources were 43.6 h and 49.9 h respectively, which includes both the calibration and science observations. A full list of observations used is available in the GitHub repository.

Calibration (JCMTCAL) scan maps for IRC+10216 were available for the years 2010, 2011, 2014, and 2015 totalling to 2.8 h of observations and for \textit{o} Cet for the years 2014 and 2015 (M15BI047) totalling 2.1 h. These maps were all downloaded simultaneously and reduced together using the same reduction method as described in Sect. \ref{sec:obs-SCUBA-2_reduction}.  

The observation times for the pointing observations ranged from a couple of seconds to a couple of minutes. These calibration pointing observations were reduced using the {\sc reduce scan} pipeline. This pipeline worked best for the pointing observations as it was a good balance between the {\sc reduce scan extended sources} pipeline and the {\sc reduce scan isolated source} pipeline; it reduces the effects of both \textit{negative bowling} and \textit{blooming}. The pipeline utilises the same reduction parameters as the {\sc reduce scan isolated source} pipeline however, it does not have the constraining radius and is also optimised to remove any artificial large-scale emission. Since we are co-adding a large number of pointing-only observations there is no risk of background emission dominating when this pipeline is used (unlike the scan observations for the other sources in our sample where the background is mapped to a certain degree as well).

Once the pipeline has processed all the individual observations, the reduced observations are registered to a single position preparing them for co-addition. The position is chosen by fitting a 2D Gaussian to a single image to identify its peak. Once we identify the peaks for all the other images in a similar manner they are registered to the initial image's peak position. Once this process is done all the images are co-added to make a single map for each year. The entire process is repeated for the observations for all years. The resulting maps are again registered and co-added to make two single deep maps of all the pointing observations for both the $450 \micron$ and $850 \micron$ data sets.

\subsection{Herschel PACS data}

The data from the MESS survey were available for public access from the \emph{Herschel} Science Archive\footnote{\url{http://www.cosmos.esa.int/web/herschel/science-archive}} (HSA). We downloaded pre-reduced data for observations carried out by the MESS survey using the PACS instrument \citep{Poglitsch2010}. The MESS survey carried out PACS photometer observations using the \textit{scan map} observing mode at a medium scan speed of $20\arcsec$s$^{-1}$ in the blue (70 \micron) and red ($160 \micron$) wavelength filters covering scan lengths ranging from 6$^\prime$ to 34$^\prime$ depending on target \citep{Groenewegen2011}. We downloaded level 2.5 data products  \footnote{\url{https://www.cosmos.esa.int/web/herschel/data-products-overview}} for the MESS PACS observations used in this study. This is the highest level of publicly-available reduced MESS data products in the HSA. The data products were calibrated using PACS calibration version \emph{PACS$\_$CAL$\_$72$\_$0}.

The MESS survey also observed a sample of galactic AGB stars using the \emph{Herschel} Spectral and Photometric Imaging REceiver \citep[SPIRE;][]{Griffin2010}. These observations cover three wavebands centred at 250, 350, and 500$~\micron$. We found that only seven of the fifteen sources in our sample were observed using SPIRE by the MESS survey and including only those sources would affect the homogeneity of the sample and methodology. Additionally the SPIRE beam FWHM are $18\arcsec$ at 250$~\micron$, $24\arcsec$ at 350$~\micron$, and $42\arcsec$ at 500$~\micron$. In this study, we are most interested in tracing circumstellar structure with typical angular diameters of $\sim 20\arcsec$ to $\sim 60\arcsec$ at sub-millimetre wavelengths. The smallest beam size of $18\arcsec$ at $250\micron$ it self is $\sim 40 \%$ larger than the SCUBA-2 $850\micron$ beam, the lowest resolution data. Hence including the SPIRE observation would reduce the angular resolution at which the data could be analysed as the data must be interpreted on the scales corresponding to the lowest resolution observations (see Sec.\ref{sec:sedfit}). Considering these factors we do not consider the SPIRE data in this study.

The \textit{AKARI} Far-Infrared Surveyor's (FIS) MLHES survey (\textit{Excavating Mass Loss History in Extended Dust Shells of Evolved Stars}; \citep{Izumiura2009}), observed the sources in our sample at $65\micron$, $90\micron$, $140\micron$ and $160\micron$. However the data are not yet publicly available. Therefore we could not include them. The AKARI FIS All Sky Survey \citep{Doi2009_AKARI_AFllSkySurvey} observations were not deep enough in order to detect extended emission and thus not useful for the objectives of this study.

\section{Extended Dust Emission}
\label{sec:extended-emission}

\subsection{Surface-brightness profiles}
\label{sec:surf-bright}

We determined azimuthally-averaged stellar surface-brightness profiles assuming spherical symmetry for each source at all four available wavelengths (PACS $70\micron$ and $160\micron$ and SCUBA-2 $450\micron$ and $850\micron$).

In order to determine the amount of extended emission we compared the derived surface-brightness profiles to the telescope PSF (beam) profiles. The left hand panels in Fig.~\ref{fig:irc10216_RadialProfile} depict the radial profiles for IRC+10216 at all four wavelengths. The blue dotted lines indicate the azimuthally-averaged surface-brightness profiles and the grey solid lines represent the beam profile of each instrument at the given wavelength. 

The SCUBA-2 beam at each wavelength, is comprised of two Gaussian components; a main beam component and a secondary error-beam component. The beam shape is determined to be the normalised sum of these two components, where each component is scaled by a beam amplitude factor \citep{Dempsey2013}. 

The \emph{Herschel}/PACS PSF exhibits a 2D Gaussian core and a much more complex 3-lobe structure, the shape of which was determined over a wide dynamic range using combined observations of Mars and the asteroid Vesta \citep{Bocchio2016_1}. The 3-lobe structure lies at around the 10 per cent level of the peak flux, and its orientation depends on the telescope position angle at the time of observation. It is customary to use deconvolution on imaging observations to interpret resolved structure, wherein the orientation of the PSF is critical to avoiding the creation artefacts in the process \citep[e.g.][]{Liseau2010,Marshall2011,Decin2011,Marshall2014,Ertel2014,Marshall2016,Hengst2017}. Here, we are combining imaging data across a broad range of wavelengths and signal-to-noise, sampling them at the lowest spatial resolution (JCMT SCUBA-2 $850\micron$). We therefore adopt azimuthally-averaged radial source profiles for our interpretation as the S/N of the SCUBA-2 observations is insufficient to warrant deconvolution and azimuthal averaging is the only way to recover as much extended emission as possible. To ensure homogeneity we therefore also azimathally average the PACS source and PSF data. This is vital as statistical methods for interpreting heterogeneous data are prohibitively complex and not yet well established. Further azimuthal averaging and re-sampling smooths out the complex structure of the PACS PSF making the precise orientation at the time of observation immaterial.

As the PSF is also dependent on the scan speed of the observation, with faster scan speeds exhibiting more elongated ellipsoidal PSFs, this facet of the PSF structure must be accounted for. We have therefore used PSFs observed under the same scan speed as the PACS observations we wish to examine (i.e. observations taken at $20 \arcsec s^{-1}$). Furthermore, the FWHM of the PACS PSF has been shown to be time-variable at levels of 3-10 per cent, depending on the waveband of observation \citep{Kennedy2012}. The 70~$\micron~$ band is most critically affected, with an uncertainty in the FWHM of $10 \%$. However, this uncertainty is negligible for our sample as we do not observe any extended emission at $\sim1$ FWHM from the central PSF. Additionally when converting the radii into physical distances and look-back times the uncertainty on the source distance far outweighs any uncertainties introduced on the radii measured by the well calibrated PACS PSF.

Appropriate PACS PSFs were downloaded from the Deep \emph{Herschel}/PACS point spread functions VizieR online data catalogue \citep{Bocchio2016_2}. The beam FWHMs and parameters of both telescopes are summarised in Tab.~\ref{Telescope}. 

\begin{table*}
  \centering
  \caption{Telescope Instrumental Information}
  \begin{threeparttable}
    \begin{tabular}{lllll}
    \hline
    \hline
          & \multicolumn{2}{c}{Herschel PACS photometer\tnote{1}} & \multicolumn{2}{c}{JCMT SCUBA-2\tnote{2}} \\
          \hline
          & 70\micron & 160\micron & 450\micron & 850\micron \\
          \hline
    Primary beam FWHM (arcsec) & 5.46 $\times$ 5.76 & 10.65 $\times$ 12.13 & 7.9   & 13 \\
    Secondary beam FWHM (arcsec) & $-$   & $-$   & 25    & 48 \\
    Main beam rel. amplitude & $-$   & $-$   & 0.94  & 0.98 \\
    Secondary beam rel. amplitude & $-$   & $-$   & 0.06  & 0.02 \\
    \hline
    \hline
    \end{tabular}%
    
    \begin{tablenotes}
    
    \item [1] http://herschel.esac.esa.int/Docs/PACS/html/pacs\_om.html
    \item [2] \citet{Dempsey2013}
    
    \end{tablenotes}
    
    \end{threeparttable}
  \label{Telescope}%
\end{table*}%

As depicted in Fig.~\ref{fig:irc10216_RadialProfile} the PSF profiles were scaled to the peaks of the surface-brightness profiles in order to compare to our data. Knowing that we have a very bright central source and a small amount of extended emission beyond it we fit the PSF in order subtract as much of the PSF flux as possible and emphasise the extended emission, hence deriving residual profiles of the extended circumstellar shell. This is a computationally-simple, conservative, zeroth-order approximation to deconvolution. The right hand panels in Fig.~\ref{fig:irc10216_RadialProfile} show the residual profiles for IRC+10216 at all four wavelengths. 

Using the residual profiles we determine the maximum radius out to which extended circumstellar shell emission is detectable at at least 3 times the noise level, i.e: the 3$\sigma$ radius ($\theta_{3\sigma}$). Hereafter we take this be the maximum verifiable circumstellar shell extent of the source at the given wavelength for our observations. We also derive the maximum physical circumstellar shell extension at $3\sigma$ brightness level ({\textit{R}$_{3\sigma}$}); the age of the circumstellar shell at the measured $3\sigma$ brightness extension (look-back time). The derived values are listed in Tab.~\ref{table:RadialProfile_Results_Final}.

The average sky-subtracted total stellar fluxes (\textit{F}$_{\rm total}$) for the fifteen sources were measured using the \texttt{python Photutils} tools \citep{Bradley2016}. The $\theta_{3\sigma}$ values were used as the source aperture size and the sizes of the sky annuli were determined using several factors. The source brightness, emission beyond $3\sigma$ levels and visible surrounding background emission affected the choice of sky annuli. Brighter, more extended sources called for a larger sky annulus in order to obtain a better averaged background value. Further, the annuli were set to avoid any bright background structure or noise spikes. At all times we took care to not include the high noise background regions found at the edge of the observations. 

By integrating over the PSF radial profiles we were also able to determine the PSF flux (\textit{F}$_{\rm PSF}$) for each observation. We then determined the flux of the extended circumstellar shell component (\textit{F}$_{\rm ext}$) by subtracting the calculated PSF flux from the measured total stellar flux. We also calculated the flux percentage of the extended component \mbox{(${\%}$ \textit{F}$_{\rm ext}$)} when compared to that of the total star. These results are also presented in Tab.~\ref{table:RadialProfile_Results_Final}.

\subsection{Spectral Energy Distribution fits}
\label{sec:sedfit}

In order to combine the residual profiles and fit the SEDs we regridded the residual profiles onto the $850 \micron$ radial grid, the one with the lowest resolution (1 pix $=$ 4$\arcsec$). By fitting a modified black body to the SED at each radial point of the extended component, we derive the temperature ($T$), spectral index of dust emissivity ($\beta$), and dust column density ($\Sigma$) profiles. The model chosen to predict the surface brightness assumes a single-temperature dust population represented by a black-body modified by a single power-law emissivity \citep{Gordon2014}. The following modified black-body equation was adopted for this purpose:

\begin{equation}
\centering
\label{eq:modified-bb}
F_{\nu} \ \propto \ \lambda^{-\beta}B_{\nu}(T).
\end{equation}

The temperature-dependent surface brightness of the dust is as follows

\begin{equation}
\centering
\label{eq:surf-bright}
S_{\nu} = \Sigma_{\nu}\kappa B_{\nu},
\end{equation}

where $\Sigma_{\nu}$ is the dust column density, B$_{\nu}$ is the Planck function and $\kappa$ is the emissivity law

\begin{equation}
\centering
\label{eq:emissivity}
\kappa = \kappa_{\rm eff,160}^{S} \left({\lambda \over 160~\micron}\right)^{-\beta_{\rm eff}},
\end{equation}

where ${{\kappa_{\rm eff,160}^{S}}}$ is the effective emissivity at $160\micron$. The value of ${{\kappa_{\rm eff,160}^{S}}}$ depends on the properties of the dust grains in the circumstellar shell. In this paper, we use ${{\kappa_{\rm eff,160}^{S}}} = 26$ cm$^{2}$g$^{-1}$ for C--rich sources and ${{\kappa_{\rm eff,160}^{S}}} = 8.8$ cm$^{2}$g$^{-1}$ for O--rich sources (including the RSG and S-type star). 

To obtain these values, we computed cross sections for spherical grains of average size $\sim$0.1 \micron\ from the Mie theory, using the ACAR optical constants for amorphous carbon grains from \citet{Zubkoetal1996} and the astronomical silicate optical constants from \citet{Ossenkopfetal1992}. We then extrapolated these cross-sections from the mid-IR to longer wavelengths as required by this study. However, the value of $\kappa$ depends on the composition and shape/size distribution of the grains in the envelope; in particular, amorphous carbon grains are more efficient absorbers in the IR than silicate grains, and CDE grains are able to reproduce the observed sub-mm flux with much lower total shell mass. 

Our chosen values for ${{\kappa_{\rm eff,160}^{S}}}$ translate to $\sim$0.9 and $\sim$0.3 cm$^{2}$g$^{-1}$ at $870\micron$ for carbonaceous and oxygen-rich dust respectively, with the extrapolation curve having a slope of -1.4. \citet{Ladjal2010} find $\kappa_{\rm 870 \mu m}$ values ranging from 2 cm$^2$ g$^{-1}$ for O--rich AGB stars to 35 cm$^2$ g$^{-1}$ for C--rich AGB stars. We therefore expect our dust masses to be larger than the \citet{Ladjal2010} values for the same stars by a factor of $\sim 6$ for O-rich stars and $\sim 40$ for C-rich stars.

In order to carry out SED fitting we utilised the \emph{emcee} python package \citep{Foreman-Mackey2013}. This package uses affine-invariant Markov Chain Monte Carlo (MCMC) algorithms to carry out Bayesian inference. We apply this to the SEDs derived from our observations in order to determine the the most probable values of $T$, $\Sigma$ and $\beta$. 
The full description of this method is described in appendix \ref{appendix:emcee_method}.

\section{Results}
\label{sec:results}

\subsection{Extent and flux levels of the thermal dust emission}
\label{sec:emission}

The radial profiles and SCUBA-2 $850\micron$ images for IRC+10216 and U Hya are shown in Figs.~\ref{fig:irc10216_All}, \ref{fig:uhya_All} respectively. Similar figures for the rest of the sample are presented in the appendix \ref{appendix:figures} (see online supplementary material). The derived results are presented in Tab.~\ref{table:RadialProfile_Results_Final}. The physical size for each source was determined using the measured projected radius and distances from the literature given in Tab.~\ref{table:Targets}. This was then converted to an age for the circumstellar shell by combining it with terminal velocities determined by \citet{DeBeck2010}.

\begin{table*}
\caption{Surface Brightness Profile Results}
\begin{threeparttable}
    \begin{tabular}{llllllll}
    \hline
    \hline
    \multicolumn{1}{c}{\multirow{2}[0]{*}{Source }} & \multicolumn{1}{c}{Wavelength} & \multicolumn{1}{c}{$\theta_{3\sigma}$} & \multicolumn{1}{c}{\textit{R}$_{3\sigma}$} & \multicolumn{1}{c}{lookback time} &  \multicolumn{1}{c}{\textit{F}$_\mathrm{total}$\tnote{1}} & \multicolumn{1}{c}{\multirow{2}[0]{*}{\% \textit{F}$_\mathrm{ext}$}} &  \multicolumn{1}{c}{RMS}
    
    \\  
    & \multicolumn{1}{c}{ ($\mu$m)} & \multicolumn{1}{c}{($''$)} & \multicolumn{1}{c}{(pc)} & \multicolumn{1}{c}{(yr)} & \multicolumn{1}{c}{(Jy)} &  & \multicolumn{1}{c}{($\times 10^{-4}$ Jy arcsec$^{-2}$)}
    \\
       \hline


CIT 6 & 70    & 91 & 0.19 $\pm$ 0.06  & 9200 $\pm$ 2700 & 173.9 $\pm$ 0.3 & 47 $\pm$ 1 & 5.3 \\
          & 160   & 70  & 0.15 $\pm$ 0.05 & 7100 $\pm$ 2100 & 31.4 $\pm$ 0.1 & 50 $\pm$ 5 & 8.6\\
          & 450   & 12.0    & 0.03 $\pm$ 0.01  & 1200 $\pm$ 700 & 1.56 $\pm$ 0.02 & 28 $\pm$ 2 & 5.4 \\ \vspace{0.1cm}
          & 850   & 44.0  & 0.09 $\pm$ 0.03 & 4400 $\pm$ 1300 & 1.190 $\pm$ 0.001 & 41 $\pm$ 1 & 0.2 \\

    EP Aqr & 70    & 139 & 0.08 $\pm$ 0.01 & 6500 $\pm$ 500 & 33.6 $\pm$ 0.1 & 49 $\pm$ 1 & 3.5 \\
          & 160   & 26 & 0.014 $\pm$ 0.001  & 1200 $\pm$ 100  & 3.5 $\pm$ 0.02 & 22 $\pm$ 9 & 7.8 \\
          & 450   & --   & --  & --  & --   & -- & 5.9 \\ \vspace{0.1cm}
          & 850   & --   & --  & -- & --     & --  & 0.2 \\

    IK Tau & 70    & 83  & 0.105 $\pm$ 0.004 & 5500 $\pm$ 200 & 181.8 $\pm$ 0.4 & 26 $\pm$ 1 & 10.8 \\
          & 160   & 99  & 0.125 $\pm$ 0.005 & 6600 $\pm$ 300 & 27.5 $\pm$ 0.1 & 47 $\pm$ 5 & 8.9 \\
          & 450   & 38.0    & 0.048 $\pm$ 0.002 & 2500 $\pm$ 100 & 1.37 $\pm$ 0.07 & 32 $\pm$ 6 & 5.8 \\ \vspace{0.1cm}
          & 850   & 24.0    & 0.030 $\pm$ 0.001 & 1600 $\pm$ 100 & 0.390 $\pm$ 0.003 & 26 $\pm$ 1  & 0.2\\

    IRC+10011 & 70 & 48 & 0.17 $\pm$ 0.05 & 8500 $\pm$ 2500 & 93.0 $\pm$ 0.2 & 25 $\pm$ 1 & 2.5 \\
          & 160   & 67  & 0.24 $\pm$ 0.07 & 12000 $\pm$ 3600  & 17.70 $\pm$ 0.04 & 44 $\pm$ 3 & 7.9 \\
          & 450   & 10.0 & 0.04 $\pm$ 0.01 & 1800 $\pm$ 500 & 0.50 $\pm$ 0.03 & 29 $\pm$ 11 & 8.9\\ \vspace{0.1cm}
          & 850   & 20.0 & 0.07 $\pm$ 0.02 & 3500 $\pm$ 1000 & 0.200 $\pm$ 0.003 & 35 $\pm$ 2 & 0.2 \\

    IRC+10216 & 70.0    & 285 & 0.18 $\pm$ 0.02 & 12100 $\pm$ 1200 &  3482 $\pm$ 3 & 57 $\pm$ 1 & 3.7 \\
          & 160   & 266 & 0.17 $\pm$ 0.02 & 11300 $\pm$ 1100 & 556 $\pm$ 1 & 59 $\pm$ 2 & 6.1  \\
          & 450   & 60    & 0.038 $\pm$ 0.004 & 2600 $\pm$ 300 & 25.23 $\pm$ 0.03 & 63.5 $\pm$ 0.2 & 1.7 \\ \vspace{0.1cm}
          & 850   & 64    & 0.040 $\pm$ 0.004 & 2700 $\pm$ 300 &  10.830 $\pm$ 0.003 & 55.90 $\pm$ 0.03 & 0.1 \\

    LP And & 70    & 51  & 0.16 $\pm$ 0.05 & 11000 $\pm$ 3300 & 65.7 $\pm$ 0.1 & 32 $\pm$ 1  & 3.0\\
          & 160   & 45  & 0.14 $\pm$ 0.04 & 9600 $\pm$ 2900 & 11.50 $\pm$ 0.03 & 40 $\pm$ 4 & 8.4	 \\
          & 450   & --     & -- & -- & --  & - & 4.3 \\                              \vspace{0.1cm}
          & 850   & 20   & 0.06 $\pm$ 0.02 & 4300 $\pm$ 1300 & 0.350 $\pm$ 0.002 & 21 $\pm$ 1 & 0.2\\

    NML Cyg & 70    & 94  & 0.74 $\pm$ 0.05 & 21800 $\pm$ 1600 & 742 $\pm$ 2 & 16 $\pm$ 2 & 14.1 \\
          & 160   & 38  & 0.30 $\pm$ 0.02 & 8900 $\pm$ 700 & 111.9 $\pm$ 0.3 & 27 $\pm$ 7 & 128.7\tnote{2} \\
          & 450   & --    & --  & -- & --   & -- & 7.2 \\                           \vspace{0.1cm}
          & 850   & 20 & 0.16 $\pm$ 0.01 & 4600 $\pm$ 300  & 1.780 $\pm$ 0.003 & 19.9 $\pm$ 0.2 & 0.2 \\

    \textit{o} Ceti & 70    & 147 & 0.07 $\pm$ 0.01 & 7900 $\pm$ 900 & 229.2 $\pm$ 0.4 & 51 $\pm$ 1 & 4.0 \\
          & 160   & 93  & 0.041 $\pm$ 0.005  & 5000 $\pm$ 600 & 33.5 $\pm$ 0.1 & 50 $\pm$ 5 & 9.9\\
          & 450   & 50    & 0.022 $\pm$ 0.003 & 2700 $\pm$ 300 & 2.64 $\pm$ 0.02 & 55 $\pm$ 1 & 1.3 \\ \vspace{0.1cm}
          & 850   & 40    & 0.018 $\pm$ 0.002 & 2100 $\pm$ 200 & 0.570 $\pm$ 0.001 & 37.7 $\pm$ 0.3 & 0.1\\

    R Cas & 70    & 141 & 0.09 $\pm$ 0.01 & 6200 $\pm$ 800 & 93.8 $\pm$ 0.2 & 47 $\pm$ 1 & 16.9 \\
          & 160   & 122 & 0.07 $\pm$ 0.01 & 5400 $\pm$ 700 & 15.9 $\pm$ 0.1 & 42 $\pm$ 3 & 12.2 \\
          & 450   & 10  & 0.007 $\pm$ 0.001  & 400 $\pm$ 100 & 0.72 $\pm$ 0.01 & 30 $\pm$ 4 & 4.4 \\ \vspace{0.1cm}
          & 850   & 16 & 0.010 $\pm$ 0.001  & 700 $\pm$ 100 & 0.270 $\pm$ 0.002 & 21 $\pm$ 1 & 0.2 \\

    R Leo & 70    & 118 & 0.04 $\pm$ 0.01 & 4400 $\pm$ 800 &  71.9 $\pm$ 0.2 & 27 $\pm$ 2 & 4.5 \\
          & 160   & 35  & 0.012 $\pm$ 0.002 & 1300 $\pm$ 300 & 11.60 $\pm$ 0.04 & 17 $\pm$ 6 & 8.5\\
          & 450   & 12  & 0.004 $\pm$ 0.001 & 500 $\pm$ 100  & 1.09 $\pm$ 0.01 & 38 $\pm$ 2 & 3.2 \\ \vspace{0.1cm}
          & 850   & --     & --  & --  & --  & -- & 0.2 \\

    RX Boo & 70    & 67 & 0.06 $\pm$ 0.01 & 6800 $\pm$ 800 & 40.5 $\pm$ 0.1 & 45 $\pm$ 1 & 1.6 \\
          & 160   & 32 & 0.030 $\pm$ 0.003 & 3200 $\pm$ 400  & 5.80 $\pm$ 0.03 & 31 $\pm$ 8 & 11.5 \\
          & 450   & 66 & 0.06 $\pm$ 0.01 & 6600 $\pm$ 800  & 5.9 $\pm$ 0.1 & 93 $\pm$ 3 & 5.6 \\ \vspace{0.1cm}
          & 850   & 32 & 0.030 $\pm$ 0.004 & 3200 $\pm$ 400   & 0.230 $\pm$ 0.003 & 50 $\pm$ 2 & 0.2 \\

    TX Cam & 70    & 73 & 0.14 $\pm$ 0.04 & 6300 $\pm$ 1200  & 61.8 $\pm$ 0.1 & 32 $\pm$ 1 & 5.0 \\
          & 160   & 28 & 0.05 $\pm$ 0.02 & 2400 $\pm$ 700  & 7.80 $\pm$ 0.03 & 17 $\pm$ 4 & 10.3 \\
          & 450   & --  & --   & -- & --     & -- & 3.9\\                         \vspace{0.1cm}
          & 850   & 56 & 0.10 $\pm$ 0.03 & 4800 $\pm$ 1400  & 0.46 $\pm$ 0.01 & 51 $\pm$ 2 & 0.2 \\

    U Hya & 70    & 130 & 0.13 $\pm$ 0.01 & 15100 $\pm$ 700 & 37.1 $\pm$ 0.1 & 83 $\pm$ 1 & 4.4 \\
          & 160   & 125 & 0.13 $\pm$ 0.01 & 14500 $\pm$ 700 & 15.2 $\pm$ 0.1 & 92 $\pm$ 1 & 9.7 \\
          & 450   & --     & -- & --  & --  & - & 6.2\\                         \vspace{0.1cm}
          & 850   & 20  & 0.020 $\pm$ 0.001 & 2300 $\pm$100 & 0.080 $\pm$ 0.002 & 28 $\pm$ 4 & 0.2 \\

    W Aql & 70    & 69 & 0.11 $\pm$ 0.03 & 5500 $\pm$ 1700 & 55.9 $\pm$ 0.1 & 48 $\pm$ 1 & 2.9 \\
          & 160   & 54 & 0.09 $\pm$ 0.03 & 4400 $\pm$ 1300 & 10.3 $\pm$ 0.03 & 54 $\pm$ 3 & 9.2 \\
          & 450   & 64 & 0.11 $\pm$ 0.03 & 5200 $\pm$ 1500  & 17.6 $\pm$ 0.3 & 97 $\pm$ 3 & 9.8 \\  \vspace{0.1cm}
          & 850   & 80  & 0.13 $\pm$ 0.04 & 6500 $\pm$ 1900  & 1.07 $\pm$ 0.01 & 77 $\pm$ 1 & 0.2 \\

    W Hya & 70    & 141 & 0.07 $\pm$ 0.01 & 8200 $\pm$ 1000 & 183.7 $\pm$ 0.4 & 38 $\pm$ 1 & 4.1 \\
          & 160   & 86 & 0.04 $\pm$ 0.01  & 5000 $\pm$ 600 & 32.7 $\pm$ 0.1 & 36 $\pm$ 5 & 10.3 \\
          & 450   & --  & --   & -- & --    & -- & 21.9 \\                                  \vspace{0.1cm}
          & 850   & 56  & 0.028 $\pm$ 0.003 & 3300 $\pm$ 400 & 1.01 $\pm$ 0.01 & 50 $\pm$ 1 & 0.3\\

   \hline
    \hline
    \end{tabular}%
    
        \begin{tablenotes}
          \item [1] All uncertainty values of SCUBA2 F$_{total}$ carry an absolute calibration uncertainty of 10$\%$ in addition to the presented uncertainties \citep{Dempsey2013, Mairs2017}.
          \item [2] The large RMS at $160\micron$ for NML Cyg is a result of the bright background emission by the Cygnus X superbubble.
        \end{tablenotes}    
\end{threeparttable}    
  \label{table:RadialProfile_Results_Final}%
\end{table*}%

As seen in Figs.~\ref{fig:irc10216_RadialProfile}, \ref{fig:uhya_RadialProfile} and \ref{fig:cit6_RadialProfile} -- \ref{fig:whya_RadialProfile}, all PACS 70 \micron\ and 160 \micron\ residual profiles show clear extended emission and circumstellar shell structure. The PACS 70 \micron\ $\theta_{3\sigma}$ can be traced out to a range from $48\arcsec$ for IRC+10011 to $285\arcsec$ for IRC+10216. The resulting \textit{R}$_{3\sigma}$ ranged from 0.06 $\pm$ 0.01 pc for RX Boo to 0.74 $\pm$ 0.05 for NML Cyg. The circumstellar shell of R Leo was traced out an age of 4400 $\pm$ 800 yrs and was the minimum age tractable at this wavelength. The maximum tractable look-back time nearly five times greater was for NML Cyg. All values are listed in Tab.~\ref{table:RadialProfile_Results_Final}.

In all sources except IK Tau and IRC+10011, the emission appeared to be less extended in PACS $160 \micron$ 
with \textit{R}$_{3\sigma}$ traceable from 0.012 $\pm$ 0.002  pc for R Leo to 0.30 $\pm$ 0.02 pc for NML Cyg. The minimum and maximum traceable look-back times at $160 \micron$ were for IRC+10011 and U Hya respectively. At both PACS wavelengths a significant portion of the total flux was found to be emitted from the extended component. We measure the weighted average of $\%$ F$_{ext}$ at $70\micron$ as ($52.6 \pm 0.2$)$\%$ with a standard deviation of $17\%$. At $160\micron$, these results were ($64 \pm 1$)$\%$ and $19\%$ respectively. 

In the SCUBA-2 observations, the sources appear to be only marginally extended. This due to a combined effect of several factors: mainly the sensitivity differences between SCUBA-2 and PACS and beam size difference between the two filters. The larger beam size of SCUBA-2 could cause the filamentary extended emission to be blurred out and blended with the background. 

While unlikely, the lower circumstellar shell extensions seen at SCUBA-2 wavelengths may also be due to the aperture defining the source region being set to $1.5\arcmin$ during SCUBA-2 data reductions. As described in section \ref{sec:obs-SCUBA-2_reduction} we set the radius to $1.5^\prime$ as it was the value at which the combined effect of \textit{blooming} and \textit{negative bowling} was minimised.

However circumstellar shell emission at $3\sigma$ levels beyond the constraining radius for this sample is unlikely given the results for IRC+10216 and \textit{o} Ceti. Both IRC+10216 and Mira have extensions greater than $1.5^\prime$ at PACS wavelengths and both sources have observation times $\geq 21$ times the other 13 sources. IRC+10216 is also the brightest source in our sample at all wavelengths. The reduction process used for these two sources as described in section \ref{sec:CWLeo+Mira_Reduction} does not have the constraining radius set and therefore no emission is ignored. Considering all these factors we still see $\theta_{3\sigma}$ less than $1.5^\prime$ allowing us to surmise that the 13 sources reduced with the constraining radius do not have any emission beyond the $1.5^\prime$ radial limit in excess of the 3$\sigma$ level.

However, in case there is circumstellar shell emission beyond the limiting aperture radius which is unaccounted for we assume the results derived for $\theta_{3\sigma}$ and \textit{R}$_{3\sigma}$ are lower limits. This assumption also allows us to account for the fact that these two radii are dependent on the noise levels. 

While 13 of the 15 sources in our sample showed extended circumstellar shell emission at $3\sigma$ levels at $850\micron$, only 9 sources showed circumstellar shell extensions at $450\micron$, due to the relatively higher noise levels at this wavelength. The traceable projected radii for these sources varied from 16$\arcsec$ for R Cas to 80$\arcsec$ for W Aql at 850\micron. At 450\micron\ they varied from 10$\arcsec$ for IRC+10011 and R Cas to 66$\arcsec$ for RX Boo. In the case of W Aql we suspect that the detected larger extent may be a result of the background within the source aperture being enhanced during the reduction process (as described in section \ref{sec:obs-SCUBA-2_reduction}). The $\theta_{3\sigma}$ values for SCUBA-2 observations for this source are comparable to those of the PACS observations which is unlikely. Furthermore, W Aql is less extended than IRC+10216 at \emph{Herschel} wavelengths but more extended in SCUBA-2 wavelengths, providing possible evidence for the circumstellar shell being merged with the enhanced background making it difficult to separate the two. The same conditions may also result in the extensions seen for RX Boo at $450\micron$ which also has extensions greater than that of IRC+10216 at $450\micron$. 

The APEX LABoCa $870\micron$ radial profiles presented in the previous study by \citet{Ladjal2010} show that only 4 out of 9 sources are extended at $3\sigma$ brightness levels, while we see extended emission at $3\sigma$ brightness levels in the $850\micron$ SCUBA-2 observations in 13 out of 15 sources. Furthermore, for the three sources in common with \citet{Ladjal2010}; IRC+10216 (Fig.~\ref{fig:irc10216_RadialProfile}), IRC+10011 (Fig.~\ref{fig:irc10011_RadialProfile}) and \textit{o} Ceti (Fig.~\ref{fig:ocet_RadialProfile}), we are able to detect extended emission $\sim 10$\% -- $\sim 25$\% further out than \citet{Ladjal2010}. This is due to a factor of $\sim 1.5$  lower noise levels of the SCUBA-2 observations compared to that of the APEX LABoCa instrument. It is therefore clear that this factor of $\sim 1.5$ difference is significant when considering the faint outer extended emission of these sources.  

The physical radii of the circumstellar shell at 850\micron\ ranged from as low as 0.010 $\pm$ 0.001 pc in the case of R Cas to 0.16 $\pm$ 0.01 pc for NML Cyg. We trace a minimum look-back time of 700 $\pm$ 100 years for R Cas at this wavelength, while the maximum age traced out to at $3\sigma$ surface brightness level is 6500 $\pm$ 1900 years for W Aql. At 450\micron\ the least extended physical radius, 0.004 $\pm$ 0.001 pc, was observed for R Leo and the most extended source at this wavelength was W Aql with {\textit{R}$_{3\sigma}$} equal to 0.11 $\pm$ 0.03 pc. At 450 \micron, the emission ages traced out to a range of 400 $\pm$ 100 years for R Cas to 6600 $\pm$ 800 years for RX Boo.

For the sample of objects studied, we calculate a weighted averaged $\%$ F$_{ext}$ of $(54.9 \pm 0.03) \%$ at 850\micron\ with a standard deviation of $17\%$ and for 450\micron\ it was found to be  $(63.1 \pm 0.2) \%$ with a standard deviation of $28\%$. This shows that while there is little detected extended emission a significant portion of the total flux is still emitted by the extended region at these wavelengths. 

The stellar radial and circumstellar shell residual profiles also reveal sources which deviated from uniform mass-loss as well as several circumstellar shell features. They are seen in the form of bumps or enhancements in an otherwise smoothly decreasing surface-brightness profile. Sources observed to have non-uniform mass-loss or showing a significant circumstellar shell feature at PACS wavelengths, such as EP Aqr (Fig.~\ref{fig:epaqr_RadialProfile}), \textit{o} Ceti (Fig.~\ref{fig:ocet_RadialProfile}), R Leo (Fig.~\ref{fig:rleo_RadialProfile}), U Hya (Fig.~\ref{fig:uhya_RadialProfile}, W Aql (Fig.~\ref{fig:waql_RadialProfile}) and W Hya (Fig.~\ref{fig:whya_RadialProfile}) all show such enhancements. At SCUBA-2 wavelengths all 15 sources in our sample were observed to be nearly spherically symmetric and hence show no such enhancements.

\subsection{Radial variation in dust properties}
\label{sec:dust-props}

The radially dependent T, $\beta$, $\Sigma$ profiles derived for IRC+10216 and U Hya are presented in Figs.~\ref{fig:irc10216_TempDensBeta_Profile} and \ref{fig:uhya_TempDensBeta_Profile} and are discussed in detail in Sect.~\ref{sec:IRC10216} and \ref{sec:UHya}. Figs.~\ref{fig:cit6_TempDensBeta_Profile} -- \ref{fig:whya_TempDensBeta_Profile} in appendix \ref{appendix:figures} present the parameter profiles for the rest of the sample (see online supplementary material). The profiles are plotted against both radius as well as time, with the time being the number of years elapsed since the corresponding mass has been ejected by the star calculated assuming a constant outflow velocity. Due to the PSF subtraction in the central position, the innermost $\sim 16\arcsec - \sim 20\arcsec$ of the parameter profiles are unreliable.

The shape (curvature) of the fitted modified blackbody depends on both temperature and $\beta$. It has been shown that $\beta$ is well constrained by the Rayleigh-Jeans tail of the SED, i.e., by longer wavelengths ($\lambda \ \geq \ 300\micron$ ) \citep{Doty1994, Shetty2009, Sadavoy2013}. Therefore our $\beta$ results are constrained by SCUBA-2 data. The temperature at each radial point is expected to be constrained by the SED peak and Wien tail of the energy distribution \citep{Shetty2009}. The temperature profile will therefore be constrained by the shorter wavelength PACS data. The circumstellar shell dust column density could be constrained by either PACS or SCUBA-2 data.

\subsubsection{Radial Variation in $\beta$}
\label{sec:Beta_Profiles}

Outside of the region affected by PSF subtraction, we see a trend of a relatively steeply increasing $\beta$ profile for 11 out of the 15 sources, followed by the profile more or less flattening out. The sharply increasing region ranged from as low as 0.003 pc in the case of W Aql up to 0.08 pc for NML Cyg. The average physical size of this region was found to be 0.03 pc with a standard deviation of 0.02 pc. 

This steep increase in $\beta$ is most likely a result of the multiple temperature components probed along a line of sight. Along each line of sight we observe multiple dust layers and hence multiple temperature/density components. These components decrease with increasing radius meaning we probe a larger number of temperature components in the inner region and the number probed decreases as the stellar radius increases. A larger number of temperature components will result in a broader SED peak causing $\beta$ to be biased towards lower values. MCMC attempting to fit this will therefore only be able to produce a lower limit of $\beta$ within this region. As we move away from the inner region the temperature components will decrease resulting in the sharpening of the SED peak allowing MCMC to better fit $\beta$, resulting in more accurate values of this parameter with increasing radius.  

For all sources except IRC+10216  (discussed further on in Sect.~\ref{sec:IRC10216}), $\beta$ flattens out to an average of $\sim 1.95 \pm 0.01$ outside of the steep inner region. For EP Aqr, R Cas, R Leo and U Hya, the four sources without the steeply increasing inner region, the radial profile is flat from the innermost reliable radial point. It is possible that the steeply increasing region for these four sources fell within the PSF reduction affected region and hence removed resulting in a near flat profile from the start. 

At large radii, the $\beta$ profile tend to be prior dominated. As $\beta$ is is constrained by the Rayleigh-Jeans end of the SED and hence the long SCUBA-2 wavelengths, non-constraining non-detections at these wavelengths causes this effect. Therefore there are two competing effects, the multiple temperature components in the inner regions and the non-constraining non-detection in the outer regions which prevents the $\beta$ profile from being constrained accurately. 

As shown by Fig.~\ref{fig:Beta_Histogram}, the measured average of $\beta$ is consistent with the peak of the probability-density function used as our $\beta$ prior. This $\beta$ distribution is obtained from \citet{Smith2012} who found the average ISM $\beta$ of M 31 to be $1.9 \pm 0.31$, consistent with the $\beta$ values of the Milky Way ISM. Therefore as our $\beta$ profiles are prior dominated we see no evidence for the difference between dust grain size and composition in the outer circumstellar environments and the ISM dust grain sizes and compositions. Therefore in order to better constrain $\beta$ we most likely require deeper better resolved longer wavelength (i.e: SCUBA-2) observations.

\subsubsection{Temperature Radial Variation}
\label{sec:Temperature_Profiles}

The presence of detections at larger radii in the PACS observations allowed MCMC to produce well-constrained temperature radial profiles. Using these profiles we were able to discern several key features of the circumstellar shell. 

The inner region of the temperature profiles for all sources (top panels of Figs.~\ref{fig:irc10216_TempDensBeta_Profile} and \ref{fig:uhya_TempDensBeta_Profile}, and Figs.~\ref{fig:cit6_TempDensBeta_Profile} -- \ref{fig:whya_TempDensBeta_Profile} in the online supplementary material show a trend of decreasing temperature gradient with increasing radius. This is followed by a region of well constrained temperatures with an almost flat gradient. These two regions are then followed by a final region of noise dominated radial points. 

The decreasing gradient trend seen in the inner $\sim 40\arcsec$ region is consistent with temperature gradient of an optically thin centrally heated spherical dust shell. The following near flat temperature gradient region has an average temperature of  $(37 \pm 3)$ K, consistent with dust grains heated by a uniform Interstellar radiation field (ISRF). Dust within this region has the same temperature distribution as it is heated by the same uniform ISRF, giving rise to a single temperature dust component.
Therefore the temperature radial profiles allow us to estimate the region at which stellar radiation is the dominant heating source of circumstellar shell dust and at which point the ISRF takes over and becoming the dominant source. We estimate that ISRF domination begins from $\sim 0.06 \pm 0.02$ pc for our sample translating to an average circumstellar shell age of $\sim 3300 \pm 500$ years. The point at which grain heating by the central source is balanced out by grain heating due to the ISRF is dependent primarily on the stellar luminosity. Assuming a constant ISRF, brighter sources will have larger transition radii. The transition radii are presented in Tab.~\ref{table:TempTransRegions}, which are in qualitative agreement with the above hypothesis. 

\begin{table}
  \centering
  \caption{ISRF transition region}
    \begin{tabular}{llll}
    \hline
    \hline
   \multirow{2}[0]{*}{Source} & \multicolumn{3}{c}{ISRF transition region}
    \\
    
     & (arcsec) & (pc) & (yr)\\
    
    \hline 
     
CIT 6	&	44	&	0.09	&	4400	\\
EP Aqr	&	-	&	-	&	-	\\
IK Tau	&	28	&	0.04	&	1900	\\
IRC+10011	&	24	&	0.09	&	4300	\\
IRC+10216	&	68	&	0.04	&	2900	\\
LP And	&	24	&	0.07	&	5100	\\
NML Cyg	&	24	&	0.19	&	5600	\\
\textit{o} Ceti	&	36	&	0.02	&	1900	\\
R Cas	&	28	&	0.02	&	1200	\\
R Leo	&	-	&	-	&	-	\\
RX Boo	&	-	&	-	&	-	\\
TX Cam	&	-	&	-	&	-	\\
U Hya	&	52	&	0.05	&	2400	\\
W Aql	&	-	&	-	&	-	\\
W Hya	&	60	&	0.03	&	3500	\\

       \hline
    \hline
    \end{tabular}%
  \label{table:TempTransRegions}
\end{table}%

The temperature profile of five of the fifteen sources (EP Aqr, R Leo, RX Boo, TX Cam and W Aql) is noise dominated,  meaning we were unable to discern the point at which the ISRF domination occurred. The temperature profile of NML Cyg, the only RSG in the study deviates significantly from the rest of the sample and is discussed in detail in Sect.~\ref{sec:NMLCyg}. 

\subsubsection{Radial variation in the dust column density}
\label{sec:Density_Profiles}

When inspecting $\Sigma$ radial profiles we found them to be the best indicator of historical mass-loss variations. We compared these profiles to a $\Sigma$ profile expected for uniform and constant mass-loss ($\Sigma_{\rm um}$), which was generated by projecting a 3-dimensional model onto 1 dimension. The $\Sigma$ profile of all 15 sources in our sample noticeably deviated from $\Sigma_{\rm um}$, indicating that none of them experienced uniform and constant mass-loss in their recent evolutionary history. 

Ten of the 15 sources show an increase in mass-loss rate (MLR) with increasing radius and hence look-back time, meaning these AGB stars underwent higher mass-loss in the past when they were younger. While this is what we observe for the overall $\Sigma$ profile trend (i.e: large time scales) all sources also show minor modulations in their profiles indicating short time scale mass-loss variations. This increase in MLR with increasing radius could also mean that we maybe including emission from the astrosphere where AGB winds sweep up the ISM \citep{Wareing2007}. 

IRC+10216 and U Hya deviate from the increasing MLR trend with the gradients of their $\Sigma$ profiles being more steep than the $\Sigma_{\rm um}$. This points to an increasing mass-loss rate as the source evolved and it had lower mass-loss in the past (see Sect.~\ref{sec:IRC10216} and \ref{sec:UHya}). 

$\Sigma$ profiles LP And, RX Boo and TX Cam were noise dominated and therefore we could not determine the exact nature of their MLR variation. Due to the limitation of the observations and the resulting MLRs it is difficult to determine any further parameters such as the acceleration parameter as it would require much better constrained data. In order to overcome this, higher resolution data at longer wavelengths and better sampling are required in order to constrain the SED and hence the temperature. 

The density enhancements seen in these parameter profiles allowed us to determine the presence several features for several AGB stars: 

\begin{enumerate}
\item As seen by Fig.~\ref{fig:cit6_TempDensBeta_Profile}, the wave-like pattern between $20\arcsec - 70\arcsec$ in CIT 6 could be indicative of a period of $\sim 5000$ years where the MLR varied more rapidly instead of a smooth MLR variation. 
\item The density enhancement between $48\arcsec - 128\arcsec$ is clear evidence for the detached shell of U Hya which is discussed later on in the paper. 
\item The enhancement seen in Fig.~\ref{fig:waql_TempDensBeta_Profile} up to $60\arcsec$ corresponds well to the observed denser Eastern region of the circumstellar shell of W Aql confirmed by \citet{Mayer2013}. 
\end{enumerate}

By studying these density enhancements and the circumstellar shell age at which they occur, we were also able to determine an average of the maximum time scale for events of large scale MLR variation. We determined large scale MLR variations is observed up to $\sim 7400 \pm 1000$ years for the stars in our sample. These time scales are comparable to thermal pulse time scale as seen by \citet{Olofsson1990} but are much greater than the mass-loss modulations seen by \citet{Marengo2001}. The variations in total dust masses, MLRs and resulting dust-to-gas ratios are discussed further in Sect.~\ref{sec:dust-to-gas_ratios}.

\subsection{Notes on selected sources}
\label{sec:individual}

\subsubsection{NML Cyg}
\label{sec:NMLCyg}

The only Red Supergiant in our sample, NML Cyg, is understood to have its circumstellar shell shaped  by the nearby Cyg OB2 association within the Cygnus X superbubble \citep{Schuster2006NMLCyg}. The strong emission from the background superbubble has much substructure at all wavelengths and therefore blends with the circumstellar shell emission making it near impossible to discern the difference between the two. This emission is a result of the large dust content of the Cyg X superbubble \citep{Zhang2012NMLCyg}. \citet{DeBeck2010} derived a constant gas MLR of $8.7 \times 10^{-5}$ M$_{\odot}$ yr$^{-1}$ for this RSG and an outer circumstellar shell radius of 0.07 pc based on this constant MLR.  

While NML Cyg does show $3\sigma$ brightness level extensions of $\sim 1.5^\prime$ in the PACS 70 \micron\ observations, it shows little to no extension in all three of the other wavelengths. This is shown by Fig.~\ref{fig:nmlcyg_RadialProfile}. The converted circumstellar shell physical radius at 70 \micron\ is 0.74 $\pm$ 0.05 pc which is $\sim 10$ times greater than the outer circumstellar shell predicted by \citet{DeBeck2010}. Similar to our results, \citet{Cox2012} also find the MESS PACS observations of NML Cyg to not be well resolved and only provide a predicted extension of $\leq 25.4$ pc. 

The T, $\Sigma$ and $\beta$ profiles as shown by Fig.~\ref{fig:nmlcyg_TempDensBeta_Profile} are all indicative of strong background domination from very inner regions of the parameter profiles. All three parameter profiles begin to comparatively flatten out at $\sim 0.19$ pc with well constrained uncertainties. The gradient of the T and $\Sigma$ profiles however do show a small change with look-back time. The T profiles appears to have a negative gradient with increasing look-back time but is still well within the expected temperature limits for ISRF dominated dust. The $\Sigma$ profile on the other hand show a slightly positive gradient. These features suggest strong ISRF and background emission domination within this region, indicating emission from the circumstellar shell of NML Cyg being blended with the surrounding Cyg X interstellar dust.
 
\subsubsection{IRC+10216}
\label{sec:IRC10216}

The extensively studied C-rich AGB star IRC+10216 (CW Leo) is one of the brightest sources in the near-IR sky. It is also a very bright sub-mm source making it a good pointing calibrator at SCUBA-2 wavelengths. At $130\pm13$ pc \citep{McDonald2017}, it is the nearest C-rich AGB star and has a present-day mass-loss rate of $1.6 \times 10^{-6}$ M$_{\odot}$ yr$^{-1}$ \citep{DeBeck2010}. \citet{Mauron2013} carried out optical observations of IRC+10216 and found a spherical circumstellar envelope with extended emission to outwards to an angular radius of $46\arcsec$. \citet{Cox2012} shows the presence of a wind-ISM interaction bow shock at $6.6^\prime$ (0.23 pc). Using PACS $70\micron$, $100\micron$ and $160\micron$ observations, \citet{Decin2011} observe the extended dust envelope of IRC+10216 to be composed of multiple arcs extending outwards up to $\sim 5.2\arcmin$. 

\begin{figure*}
\centering
\begin{subfigure}[b]{0.8\textwidth}
  \centering
  \includegraphics[width=\textwidth]{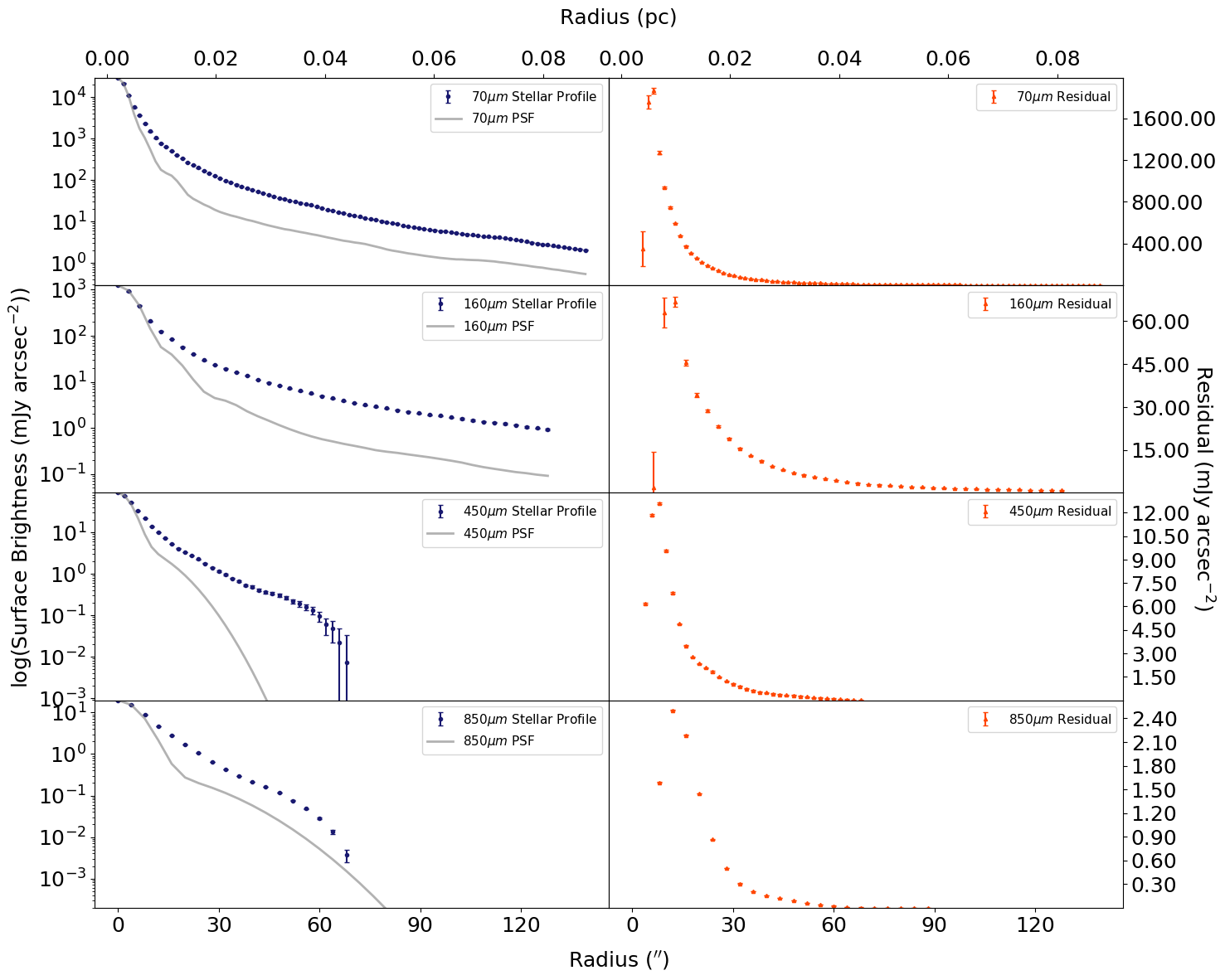}
  \subcaption{}
  \label{fig:irc10216_RadialProfile}
  \end{subfigure}
  
\begin{subfigure}[b]{0.4\textwidth}
  \centering
  \includegraphics[width=\textwidth]{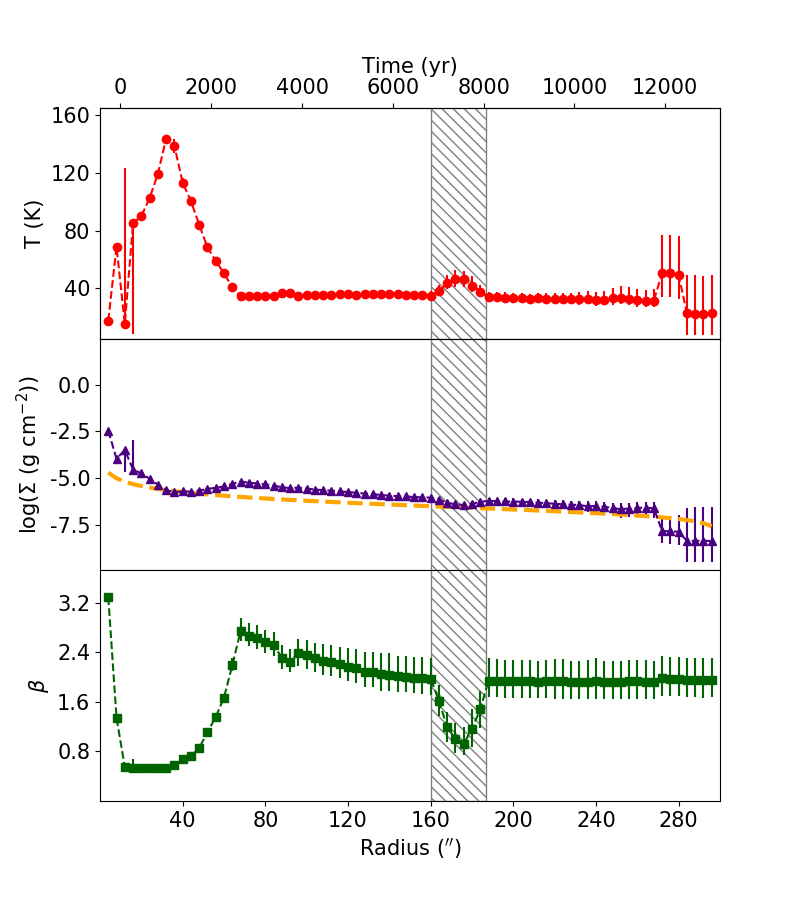}
  \subcaption{}
  \label{fig:irc10216_TempDensBeta_Profile}
  \end{subfigure}
\begin{subfigure}[b]{0.4\textwidth}
  \centering
  \includegraphics[width=\textwidth]{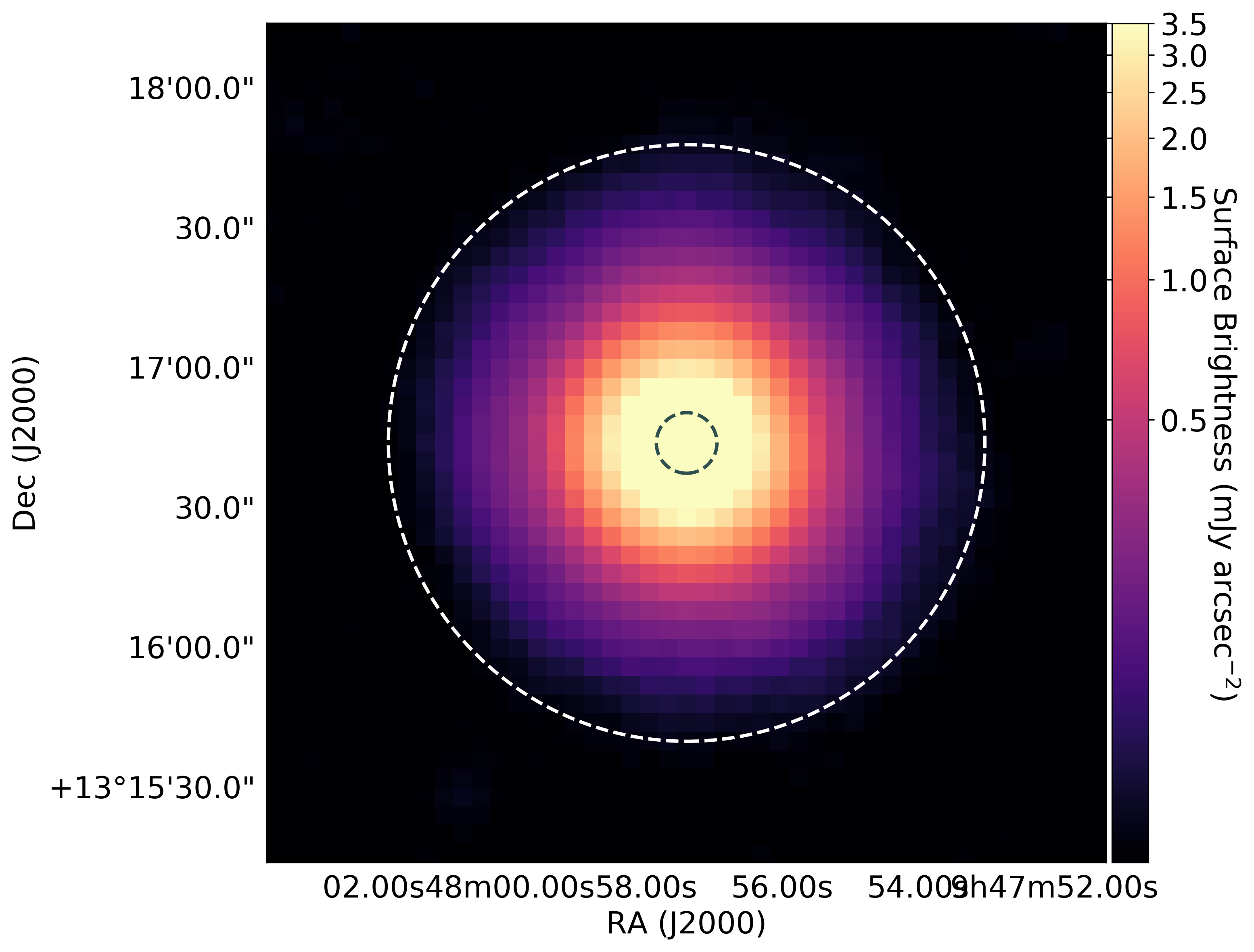}
  \subcaption{}
  \label{fig:irc10216_ContourPlot_850}
  \end{subfigure}
  \caption{Observational and modelling results for AGB star IRC+10216. (a):surface-brightness profiles (blue lines; grey lines correspond to the PSF profile) and the PSF reduced residual profiles (orange lines) as a function of projected radius ($\arcsec$) and physical size (pc) for PACS $70\micron$ and $160\micron$ and SCUBA-2 $450\micron$ and $850\micron$. The observed peak location is a result of the shape of the PSF being subtracted; (b): Profiles of the temperate (red lines), surface density (purple lines), and emissivity spectral index (green lines) as a function of projected radius and time. The surface density for a uniform mass-loss rate is also shown (yellow dashed lines). The shaded grey region highlights an artefact effect which arises due to the SCUBA-2 beam and should be ignored; (c): Reduced SCUBA-2 $850\micron$ observation of source with the FWHM of the PSF (grey circle) and the maximum extension at $3\sigma$ brightness level (white circle).}
  \label{fig:irc10216_All}
\end{figure*}

Fig.~\ref{fig:irc10216_RadialProfile} shows extended emission up to a projected radius of $\sim 1^\prime$ at the two SCUBA-2 wavelengths and $\sim 4.5^\prime$ at the PACS wavelengths. The SCUBA-2 extent seen is comparable to the optical extent seen by \citet{Mauron2013}. In addition, more than half of the emission is from the extended component at all wavelengths. The PACS extents found are smaller than those measured by \citet{Cox2012} and \citet{Decin2011} and we do not see any evidence for the dust arcs in our azimuthally averaged radial or residual profiles. This is expected as the dust arcs are non-concentric and hence will be smoothed out by azimuthal averaging.

The temperature profile shows evidence for ISRF domination from $\sim$ 0.04 pc where it flattens out to $\sim 35.2 \pm 3.2$ K. The $\Sigma$ profile has a negative overall gradient slightly steeper than $\Sigma_{\rm um}$. This is evidence for an increase in mass-loss rate over time as confirmed by both \citet{Cernicharo2014} and \citet{Dehaes2007}. 

Unlike all other sources which flatten out following the steeply increasing region discussed in Sect.~\ref{sec:Beta_Profiles}, the $\beta$ profile of IRC+10216 show a clear negative gradient prior to flattening out. This downward trend could be caused by several factors. A change in temperature of $\sim 50\%$ or more, could result in such a negative trend. As the temperature decreases the peak of the SED will shift towards the sub-mm SCUBA-2 wavelengths causing the slope of the SED, hence apparent value $\beta$, to change. However as the temperature profile is nearly flat this is unlikely. Optical depth effects could also give rise to such a trend, however again this scenario is unlikely too. If this were the case $\beta$ would evolve differently with increasing radius, i.e: the optical depth decreases with radius therefore $\beta$ would increase with radius. Furthermore, the optical depths which could cause such an effect are expected to be much higher. At sub-mm wavelengths the optical depths needs to be $\geq 1$ to result in such a change. Therefore this trend in $\beta$ could provide evidence for the dust grain properties such as size and/or composition in the circumstellar shell of IRC+10216 evolving with radius and hence changing circumstellar shell age. 

The negative gradient region of $\beta$ ranges from 2.7 -- 1.9 from the central star to the outskirts. This range differs from the beta values for amorphous carbon derived by \citet{Mennella1998}, who find values close to 1 using laboratory measurements. The dust in the outskirts of IRC+10216 is comparable to silicate dust \citep{Mennella1998} and to graphitic dust \citep{Draine2016}. The dust grain in the inner region agree with models by \citet{Jones2013} of hydrogenated amorphous carbon dust.

\subsubsection{U Hya}
\label{sec:UHya}

U Hya is a long-period semi-regular variable carbon star with a confirmed detached shell \citep{Waters1994UHya}. The detached shell of U Hya is the largest in angular size observed to date.  

It was studied in detail by \citet{Izumiura2011} using far-IR AKARI observations at 65 \micron, 90 \micron, 140 \micron\ and 160 \micron. They measure fluxes of the decomposed detached dust shell to be $45.5 \pm 0.5$ Jy at 65 \micron\ and $23.0 \pm 1.5$ Jy at 160 \micron, integrated over the total shell. \citet{Izumiura2011} utilised a detached dust shell model and derived a dust shell temperature of 40 - 50 K and a shell $\beta$ of 1.2.

\begin{figure*}
\centering
\begin{subfigure}[b]{0.8\textwidth}
  \centering
  \includegraphics[width=\textwidth]{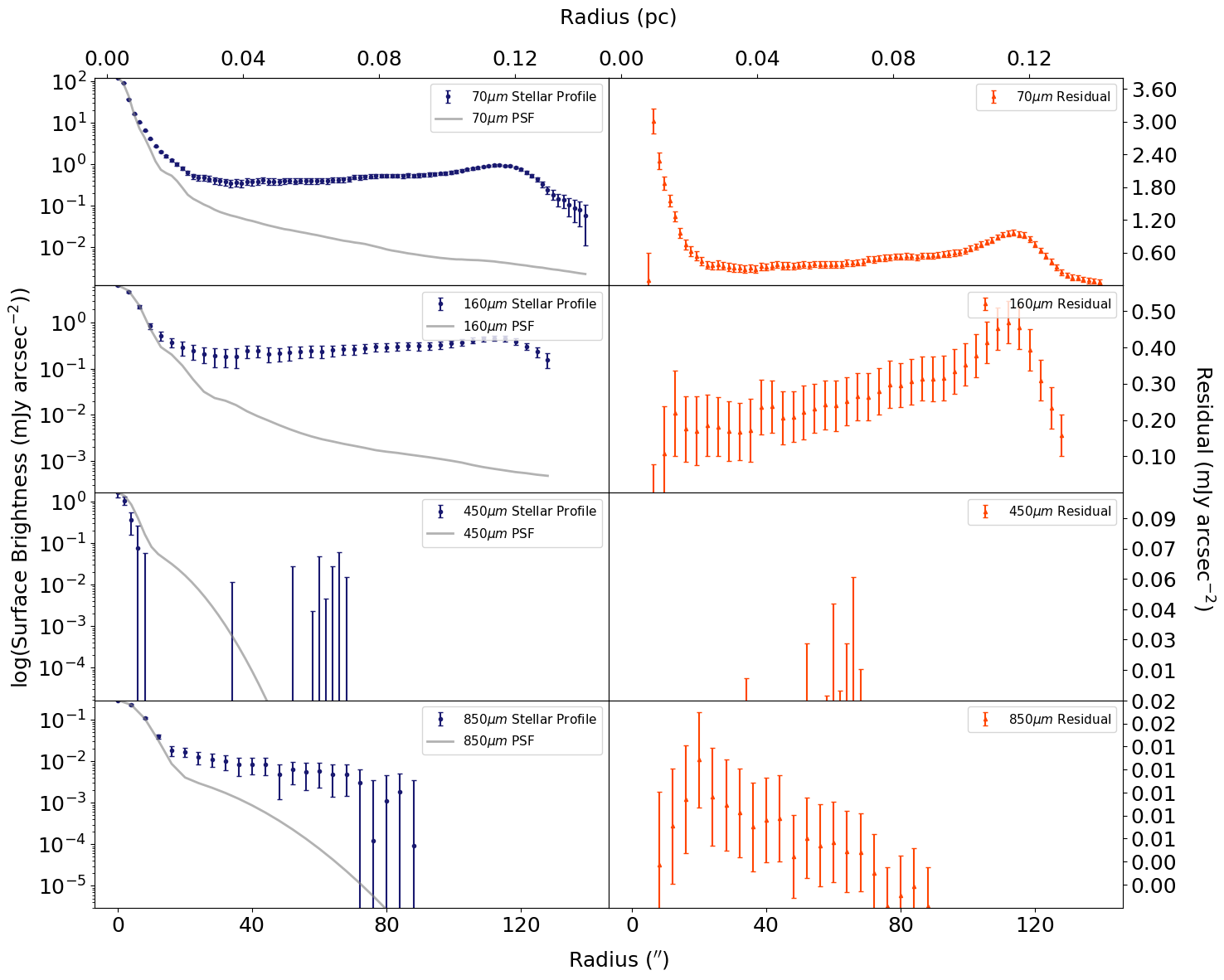}
  \subcaption{}
  \label{fig:uhya_RadialProfile}
  \end{subfigure}
  
\begin{subfigure}[b]{0.4\textwidth}
  \centering
  \includegraphics[width=\textwidth]{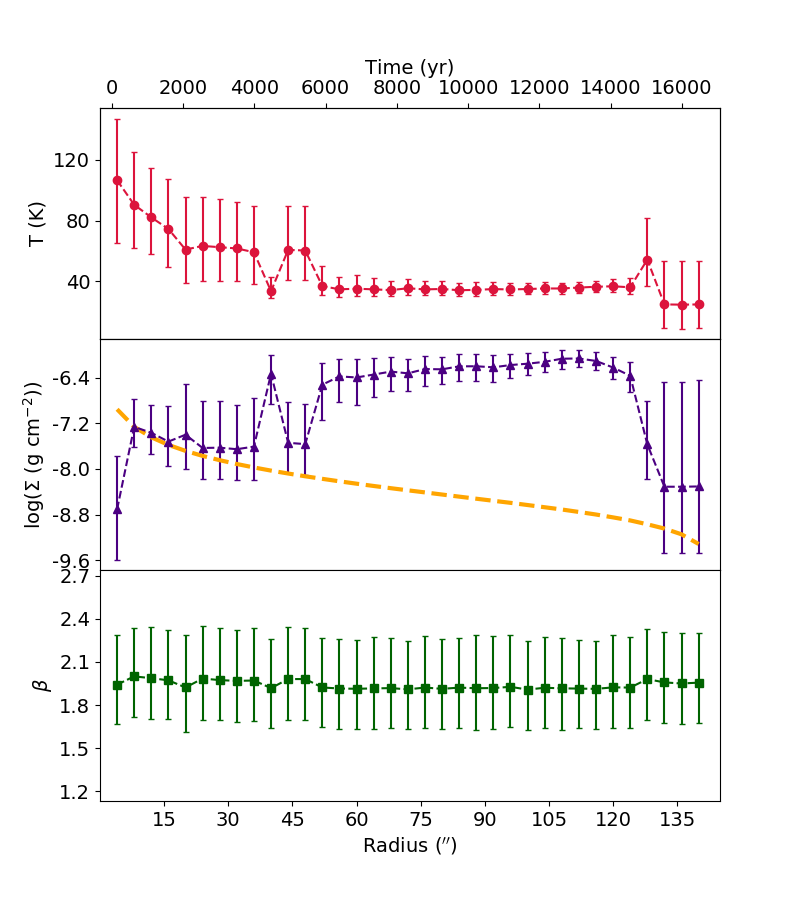}
  \subcaption{}
  \label{fig:uhya_TempDensBeta_Profile}
  \end{subfigure}
\begin{subfigure}[b]{0.4\textwidth}
  \centering
  \includegraphics[width=\textwidth]{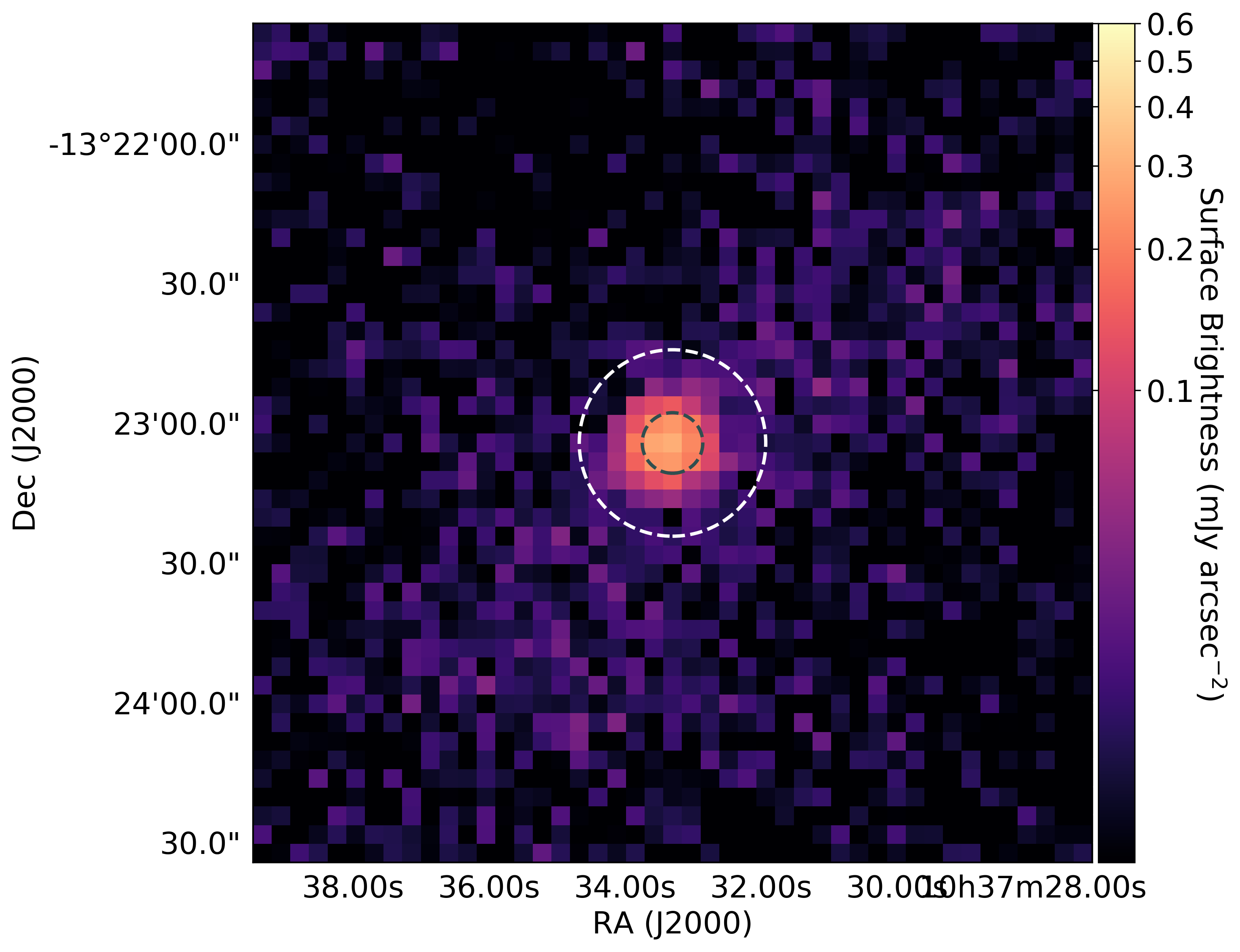}
  \subcaption{}
  \label{fig:uhya_ContourPlot_850}
  \end{subfigure}
  \caption{Observational and modelling results for AGB star U Hya. (a):surface-brightness profiles (blue lines; grey lines correspond to the PSF profile) and the PSF reduced residual profiles (orange lines) as a function of projected radius ($\arcsec$) and physical size (pc) for PACS $70\micron$ and $160\micron$ and SCUBA-2 $450\micron$ and $850\micron$. The observed peak location is a result of the shape of the PSF being subtracted; (b): Profiles of the temperate (red lines), surface density (purple lines), and emissivity spectral index (green lines) as a function of projected radius and time. The surface density for a uniform mass-loss rate is also shown (yellow dashed lines); (c): Reduced SCUBA-2 $850\micron$ observation of source with the FWHM of the PSF (grey circle) and the maximum extension at $3\sigma$ brightness level (white circle).}
  \label{fig:uhya_All}
\end{figure*}

Evidence for the detached shell is seen in the PACS stellar radial profiles and the residual circumstellar shell profile as shown by Fig.~\ref{fig:uhya_RadialProfile} as an enhancement between $60\arcsec - 132\arcsec$.  For the detached shell component we measure $\sim 30\%$ less flux at 70 \micron\ and $\sim 40\%$ less flux at 160 \micron\ compared to the AKARI 65 \micron\ and 160 \micron\ fluxes reported by \citet{Izumiura2009} (see Sect.~\ref{sec:emission}). 

The T and $\Sigma$ profiles, seen in Fig.~\ref{fig:uhya_TempDensBeta_Profile}, also provide evidence for the detached shell. The T profile has a flat single-temperature component dominated region between 0.05 pc -- 0.13 pc from the central source position. This is located in the region where the detached shell has been previously confirmed to be and is consistent with typical spherical detached shell properties. The average temperature of $35.1 \pm 0.7$ K measured in this region is slightly lower than that measured by \citet{Izumiura2011}. However it is consistent with ISRF heated dust as expected given the large distance of the detached shell from the central star. 

The $\Sigma$ profile, which deviates significantly from the $\Sigma_{\rm um}$ profile, has a clear density enhancement from 0.05 pc to 0.13 pc. This corresponds to the region of the detached shell visible in PACS observations allowing us to obtain a measured width of 0.08 pc for the detached shell which fall well within the range of shell of thicknesses determined by  \citet{Izumiura2011}. 

The MLR suddenly increased between 14000 and 15000 years into look-back time. Then the slope decreases significantly moving towards the present indicating the mass-loss rate decreased as time progressed. The scenario is well matched with the MLR variations expected for a detached shell. Using these results were are able to determine that the high mass losing event, hypothesised to be a thermal pulse, which created the detached shell, occurred $\sim 14000 - 15000$ years ago. 

The $\beta$ profile is prior dominated, and hence flat due to the lack of detections in the long-wavelength observations. Therefore we can not draw any conclusions on the grain properties of U Hya using this profile. 

\subsection{Dust mass-loss Rates and Dust-to-Gas Ratios}
\label{sec:dust-to-gas_ratios}

Assuming that the gas mass-loss rate is constant, we utilised the terminal velocities measured using CO line emission and the CO MLRs presented in \citet{DeBeck2010} to obtain the CO gas mass of the extended component for each of our sources. Combining resolved extended MLRs with point source MLRs and/or assuming constant MLRs for extended sources may not be justifiable. However applying these commonly-employed methods enables us to test the effects of such assumptions.

We employed the PACS $160\micron$ $R_{3\sigma}$ extensions, presented in Tab.~\ref{table:RadialProfile_Results_Final}, as the shell radii for this purpose. This radius was chosen over the more extended PACS $70\micron$ radius as MCMC is able to better constrain $\Sigma$ and its uncertainties with two detections at a given radial point. 

We then derived the dust mass of the extended circumstellar shell components by integrating over the $\Sigma$ profile up to the same PACS $160\micron$ radii. By dividing the integrated dust mass by the age of the circumstellar shell we also derived the dust mass-loss rate which can be compared to the gas mass-loss rate. Further, by dividing the integrated dust mass by measured CO gas mass we were able to acquire the dust-to-gas mass ratios for the sources in our sample. 

\begin{figure}
    \centering
    \includegraphics[width=0.5\textwidth]{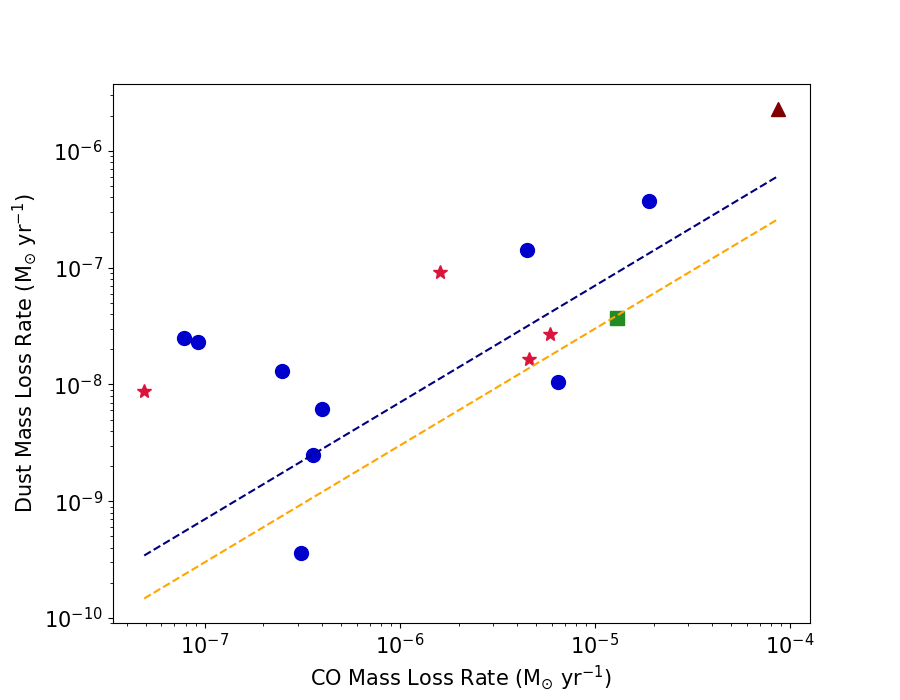}
    \caption{Gas mass-loss rate vs. Dust mass-loss rate. Blue circles: O-rich AGB stars; Crimson stars: C-rich AGB stars; Green square: S-type AGB star; Maroon triangle: RSG; Orange dashed line: the accepted dust-to-gas ratio of 1/400 for C-rich sources; Blue dashed line: the accepted dust-to-gas ratio of 1/160 for O-rich sources }
    \label{fig:MassLossRates}
\end{figure}

The resulting values are presented in Tab.~\ref{table:Distance+MassLoss} and the dust-to-gas mass-loss rate comparison is shown in Fig.~\ref{fig:MassLossRates}. When comparing the dust-to-gas ratios of sources we noticed five of the fifteen were outliers with dust-to-gas ratios $> 0.050$. The sources were IRC+10216, \textit{o} Ceti, R Leo, U Hya and W Hya, three O-rich and two C-rich sources. When calculating averaged results we therefore left these sources from consideration.  

\begin{table*}
  \centering
  \caption{Total Dust and Gas Masses and MLRs}
  \begin{threeparttable}
    \begin{tabular}{lllllll}
    \hline
    \hline
    \\
    \multicolumn{1}{c}{\multirow{2}[0]{*}{Source }} & {$\rm \dot{M}_{gas}$}  & {log (M$_{gas}$)} & {$\rm \dot{M}_{dust}$} & {log (M$_{dust}$)} & {Dust/Gas} & {$\rm \dot{M}_{GRAMS}$} \\
    
     & (M$_{\odot}$ yr$^{-1}$) & (M$_{\odot}$) & (M$_{\odot}$ yr$^{-1}$)  & (M$_{\odot}$) &  Ratio & (M$_{\odot}$ yr$^{-1}$)\\

    \hline
    CIT 6  & $5.9\times10^{-6}$ & -1.38 & $2.7\times10^{-8}$ & -3.72 & 0.005 & $1.18\times10^{-8}$ \\
    
    EP Aqr & $3.1\times10^{-7}$ & -3.43 & $3.6\times10^{-10}$ & -6.46 & 0.001 & $9.04\times10^{-10}$\\
    
    IK Tau  & $4.5\times10^{-6}$ & -1.53 & $1.4\times10^{-7}$ & -3.02 & 0.032 & $1.39\times10^{-8}$ \\
    
    IRC+10011 & $1.9\times10^{-5}$  & -0.65 & $3.7\times10^{-7}$ & -2.35 & 0.020 & $7.37\times10^{-8}$ \\
    
    IRC+10216 & $1.6\times10^{-6}$  & -1.74 & $9.2\times10^{-8}$ & -2.99 & 0.057 & $1.69\times10^{-8}$\\
    
    LP And  & $4.6\times10^{-6}$  & -1.36 & $1.7\times10^{-8}$ & -3.80 & 0.004 & $1.52\times10^{-8}$\\
    
    NML Cyg  & $8.7\times10^{-5}$  & -0.11 & $2.3\times10^{-6}$ & -1.69 & 0.026 & --\\
    
    \textit{o} Ceti & $2.5\times10^{-7}$  & -2.90 & $1.3\times10^{-8}$ & -4.19 & 0.052 & $5.28\times10^{-9}$\\
    
    R Cas  & $4.0\times10^{-7}$ & -2.67 & $6.2\times10^{-9}$ & -4.48 & 0.015 & $2.11\times10^{-9}$\\
    
    R Leo & $9.2\times10^{-8}$  & -3.91 & $2.3\times10^{-8}$ & -4.52 & 0.249 & $1.44\times10^{-9}$ \\
    
    RX Boo & $3.6\times10^{-7}$  & -2.94 & $2.5\times10^{-9}$ & -5.10 & 0.007 & $1.11\times10^{-9}$\\
    
    TX Cam & $6.5\times10^{-6}$  & -1.80 & $1.1\times10^{-8}$ & -4.59 & 0.002 & $1.21\times10^{-8}$\\
    
    U Hya & $4.9\times10^{-8}$  & -3.15 & $8.8\times10^{-9}$ & -3.89 & 0.180 & $3.00\times10^{-8}$\\
    
    W Aql & $1.3\times10^{-5}$ & -1.24 & $3.7\times10^{-8}$ & -3.79 & 0.003 & $6.43\times10^{-9}$\\
    
    W Hya & $7.8\times10^{-8}$   & -3.41 & $2.4\times10^{-8}$ & -3.90 & 0.319 & $1.20\times10^{-9}$ \\
    
    \hline
    IRC+10216 & $2-4\times10^{-5}$\tnote{(*)}  & -0.47 & $9.2\times10^{-8}$ & -2.99 & 0.003 & $1.69\times10^{-8}$\\
    \hline
    \hline
    \end{tabular}%
    
    \begin{tablenotes}
    \item The gas MLRs presented in column 2 (except $^{*}$) are from \citet{DeBeck2010}. The gas MLR of  $^{*}$ is from \citet{Cernicharo2014}. The total gas and dust masses in column 3 and 5 are measured up to PACS 160\micron\ R$_{3\sigma}$ radius given in table \ref{table:RadialProfile_Results_Final}. The dust MLR is derived by dividing the total integrated dust mass by the age of the CSE at PACS 160\micron\ R$_{3\sigma}$ radius.
    \end{tablenotes} 
    
    \end{threeparttable}
  \label{table:Distance+MassLoss}
\end{table*}%

The average dust-to-gas ratio measured for the combined sample of C-rich and O-rich is somewhat in  consistent with the canonical value of $\sim 0.005$. We measure a dust-to-gas ratio of $0.011\pm0.004$ which is however close to the canonical value at lower limits. When sources (excluding the outlying sources) are grouped by their chemistry we find the resulting dust-to-gas ratios to be similar to their respective dust-to-gas ratios within uncertainties. The canonical C-rich dust-to-gas ratio of $0.003$ measured by \citet{Knapp1985} fall well within the uncertainty limits of the averaged C-rich dust-to-gas ratio of $0.004\pm0.001$. Similarly, the \citet{Knapp1985} O-rich dust-to-gas ratio of $0.007$ is similar within error-bars to the averaged O-rich dust-to-gas ratio of $0.013\pm0.005$. 

While the average dust-to-gas ratios are somewhat consistent with their respective canonical values we also observe a significant scatter in our measured ratios, as shown by Fig.~\ref{fig:MassLossRates}. This scatter is mainly due to the uncertainties in our measurements, particularly having to do with treating the CO mass-loss rate as constant, while we allow the dust mass-loss rate to be variable. It is also a result of the varying mass-loss, grain properties and chemistry of the sources. The difference between the two canonical ratios for C and O--rich chemistry is much smaller than the scatter in the measurements. Given the intrinsic difficulty in determining the dust-to-gas ratio, and the variation from source to source, it is hard to see that the existence of two distinct values for this ratio is justified, depending on chemistry, with such a small difference between them. We conclude that it is possible to study a total population using canonical values but such values can not be applied to individual sources. 

When considering the outlying sources,  IRC+10216, \textit{o} Ceti, R Leo, U Hya and W Hya, they are all sources with complex envelopes and show clear dust mass-loss variations with time. The large variation seen in these sources and the range of ratios seen in the rest of the sample can be attributed to several reasons. Dissociation of CO could cause the gas mass measured to be significantly lower resulting in a larger than expected dust-to-gas ratio. This would particularly be a likely for sources with low MLRs. Alternatively, the \citet{DeBeck2010} gas MLRs used is only a snapshot of the central present day gas MLR as they are derived from central position CO observations, thus lacking spatial information. Further they are derived assuming uniform mass-loss. However as shown by \citet{Decin2007} and \citet{Cernicharo2015} gas mass-loss also varies with time. Considering these factors extrapolating a total gas mass from central position observations assuming uniform gas MLRs is difficult to justify. 

\citet{Cernicharo2015} measured an averaged gas mass-loss rate of $2-4 \times 10^{-5}$ M$_{\odot}$ yr$^{-1}$ for IRC+10216 over a period of 10000 years using CO observations of the circumstellar shell and hence incorporating the spatial gas MLR variations. This average rate is $\sim 20$ times greater than the central position CO MLR measured by \citet{DeBeck2012}. Replacing the \citet{DeBeck2010} MLR with this new MLR, we derive a new dust-to-gas ratio of 0.003, which is significantly lower than what was derived using the \citet{DeBeck2010} MLR and closer to the canonical dust-to-gas ratio, further proving the importance of incorporation spatial information in these calculations. It must be noted that as described before, for sources with varying expansion velocities, MLRs and symmetries (eg: detached shell source U Hya), therefore there is no reason to expect the dust-to-gas ratios to be the canonical values. IRC+10216 is observed by \citet{Cernicharo2015} to expand with a nearly constant velocity. Combined with its spherical symmetry, this means it is reasonable to expect the overall dust-to-gas ratio to match that of the central region.  

The above factors further emphasise the importance of not limiting observations of evolved stars to only the central point source, but instead to include spatial information on the source when aiming to obtain such measurements. This is the goal of a follow up study. By combining our measured dust MLRs with these new CO gas MLRs we can obtain a much more accurate and representative dust-to-gas ratio. Furthermore, the Nearby Evolved Stars Survey (NESS; \emph{Scicluna et al., in prep.}) aims to obtain accurate CO gas MLRs by observing the extended circumstellar shell component of a volume-limited subset of its sample in low-J CO lines. 

Another cause for the varying dust-to-gas ratios could be caused by the chosen ${{\kappa_{\rm eff,160}^{S}}}$ used to derive the dust masses. As discussed in Sect.~\ref{sec:sedfit} varying $\kappa_{eff}$ results in changes to the $\Sigma$ profile and hence the dust masses derived. For example, \citet{Mennella1998} observes a range of $\kappa_{eff}$ where the upper limit is twice as large as the lower for amorphous Carbon laboratory measurements at $160\micron$. Considering such a range in $\kappa_{eff}$ our dust masses will vary by a factor of two. 

We measure dust mass-loss rates ranging from $3.6\times10^{-10}$ M$_{\odot}$ yr$^{-1}$ to $2.3\times10^{-6}$ M$_{\odot}$ yr$^{-1}$ for the total sample with a standard deviation of $5.9\times10^{-7}$ M$_{\odot}$ yr$^{-1}$. We derive the dust MLRs using the GRAMS fitter for this sample and find that the dust MLRs measured by us are on average greater than the GRAMS dust MLRs. This deviation is due to GRAMS being biased by the mid-IR spectrum and therefore measuring only a very small fraction of the total dust mass. By fitting the mid-IR spectrum we are only sensitive to the hot inner region of the source and thereby we disregard the outer circumstellar shell component and only very weakly constraining and underestimating the total dust mass. Further this deviation provides possible evidence for the presence of an additional cold dust component unaccounted for by the GRAMS fitter results and which should be included when deriving more accurate dust MLRs in evolved stars.

When comparing the dust masses we derive to the results from \citet{Ladjal2010} for IRC+10011, \textit{o} Ceti and IRC+10216 we find that our dust masses are decidedly larger by factors of 26, 60 and 86 respectively. This is the result of several factors. The primary reason of this large difference is the deviation between ${{\kappa_{\rm eff,160}^{S}}}$ values used by us and \citet{Ladjal2010} as discussed in Sect.~\ref{sec:sedfit}. Further \citet{Ladjal2010} uses a fixed outer radius of $10^{3}$ times the inner radius which is approximately 30 and 4.7 times smaller than the \textit{R}$_{3\sigma}$ radii used by us as the outer for IRC+10216 and IRC+10011 respectively. This will result in a lower dust mass estimation by \citet{Ladjal2010} when compared to our measured results.

\section{Summary} 
\label{sec:summary}

We observed a sample of 14 AGB stars and one RSG using SCUBA-2 at $450\micron$ and $850\micron$ and combined these observations with published \emph{Herschel} PACS data to study their dust mass-loss histories. By analysing the radial surface brightness and residual profiles we infer the variation of the dust properties of the circumstellar shell with radii and hence time. 

The PACS observations show extended emission at both wavelengths: At $70\micron$ we observe extended emission at $3\sigma$ brightness levels from 0.04 pc to 0.74 pc averaging at $0.16 \pm 0.04$ pc. The observations allowed us to observe the mass-loss histories up to an average of $9000 \pm 1100$ years, ranging from 4500 -- 21800 years. At $160\micron$ \textit{R}$_{3\sigma}$ and look-back time averaged $0.011 \pm 0.02$ pc and $6500 \pm 1000$ years respectively. At both PACS wavelengths, the circumstellar shell structure and extended emission we measured are consistent with the results from \citet{Cox2012}. 

The sources are typically marginally resolved in the SCUBA-2 observations. The \textit{R}$_{3\sigma}$ at $850\micron$ ranged from 0.004 pc -- 0.11 pc, averaging $0.04 \pm 0.01$ pc. Despite the small $\theta_{3\sigma}$ extents we found a significant fraction of the total stellar flux was within the extended component: $\sim (63 \pm 0.02)\%$ of the total flux at 450\micron\ and $\sim (55.0 \pm 0.03)\%$ of the total flux at 850\micron\ is within found to be in the extended component. 

Outwards of the SCUBA-2 ${3\sigma}$ radius, $\beta$ is prior dominated and flattens out $1.95 \pm 0.01$. As this value is consistent with that of the ISM prior used we see no evidence for deviations from the interstellar grain properties in our $\beta$ profiles. 

All radial temperature profiles show a flattening out when the temperature has dropped to $\sim 37 \pm 3$ K, which occurs at the point where the dust is heated in equal amounts by the central sources and the ISRF. Therefore we estimate the ISRF begins dominating at $\sim 3300 \pm 500$ years for our sample.

As we expect these AGB stars to all experience roughly the same uniform ISRF, the column that is affected by the IRSF, in terms of e.g. dust heating, is comparable for each object. Therefore the transition radius could be used as another measure of the region of influence of the AGB star.

The $\Sigma$ profiles of all 15 sources deviate decidedly from the uniform (or constant) mass-loss case over the past $\sim10,000$ years. Ten sources show an overall less steep $\Sigma$ gradient compared to that of $\Sigma_{\rm um}$ indicating that their MLRs decreased as the sources evolved. Density enhancements seen in the $\Sigma$ profiles correspond to circumstellar shell features of several sources, such as the detached shell of U Hya and the denser eastern circumstellar shell region of W Aql. We estimate a time scale of $\sim 7400 \pm 1000$ years on which large scale MLR variations took place in our sample.  

The $\beta$ profile of IRC+10216 show a negative gradient before prior domination. 
The likely scenario for this feature is the variation of dust grain sizes and properties in this region hence indicating their evolution with time and radius. The $\Sigma$ profile of this AGB star indicates an increasing MLR as the star evolves differing from majority of the sources in our sample.  

We see clear evidence for the detached shell of U Hya between 0.05 pc -- 0.13 pc. The detached shell is measured to be 0.08 pc in width which is consistent with previous studies. We estimate the thermal pulse which resulted in the detached shell occurred $\sim 14000 -- 15000$ years ago.

The derived dust-to-gas ratios showed a large amount of scatter around the canonical values for C--rich and O--rich outflows, in part due to uncertainties in our measurements where we have treated the CO mass-loss rate as constant, while allowing the dust mass-loss rate to be variable. It is important to note that the difference between the canonical values for the C--rich and O--rich chemistry is smaller than the scatter seen within our sample.  Therefore we conclude that it is possible to study a total population using canonical values however, they can not be applied to individual sources. Additionally, these results emphasise the need to take into considerations the total source including the extended regions hence the importance of the spatial information of a source when deriving such measurements.  Further this also show that extending the central, present day dust-to-gas ratio to the outer older circumstellar shell regions is in general not feasible. Especially sources with large scale circumstellar shell structure and MLR variations.

\section*{Acknowledgements}

We thank Matt Smith for kindly providing the M31 $\beta$ map.
Thavisha Dharmawardena wishes to thank Prof. Chung-Ming Ko at NCU for his support of this project.
We also want to thank Jes\'us Toal\'a, whose insights in the data encouraged us to pursue this project.
This research has been supported under grant MOST104-2628-M-001-004-MY3 from the Ministry of Science and Technology of Taiwan.
Albert Zijlstra acknowledges support from the UK Science and Technology Facility Council under grant ST/L000768/1.
Jan Cami is supported by an NSERC Discovery Grant. 
The James Clerk Maxwell Telescope is operated by the East Asian Observatory on behalf of The National Astronomical Observatory of Japan; Academia Sinica Institute of Astronomy and Astrophysics; the Korea Astronomy and Space Science Institute; the Operation, Maintenance and Upgrading Fund for Astronomical Telescopes and Facility Instruments, budgeted from the Ministry of Finance (MOF) of China and administrated by the Chinese Academy of Sciences (CAS), as well as the National Key R\&D Program of China (No. 2017YFA0402700). Additional funding support is provided by the Science and Technology Facilities Council of the United Kingdom and participating universities in the United Kingdom and Canada.
\emph{Herschel} is an ESA space observatory with science instruments provided by European-led Principal Investigator consortia and with important participation from NASA.
This research used the facilities of the Canadian Astronomy Data Centre operated by the National Research Council of Canada with the support of the Canadian Space Agency.
In addition to software cited above, this research made use of the \emph{Scipy} \citep{Scipy2001} and \emph{Astropy} \citep{Astropy2018} python packages. 
We thank the anonymous referee for refereeing this paper in a timely manner and for their suggestions which lead to the improvement in its clarity.

\bibliographystyle{mnras}
\bibliography{MassLoss_Bib}

\appendix

\section{Fitting radial SEDs with MCMC}
\label{appendix:emcee_method}

We used the convenient and fast python package \textit{emcee} \citep{Foreman-Mackey2013} which employs affine-invariant Markov Chain Monte Carlo (MCMC) algorithms to carry out Bayesian inference on our observed data set and the resulting SEDs to derive the most probable $T$, $\Sigma$ and $\beta$ at each radial point. These results can be plotted as a function of radius to derive the $T$, $\Sigma$ and $\beta$ radial profiles. 

The MCMC algorithms are optimised to carry out statistical inferences such as Bayesian inference on large data sets with multiple free parameters. They utilise Markov chains to converge via random walks towards a target distribution from which random samples are generated. In our case we assumed the $T$, $\Sigma$ and $\beta$ to be free parameters. The first step of this process was to generate synthetic photometry which will later be used by the model to compare to our observed data. The synthetic photometry were generated using the surface brightness equation \ref{eq:surf-bright} and then convolving it with filter response curves for each instrument at each of the four wavelengths, which is then normalised. The SCUBA-2 filter response curves were downloaded from the EAO JCMT SCUBA-2 filters page\footnote{\url{http://www.eaobservatory.org/jcmt/instrumentation/continuum/scuba-2/filters/}}. The PACS filter response curves were downloaded from the SVO Filter Profile Service\footnote{\url{http://svo2.cab.inta-csic.es/svo/theory/fps/index.php?mode=browse&gname=Herschel&gname2=Pacs}} \citep{Rodrigo2012-PACSfilterDownload}.

At each step, the code will generate a new set of possible initial $T$, $\Sigma$ and $\beta$ values which will be read into the synthetic photometry equation in order to generate the synthetic photometry. These initial values for the free parameter values are generated in the next step of the code where the prior is generated.

In order to carry out the MCMC algorithm, we need to define the prior, likelihood and posterior as required by Bayesian statistics \citep{Hogg2010}. The prior allows the inclusion of any previous knowledge we know about our free parameters. This will be an initial estimation of the free parameters randomly determined by MCMC given some input ranges provided by us.The $\Sigma$ prior was set to be between $10^{-10}$ g cm$^{-2}$ to $10^{10}$ g cm$^{-2}$ in log scale to include a good density range. 

In the case of the $T$ prior having a simple uniform temperature range (flat prior) gave rise to a temperature profile with no trend and large uncertainties. Some of this uncertainty also bled out in to the final density values. Therefore we set the temperature prior to be a normal distribution. The mean temperature is given by 

\begin{equation}
\centering
\label{temperature_mean}
T_{\rm mean} = T_{\rm inner} \times \sqrt{\frac{r_{\rm inner}}{r_{\rm outer}}} 
\end{equation}

where, the inner temperature (T$_{\rm inner}$) of the equation was set to 1300 K. We use the publicly available GRAMS models \citep{{Sargent2011, Srinivasan2011}} to obtain an estimate for the typical range of values for (T$_{\rm inner}$) (1300 $\pm$ 40 K). The inner radius (r$_{\rm inner}$) was set to 0.03$\arcsec$ and r$_{\rm outer}$ is the outer most radial point which is being analysed. The standard deviation defining the normal distribution was set to 40 K.

The $T$ bounds were set to be in the range 2.7 K -- 300 K. The lower limit is set according to the cosmic microwave background (CMB) which has a thermal black-body spectrum at (2.72548 $\pm$ 0.00057) K \citep{Fixsen2009}. The upper limit is an arbitrary temperature reasonably expected for AGB stellar dust shells and is high enough to guarantee that we do not resolve the region where the dust actually attains this value in any of our observations.   

When defining a $\beta$ prior we ran into many challenges. As many stellar dust properties such as temperature, grain size and composition are related to $\beta$ and therefore an accurate value of $\beta$ is required in order to obtain our other two free parameters, T and $\Sigma$, we needed a prior which is derived from a well sampled $\beta$ data set.  Simply setting upper and lower limits did produce a $\beta$ profile which was artificially forced to be constrained within the given limits. Further as $\beta$ is not yet well understood for AGB circumstellar environments at these wavelengths it was difficult to determine simple meaningful flat prior. This also meant we could not use a fixed $\beta$ value. Instead, we opted to use a realistic distribution of $\beta$ values measured for an interstellar medium environment, and identified the \emph{Herschel} maps of M31 to be useful for this purpose (Smith et al.~2012). 

\begin{figure}
    \centering
    \includegraphics[width=0.5\textwidth]{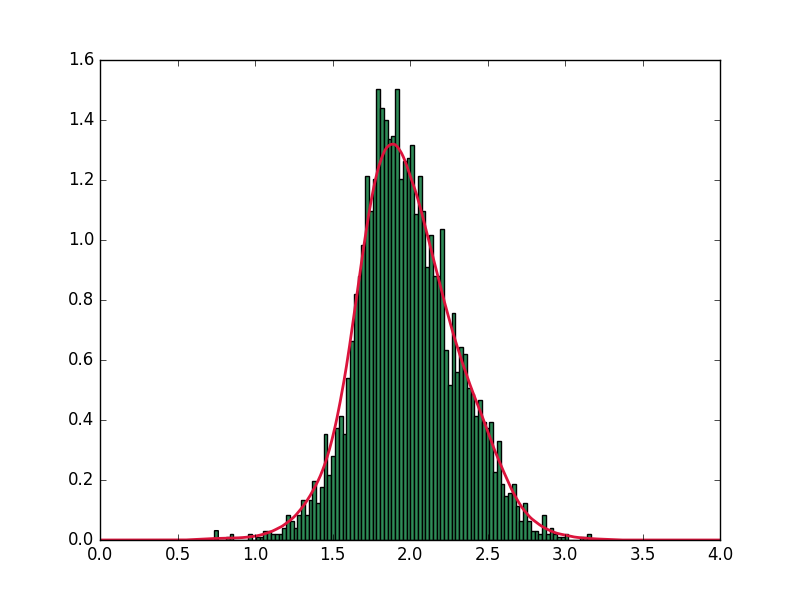}
    \caption{Normalised Histogram of the $\beta$ map of M 31 overlaid with its probability density function estimated using cosine kernel density estimation.}
    \label{fig:Beta_Histogram}
\end{figure}

The dust and gas properties of the Andromeda Galaxy (M 31) were studied at 100, 160, 250, 350 and 500 \micron\ \emph{Herschel}/PACS and SPIRE observations by \citet{Smith2012}. By fitting a single temperature modified black body model to the SED (which also included Spitzer 70 \micron\ (data) at each map pixel this study obtained the dust mass, temperature and $\beta$ results at each pixel. This then lead to a galactic $\beta$ map which we used as a well sampled and inclusive data set from which to determine our prior. By carrying out kernel density estimation on the $\beta$ sample of M31, as shown in Fig.~\ref{fig:Beta_Histogram} we produced the optimal $\beta$ probability density function which can be used as our prior. A cosine kernel was used for this purpose and the bandwidth $h$ was set according to Silverman's rule \citep{Silverman_Density_Estimation}.

We define a likelihood function which is the probability we \textit{see} our observed data is the same as the synthetic data produced by the generative model (i.e., the probability of the data set given the model parameters).  The likelihood discerns the probability that the prior generated $T$, $\Sigma$ and $\beta$ probabilities chosen by emcee for the current step will produce fluxes that match those observed once they're fed in to the generative model. Since we have data points with non-detections, especially limited by the SCUBA-2 450 observations which have no detectable extensions at $3\sigma$ levels beyond $\sim$ 30$\arcsec$ (except for the deep IRC+10216 and \textit{o} Cet maps). Therefore we defined a likelihood functions able to handle two different types of data. The first is a Gaussian function for detections, data with a well defined value (flux $\geq 3 \sigma$) with uncertainties. The second is a cumulative distribution function for non-detections, data with only an upper limit (flux $\leq 3 \sigma$). 

Finally a posterior was defined which will take the product of the prior and likelihood to produce the probability of the model given the observed data. MCMC combines the prior, likelihood and posteriors along with the observed data and run the sampler which will carry out Bayesian inference. We define $T = 28$ K, $\Sigma = 1$ g cm$^{-2}$, and $\beta$ equal to 1 plus a small random perturbation, as the initial position for the walkers required by MCMC algorithms. From this position, the walkers explore parameter space to converge on the best value for each parameter. We set the number of walkers to 100 and the number of steps with which to explore the parameter space to 10000.

An \textit{emcee} run generates a set of samples from the posterior for the model. Using the results for each of the three free parameters, we derived the median, $16^{\rm th}$ and $84^{\rm th}$ percentile values at each radial point . The median is used to approximate the centre of the distribution of samples. The $16^{\rm th}$ and $84^{\rm th}$ percentiles are its $1\sigma$ uncertainties. These final values were then used to produce the radially dependent  $T$, $\Sigma$, and $\beta$ profiles for all our sources.

\section{Figures}
\label{appendix:figures}

See online supplementary material. 


\begin{figure*}
\centering
\begin{subfigure}[b]{0.8\textwidth}
  \centering
  \includegraphics[width=\textwidth]{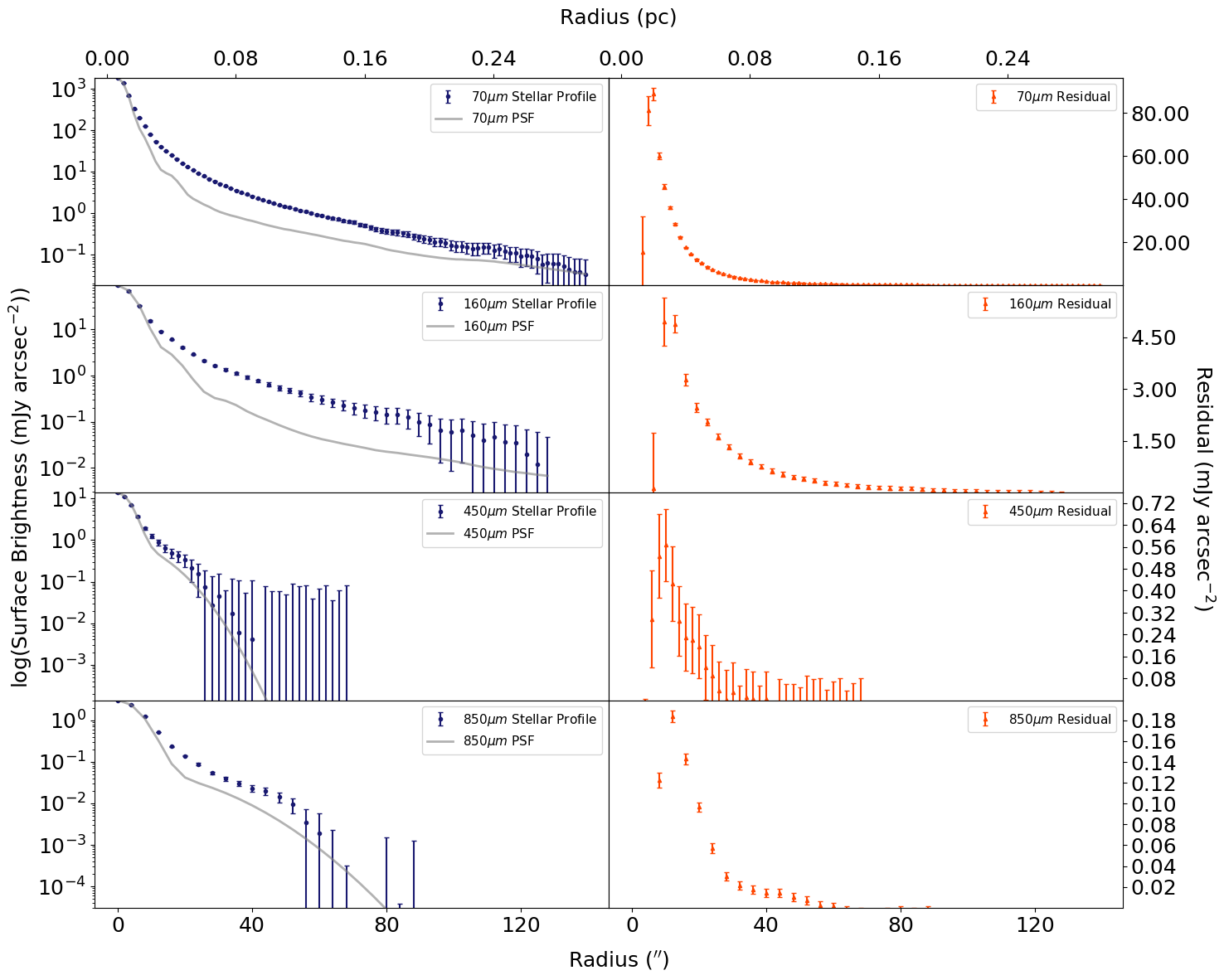}
  \subcaption{}
  \label{fig:cit6_RadialProfile}
  \end{subfigure}
  
\begin{subfigure}[b]{0.4\textwidth}
  \centering
  \includegraphics[width=\textwidth]{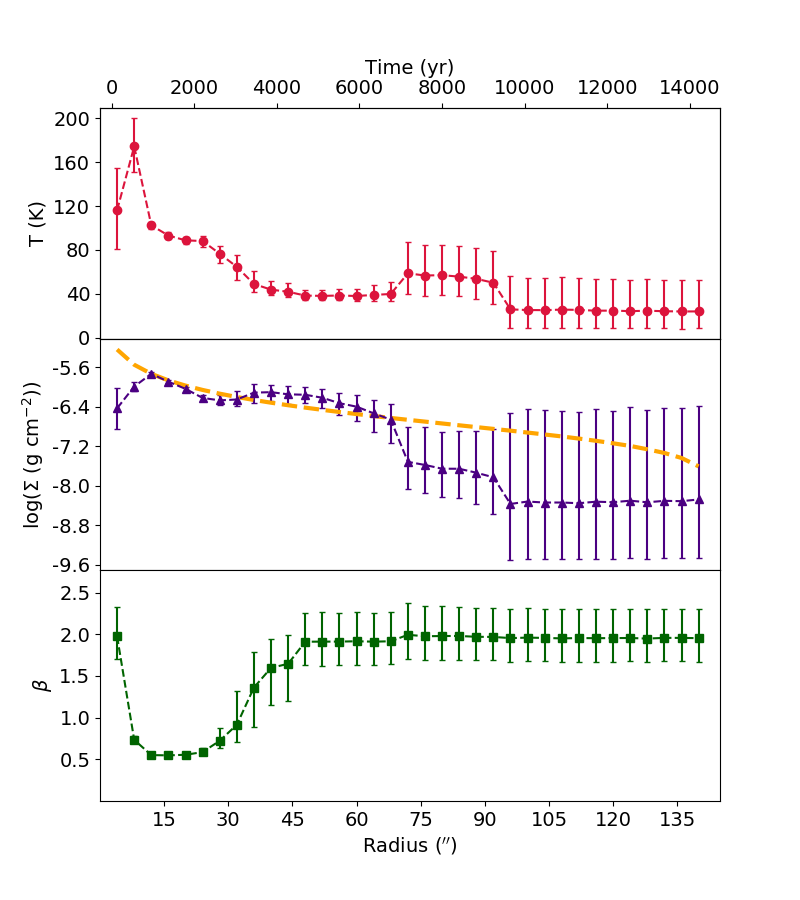}
  \subcaption{}
  \label{fig:cit6_TempDensBeta_Profile}
  \end{subfigure}
\begin{subfigure}[b]{0.4\textwidth}
  \centering
  \includegraphics[width=\textwidth]{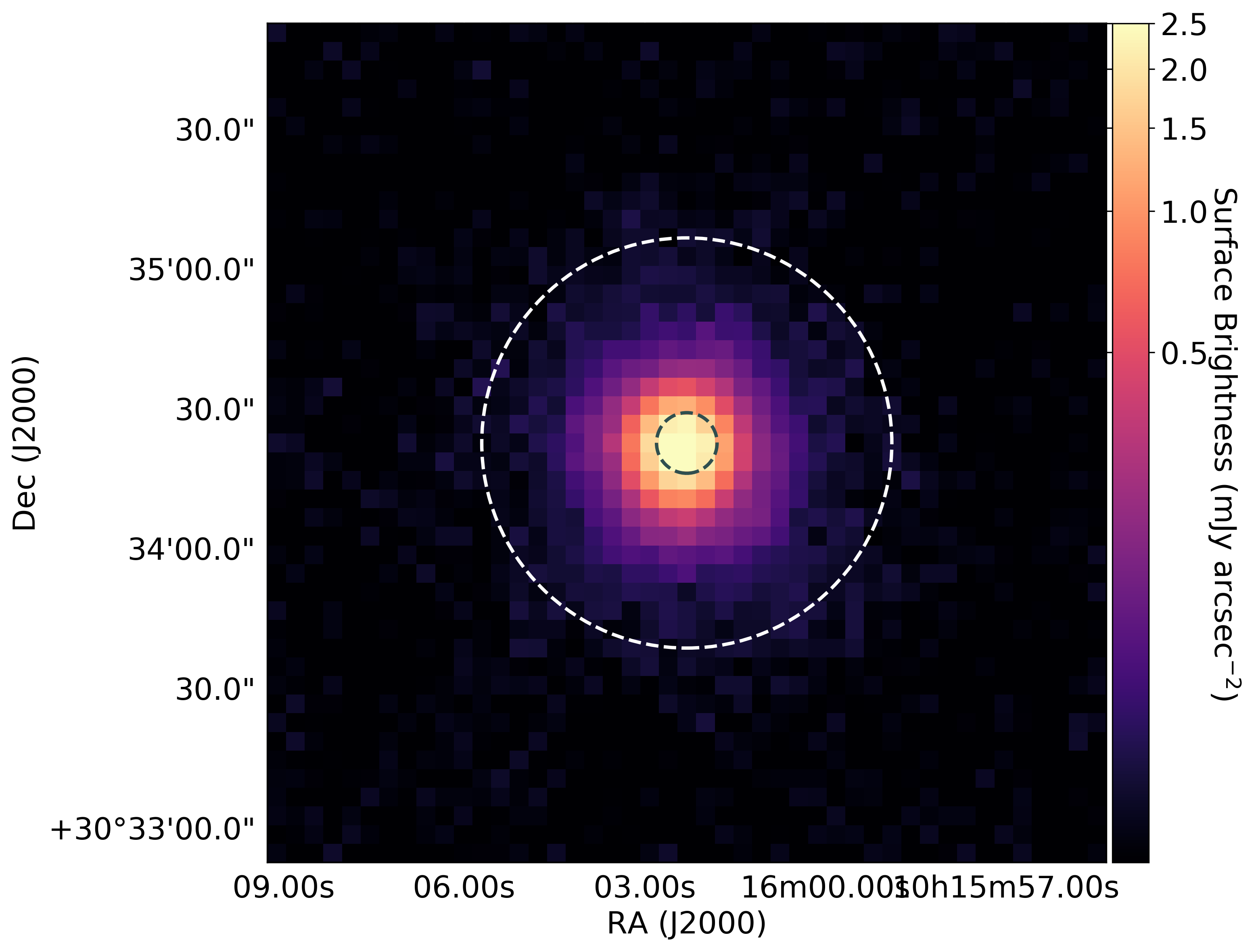}
  \subcaption{}
  \label{fig:cit6_ContourPlot_850}
  \end{subfigure}
  
  \caption{Observational and modelling results for AGB star CIT 6. (a):surface-brightness profiles (blue lines; grey lines correspond to the PSF profile) and the PSF reduced residual profiles (orange lines) as a function of projected radius ($\arcsec$) and physical size (pc) for PACS $70\micron$ and $160\micron$ and SCUBA-2 $450\micron$ and $850\micron$. The observed peak location is a result of the shape of the PSF being subtracted; (b): Profiles of the temperate (red lines), surface density (purple lines), and emissivity spectral index (green lines) as a function of projected radius and time. The surface density for a uniform mass-loss rate is also shown (yellow dashed lines); (c): Reduced SCUBA-2 $850\micron$ observation of source with the FWHM of the PSF (grey circle) and the maximum extension at $3\sigma$ brightness level (white circle).}
  \label{fig:cit6_All}
\end{figure*}

\begin{figure*}
\centering
\begin{subfigure}[b]{0.8\textwidth}
  \centering
  \includegraphics[width=\textwidth]{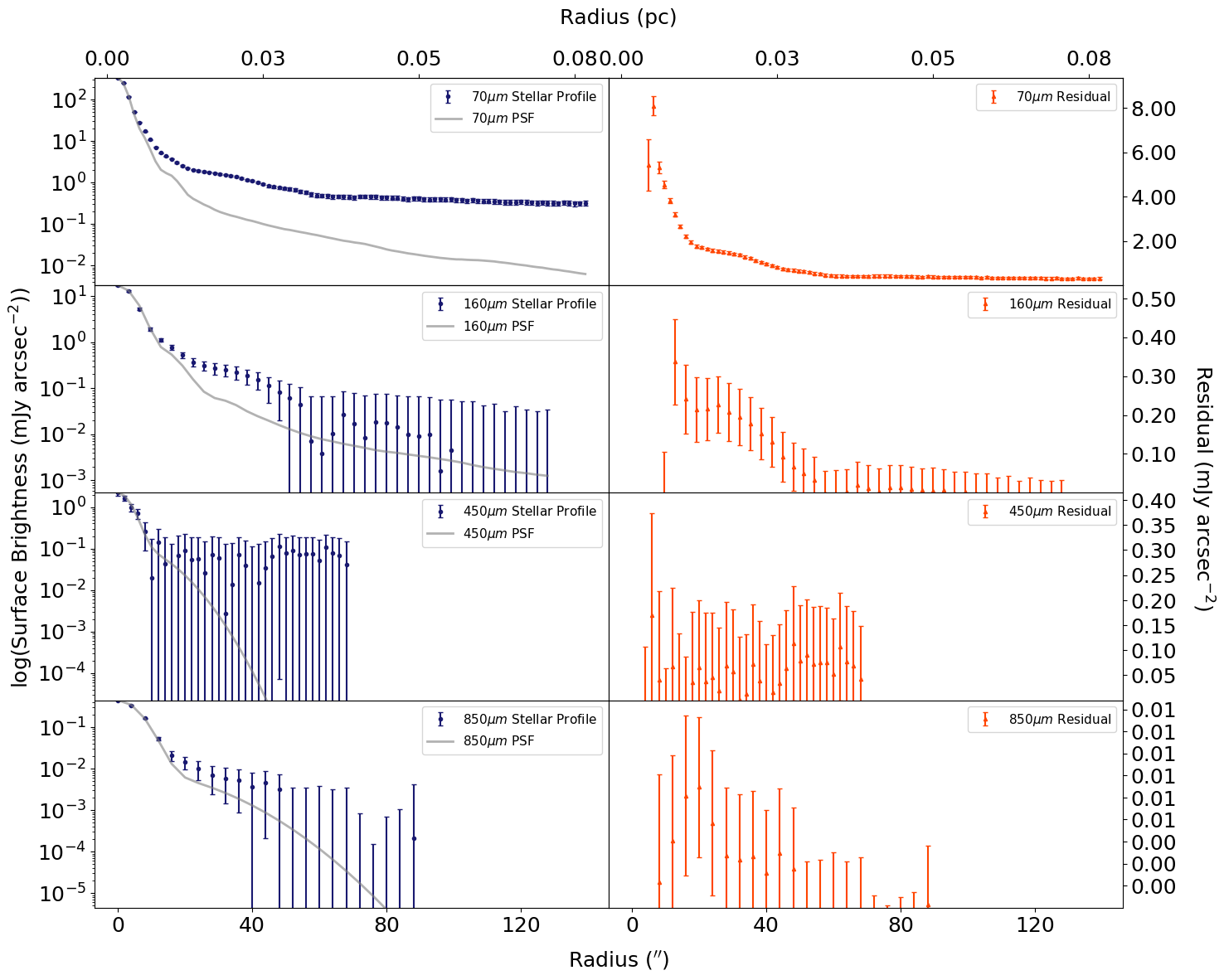}
  \subcaption{}
  \label{fig:epaqr_RadialProfile}
  \end{subfigure}
  
\begin{subfigure}[b]{0.4\textwidth}
  \centering
  \includegraphics[width=\textwidth]{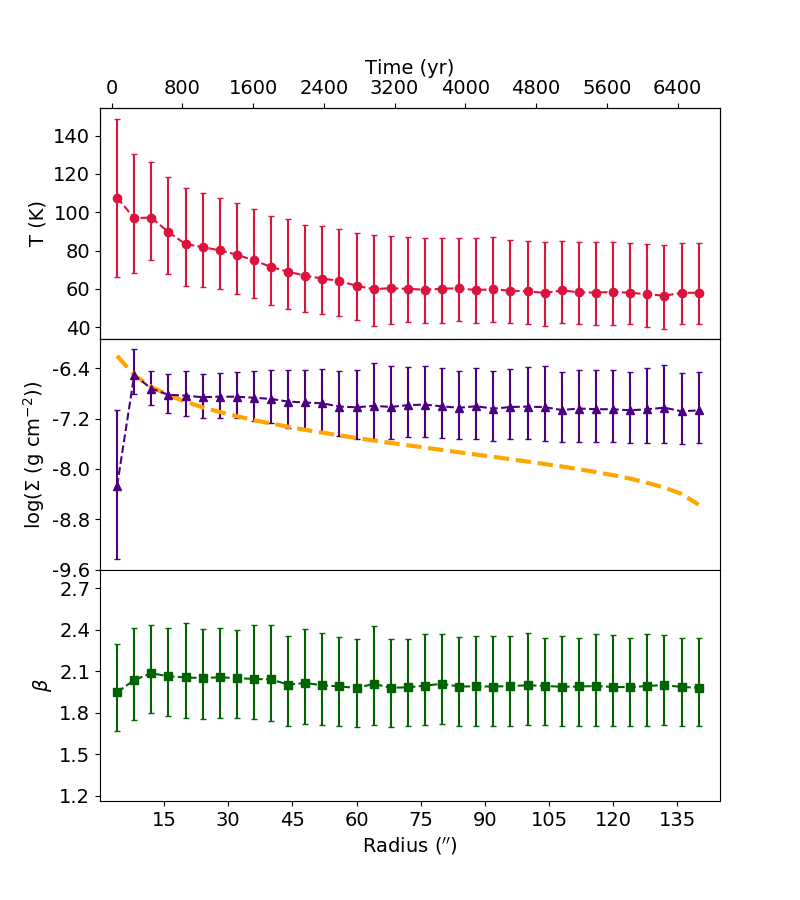}
  \subcaption{}
  \label{fig:epaqr_TempDensBeta_Profile}
  \end{subfigure}
\begin{subfigure}[b]{0.4\textwidth}
  \centering
  \includegraphics[width=\textwidth]{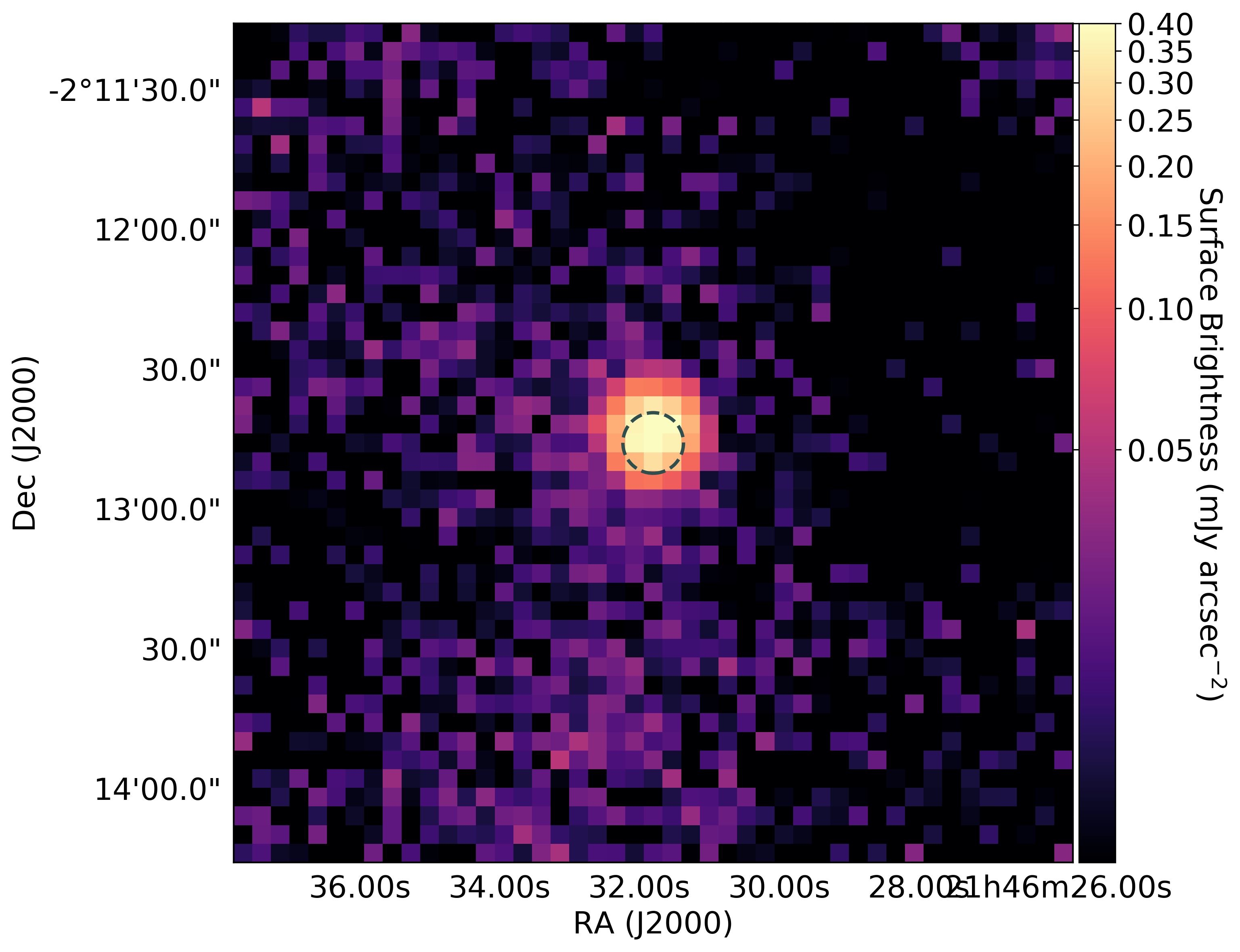}
  \subcaption{}
  \label{fig:epaqr_ContourPlot_850}
  \end{subfigure}
  
  \caption{As Fig.~\ref{fig:cit6_All} for AGB star EP Aqr}
  \label{fig:epaqr_All}
\end{figure*}

\begin{figure*}
\centering
\begin{subfigure}[b]{0.8\textwidth}
  \centering
  \includegraphics[width=\textwidth]{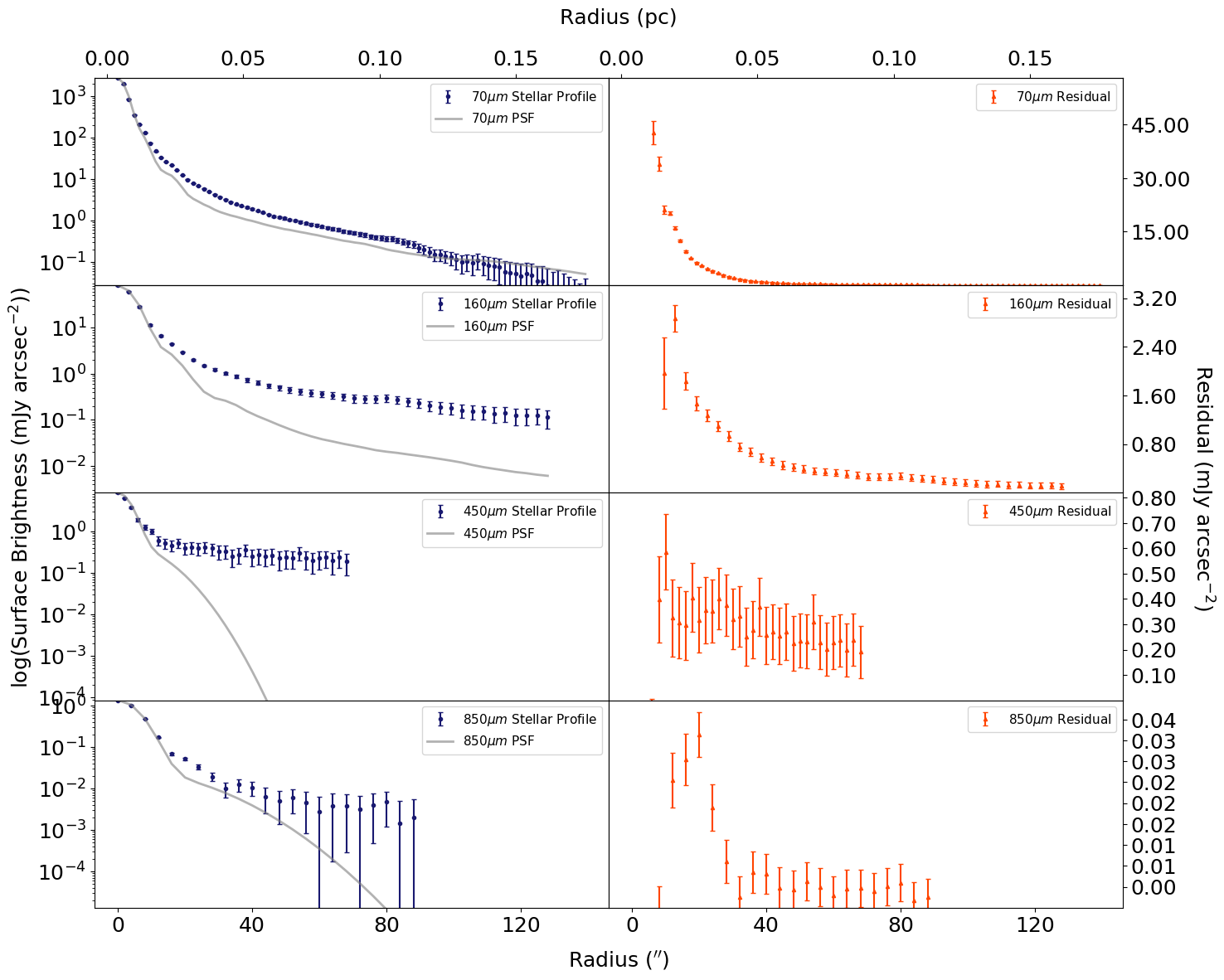}
  \subcaption{}
  \label{fig:iktau_RadialProfile}
  \end{subfigure}
  
\begin{subfigure}[b]{0.4\textwidth}
  \centering
  \includegraphics[width=\textwidth]{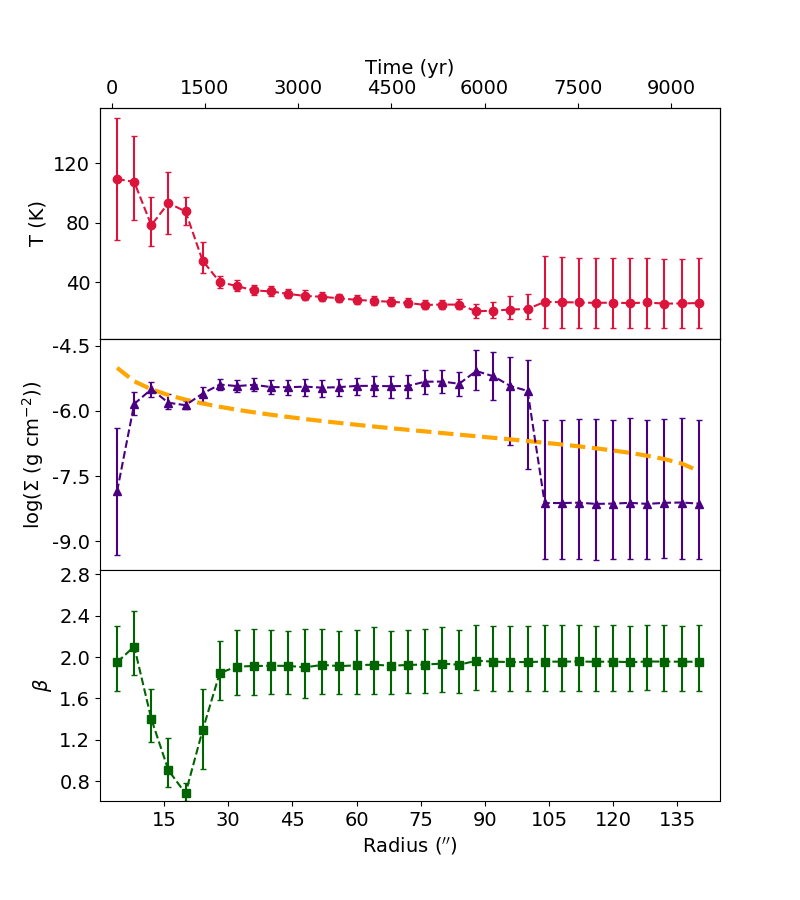}
  \subcaption{}
  \label{fig:iktau_TempDensBeta_Profile}
  \end{subfigure}
\begin{subfigure}[b]{0.4\textwidth}
  \centering
  \includegraphics[width=\textwidth]{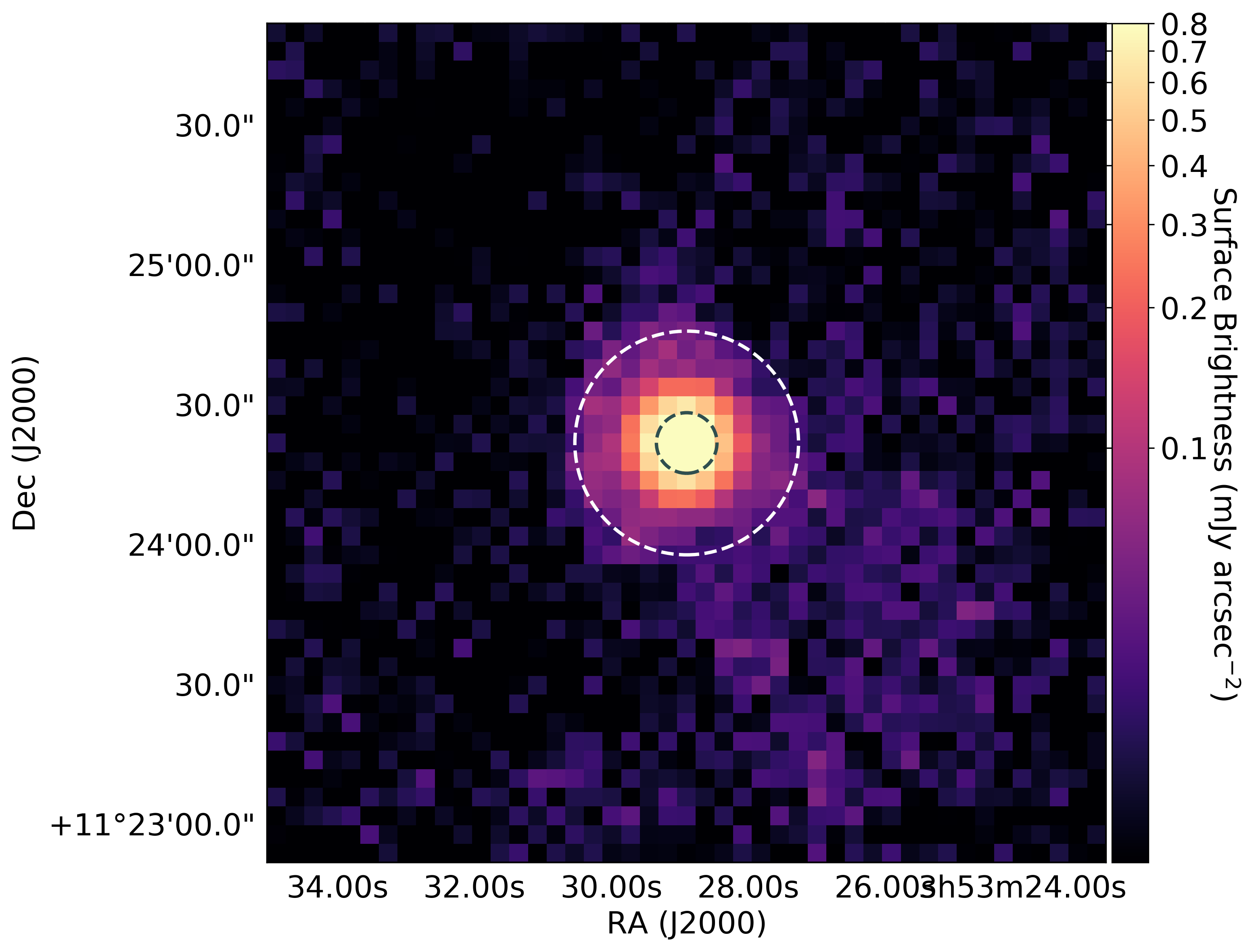}
  \subcaption{}
  \label{fig:iktau_ContourPlot_850}
  \end{subfigure}
  
  \caption{As Fig.~\ref{fig:cit6_All} for AGB star IK Tau}
  \label{fig:iktau_All}
\end{figure*}

\begin{figure*}
\centering
\begin{subfigure}[b]{0.8\textwidth}
  \centering
  \includegraphics[width=\textwidth]{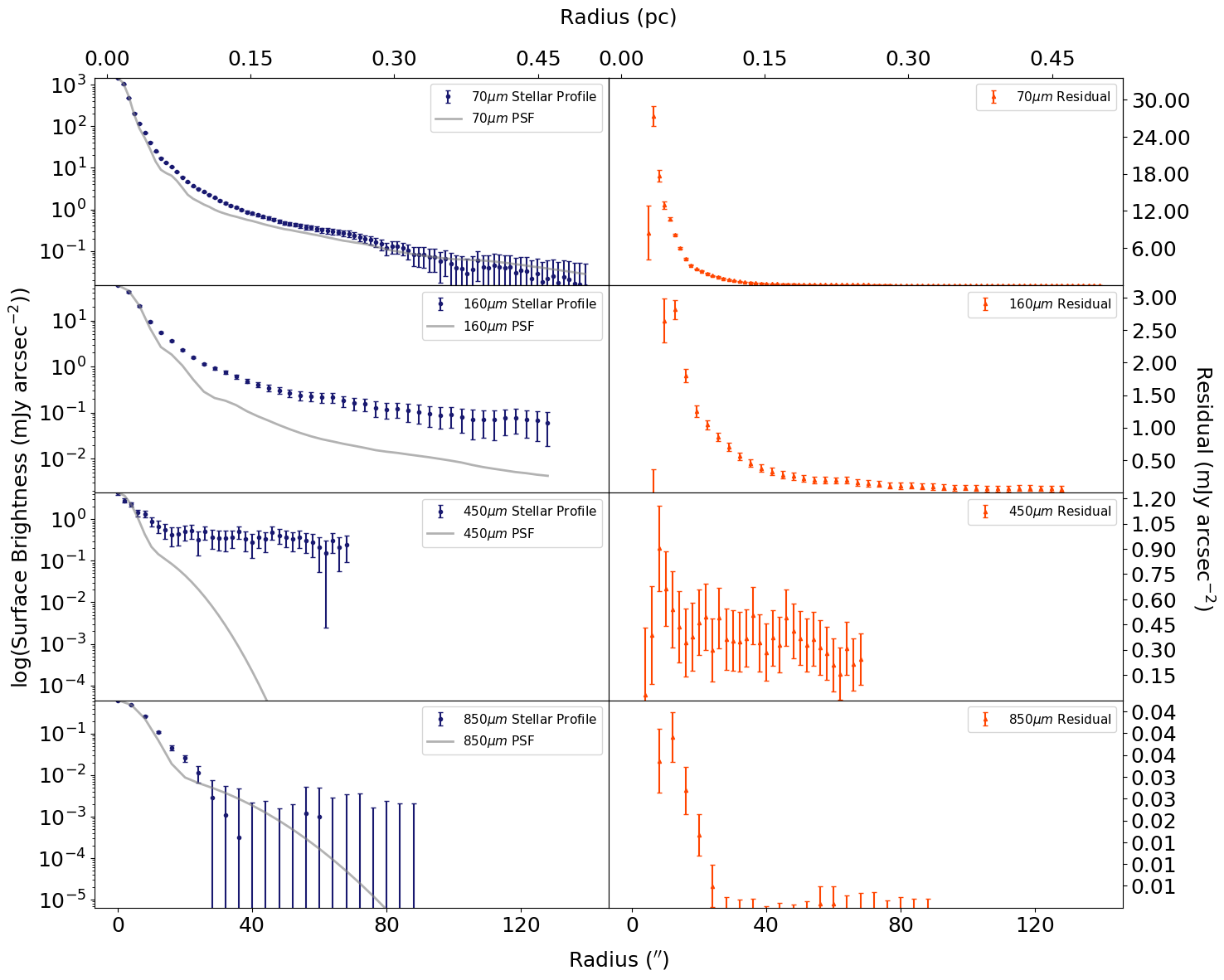}
  \subcaption{}
  \label{fig:irc10011_RadialProfile}
  \end{subfigure}
  
\begin{subfigure}[b]{0.4\textwidth}
  \centering
  \includegraphics[width=\textwidth]{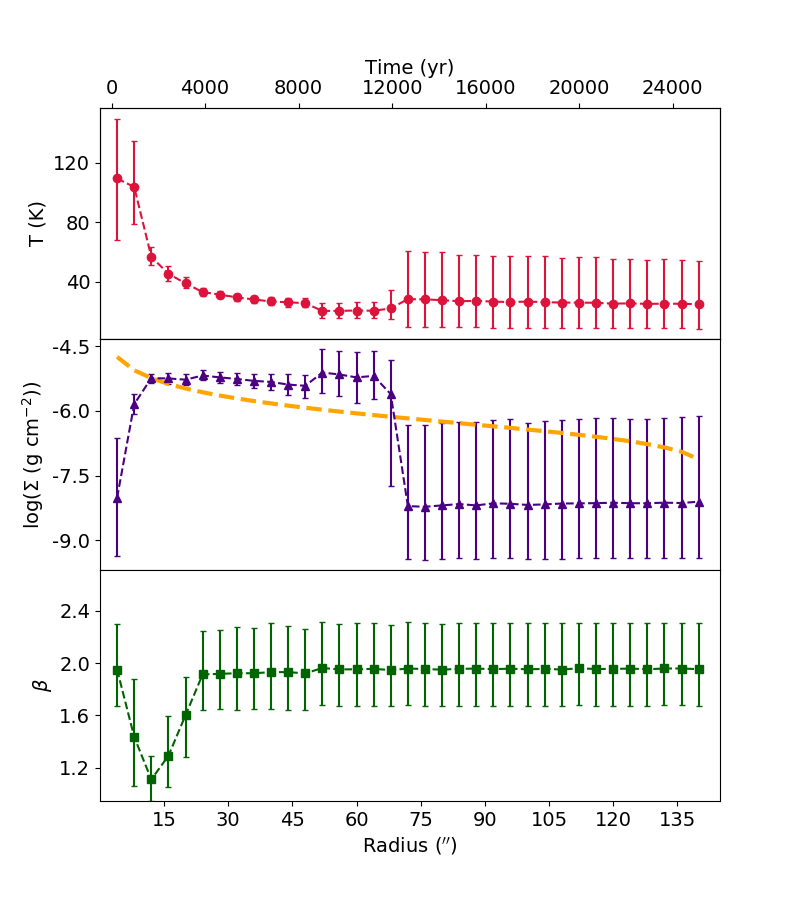}
  \subcaption{}
  \label{fig:irc10011_TempDensBeta_Profile}
  \end{subfigure}
\begin{subfigure}[b]{0.4\textwidth}
  \centering
  \includegraphics[width=\textwidth]{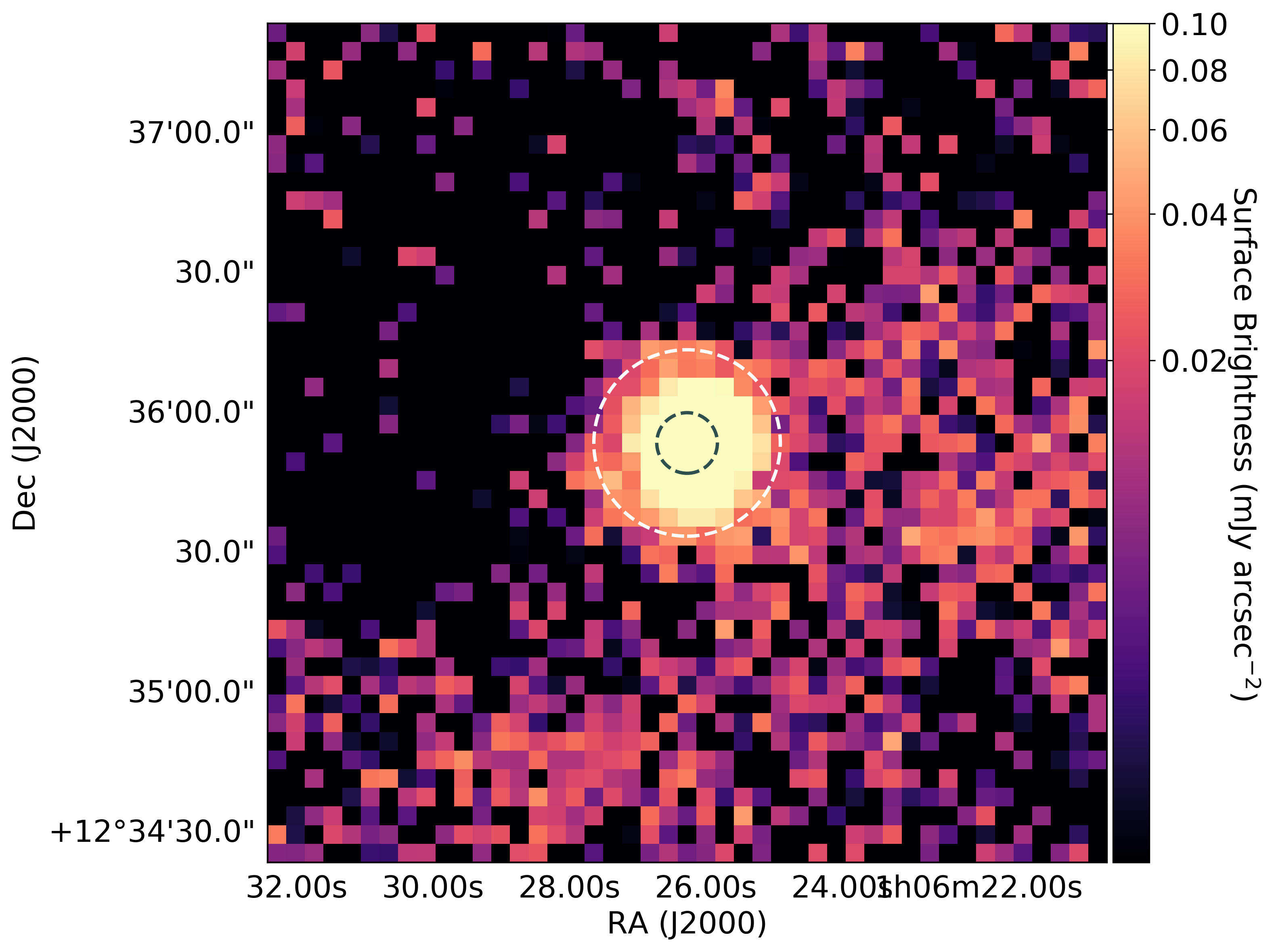}
  \subcaption{}
  \label{fig:irc10011_ContourPlot_850}
  \end{subfigure}
  
  \caption{As Fig.~\ref{fig:cit6_All} for AGB star IRC+10011}
  \label{fig:irc10011_All}
\end{figure*}

\begin{figure*}
\centering
\begin{subfigure}[b]{0.8\textwidth}
  \centering
  \includegraphics[width=\textwidth]{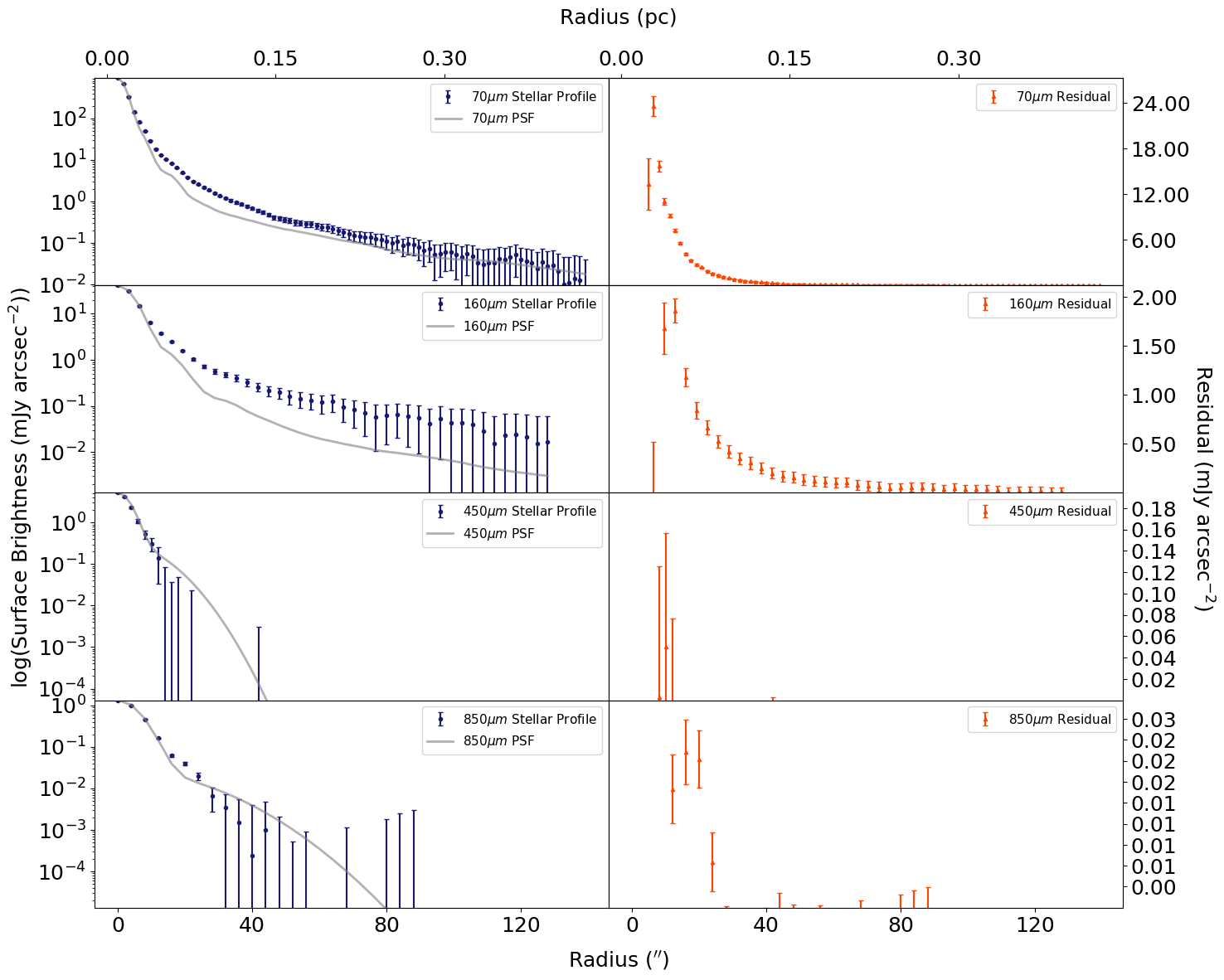}
  \subcaption{}
  \label{fig:lpand_RadialProfile}
  \end{subfigure}
  
\begin{subfigure}[b]{0.4\textwidth}
  \centering
  \includegraphics[width=\textwidth]{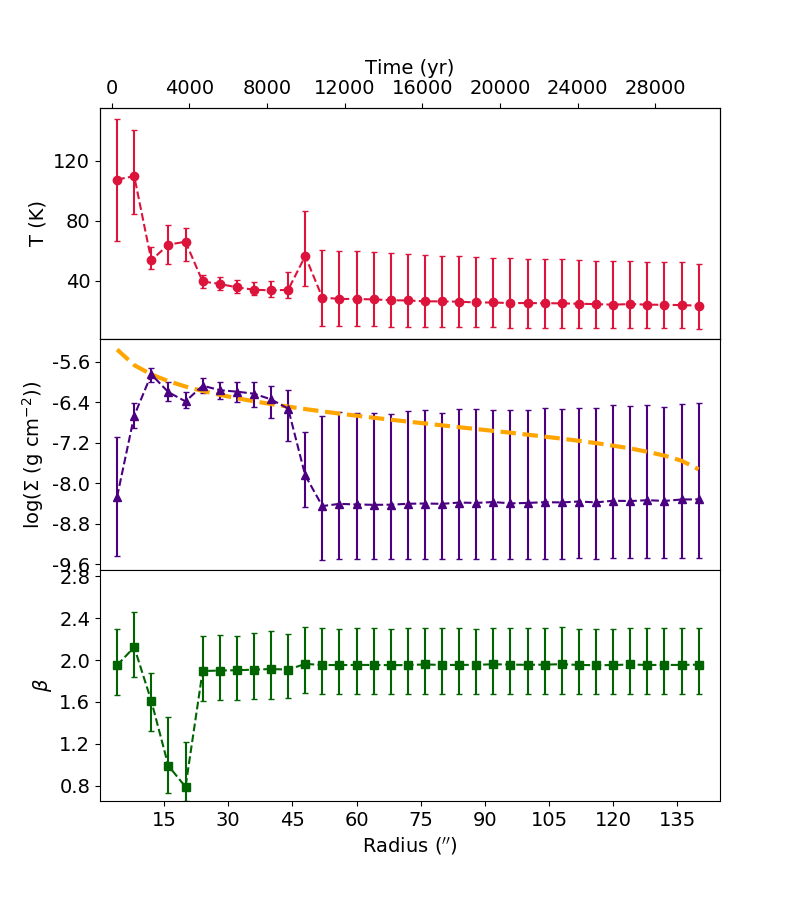}
  \subcaption{}
  \label{fig:lpand_TempDensBeta_Profile}
  \end{subfigure}
\begin{subfigure}[b]{0.4\textwidth}
  \centering
  \includegraphics[width=\textwidth]{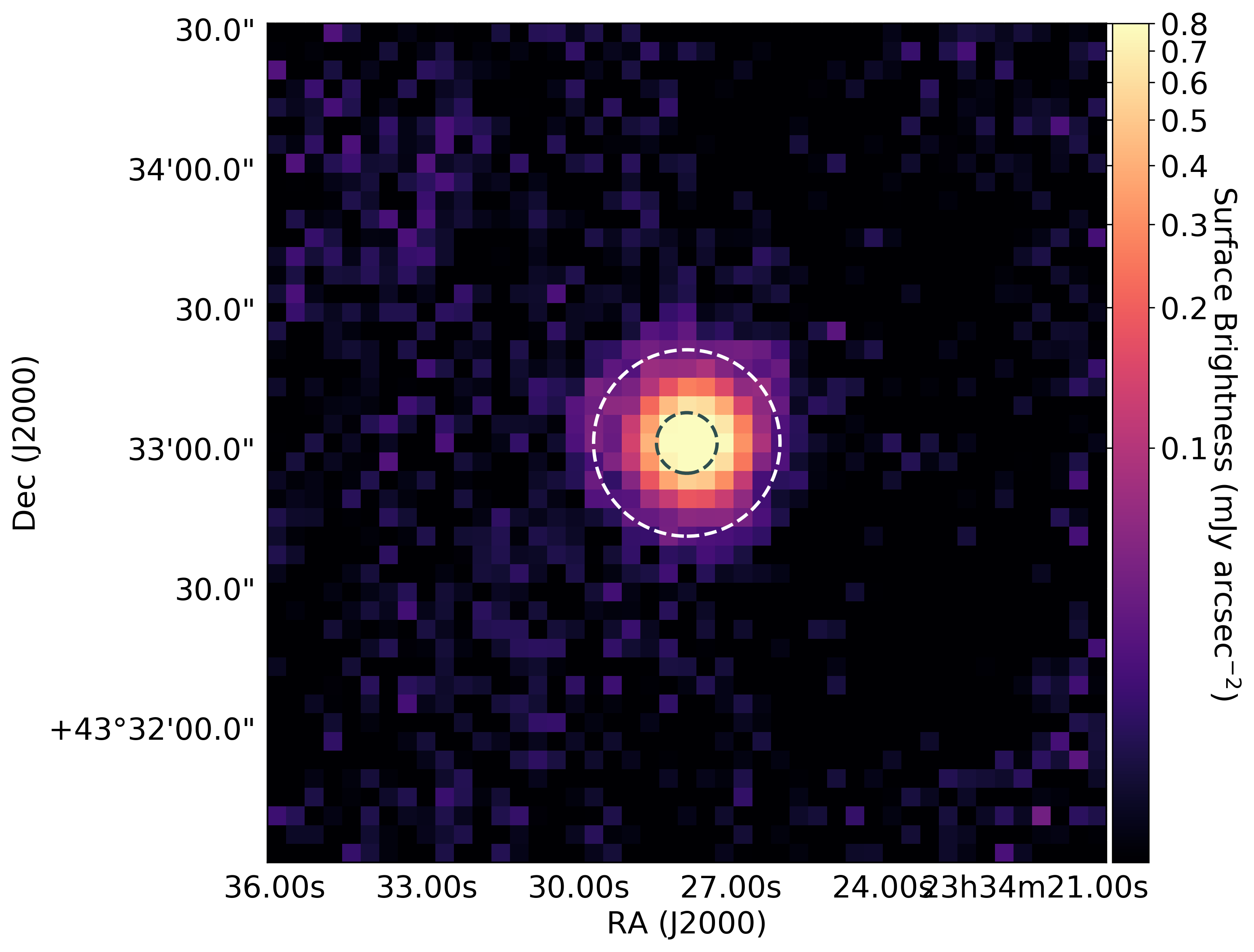}
  \subcaption{}
  \label{fig:lpand_ContourPlot_850}
  \end{subfigure}
  
  \caption{As Fig.~\ref{fig:cit6_All} for AGB star LP And}
  \label{fig:lpand_All}
\end{figure*}

\begin{figure*}
\centering
\begin{subfigure}[b]{0.8\textwidth}
  \centering
  \includegraphics[width=\textwidth]{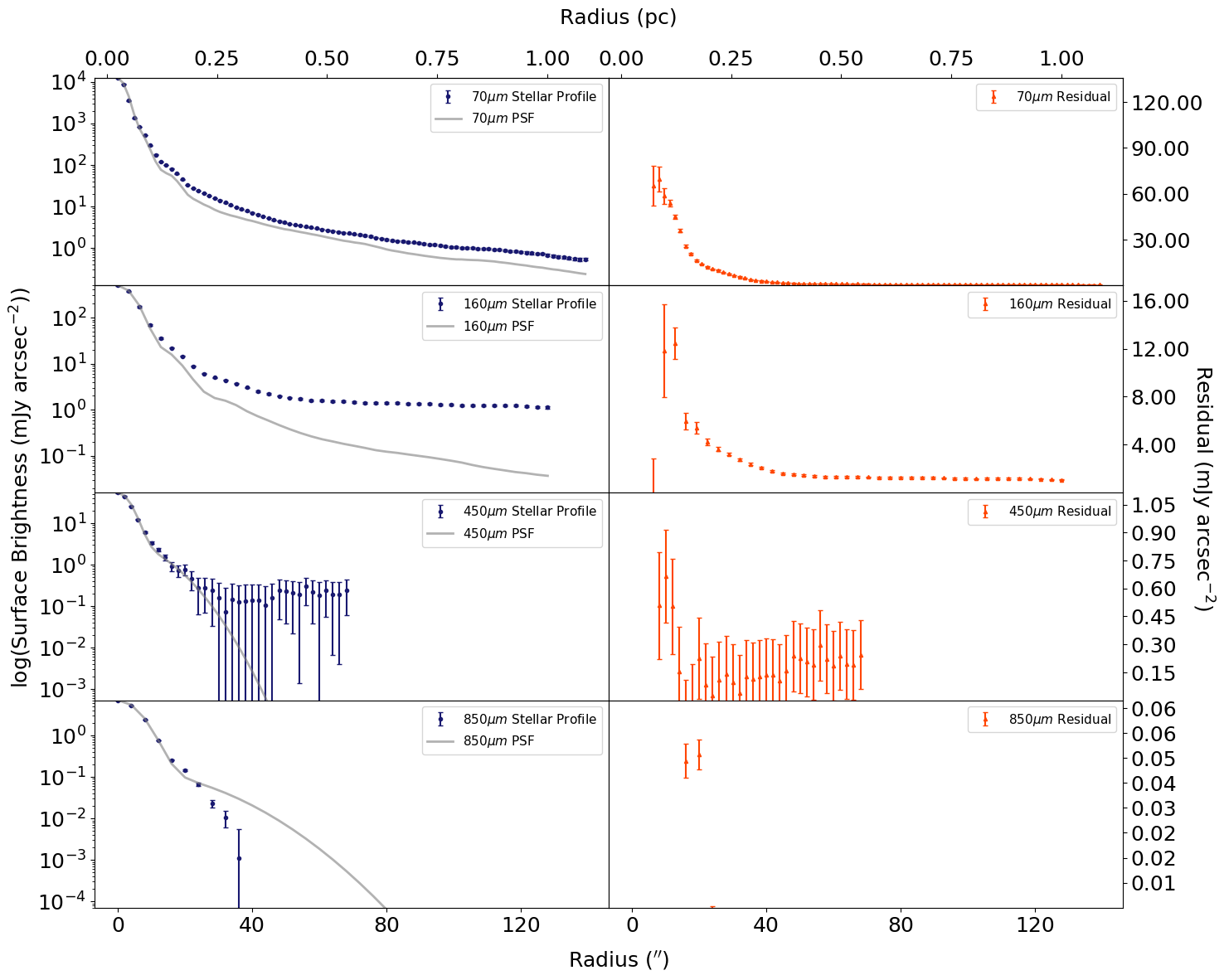}
  \subcaption{}
  \label{fig:nmlcyg_RadialProfile}
  \end{subfigure}
  
\begin{subfigure}[b]{0.4\textwidth}
  \centering
  \includegraphics[width=\textwidth]{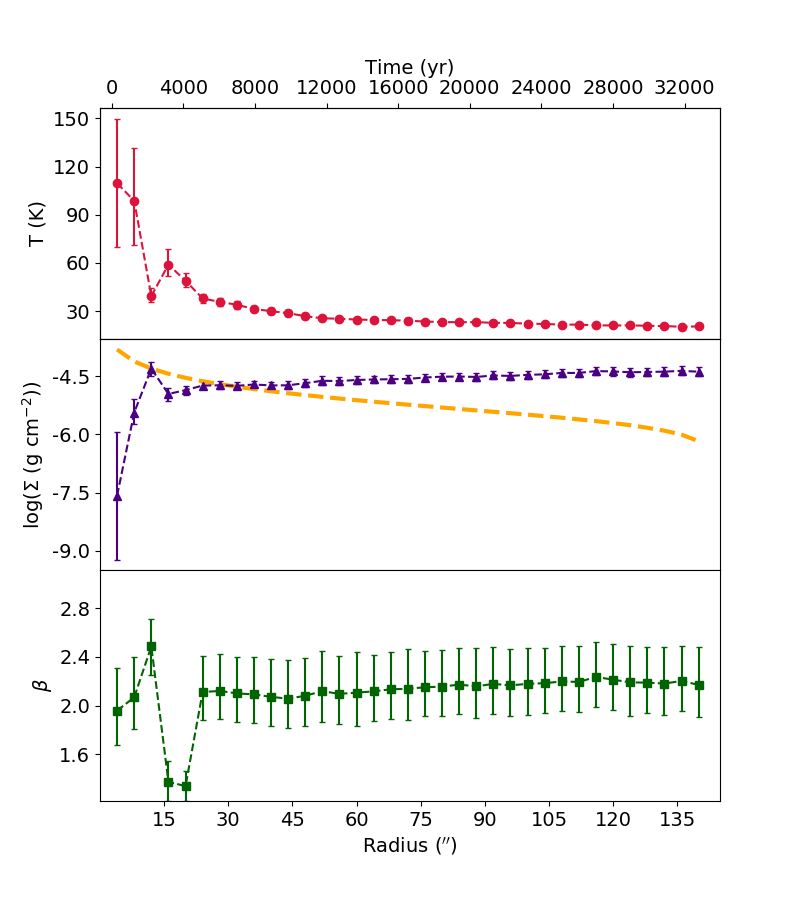}
  \subcaption{}
  \label{fig:nmlcyg_TempDensBeta_Profile}
  \end{subfigure}
\begin{subfigure}[b]{0.4\textwidth}
  \centering
  \includegraphics[width=\textwidth]{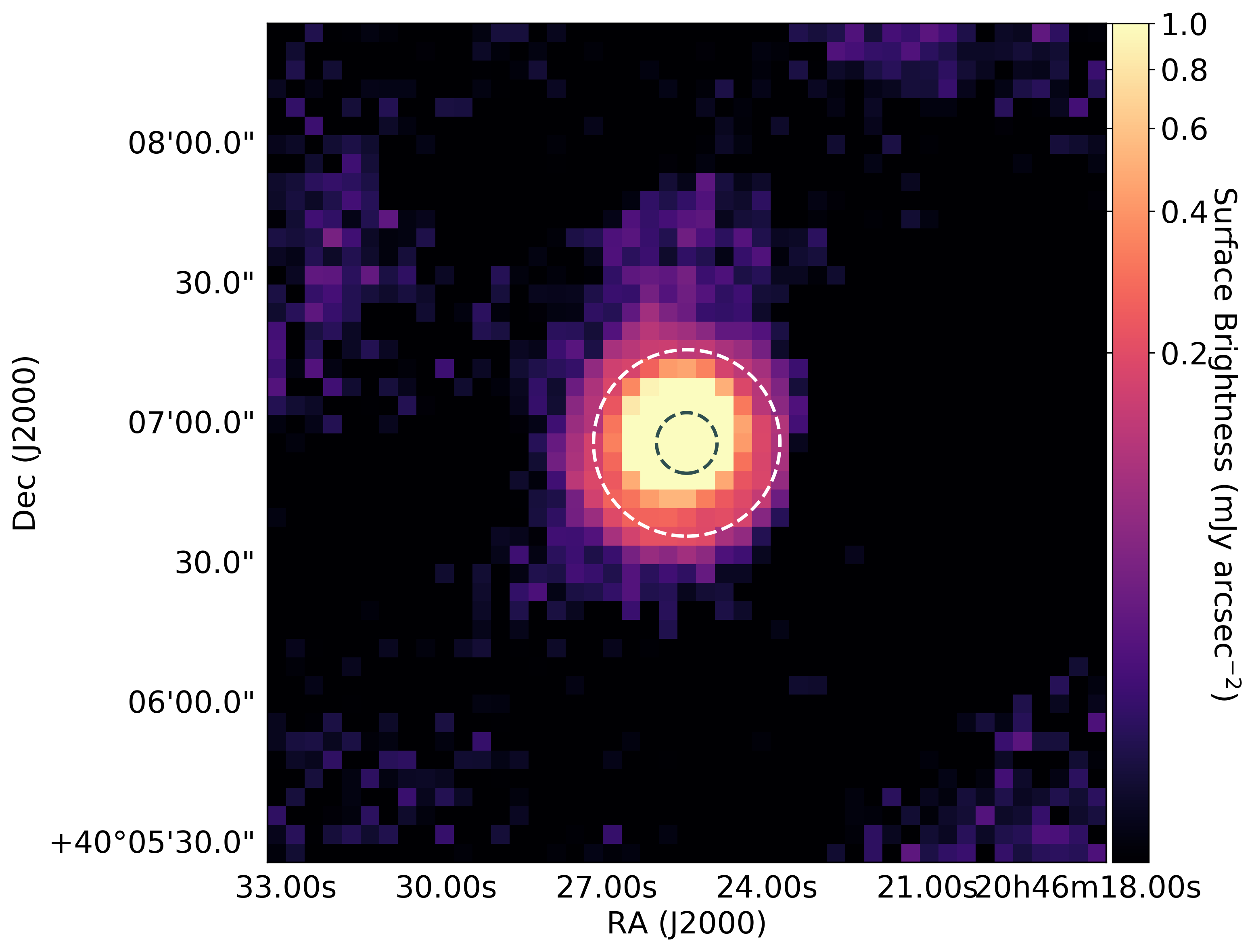}
  \subcaption{}
  \label{fig:nmlcyg_ContourPlot_850}
  \end{subfigure}
  
  \caption{As Fig.~\ref{fig:cit6_All} for RSG star NML Cyg}
  \label{fig:nmlcyg_All}
\end{figure*}

\begin{figure*}
\centering
\begin{subfigure}[b]{0.8\textwidth}
  \centering
  \includegraphics[width=\textwidth]{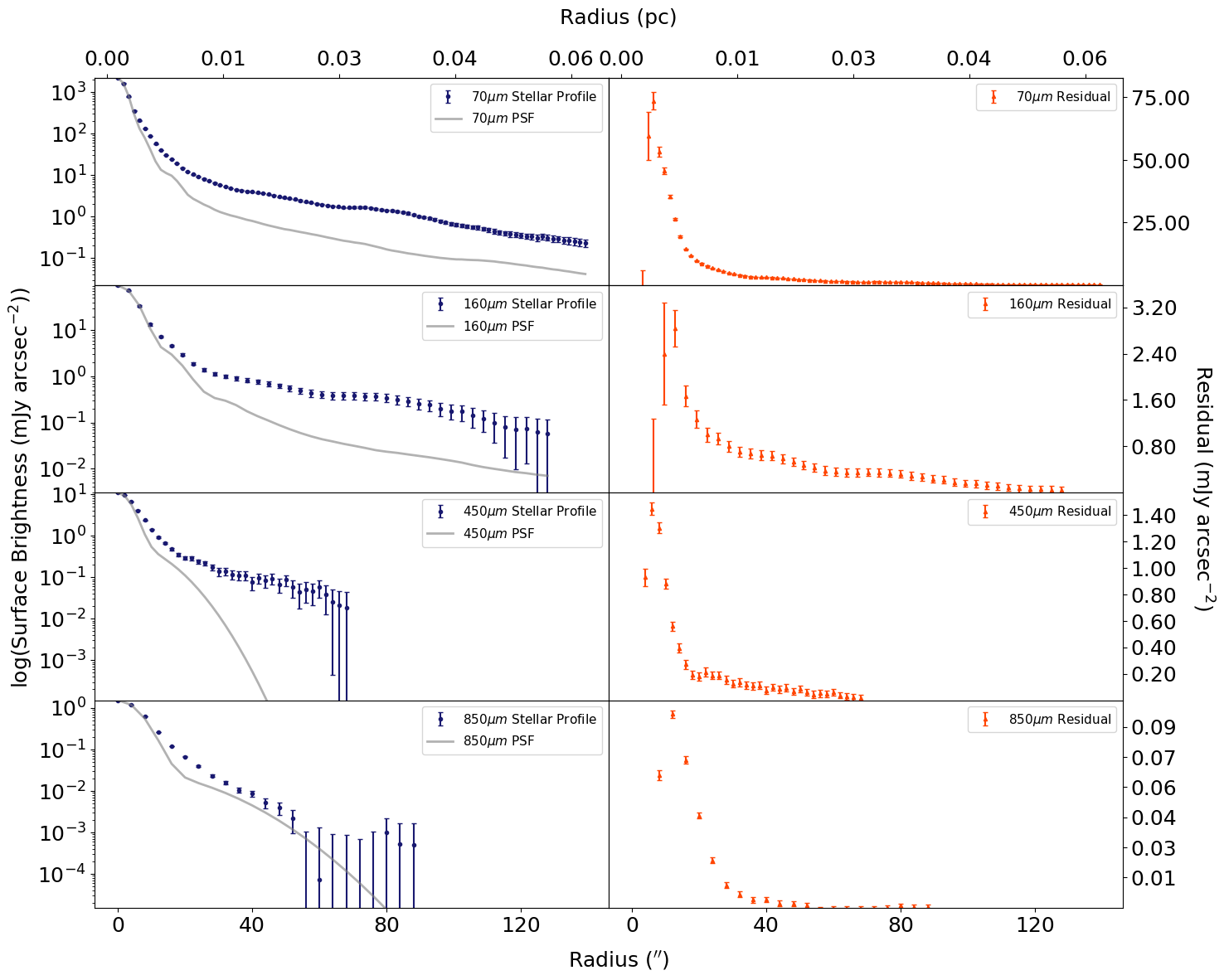}
  \subcaption{}
  \label{fig:ocet_RadialProfile}
  \end{subfigure}
  
\begin{subfigure}[b]{0.4\textwidth}
  \centering
  \includegraphics[width=\textwidth]{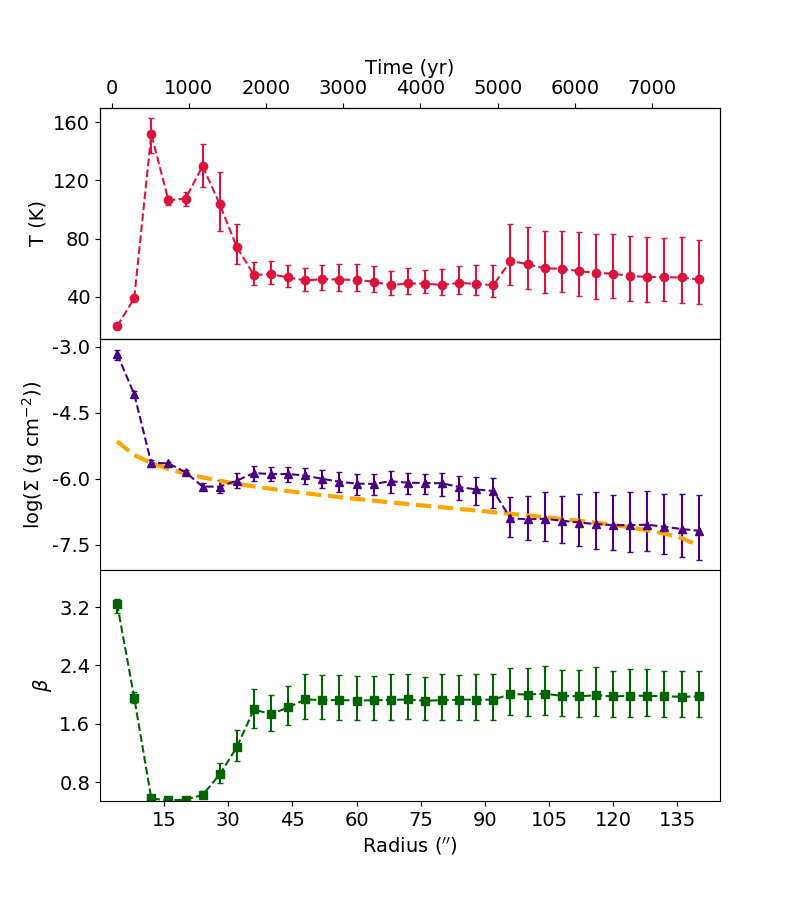}
  \subcaption{}
  \label{fig:ocet_TempDensBeta_Profile}
  \end{subfigure}
\begin{subfigure}[b]{0.4\textwidth}
  \centering
  \includegraphics[width=\textwidth]{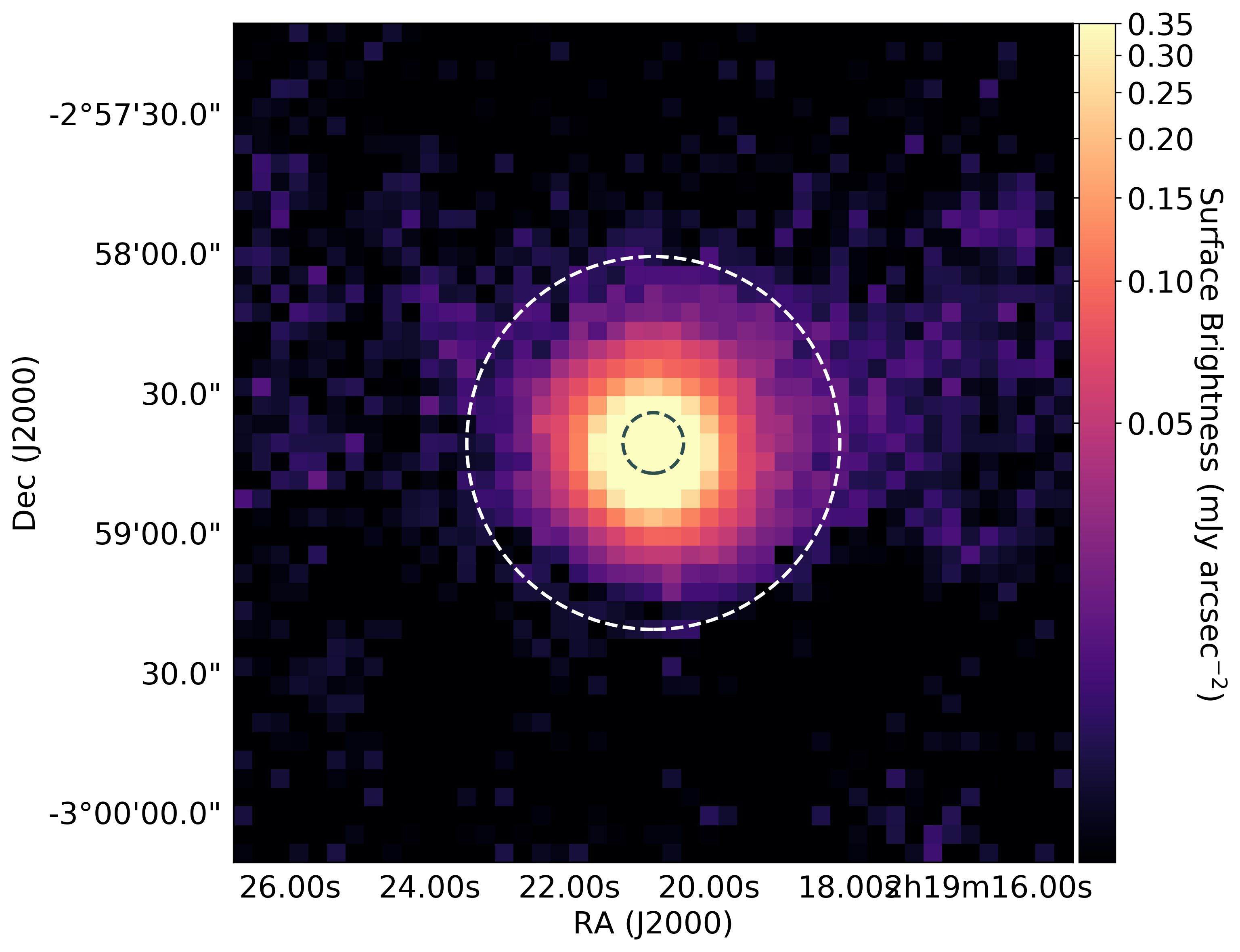}
  \subcaption{}
  \label{fig:ocet_ContourPlot_850}
  \end{subfigure}
  
  \caption{As Fig.~\ref{fig:cit6_All} for AGB star \textit{o} Ceti}
  \label{fig:ocet_All}
\end{figure*}

\begin{figure*}
\centering
\begin{subfigure}[b]{0.8\textwidth}
  \centering
  \includegraphics[width=\textwidth]{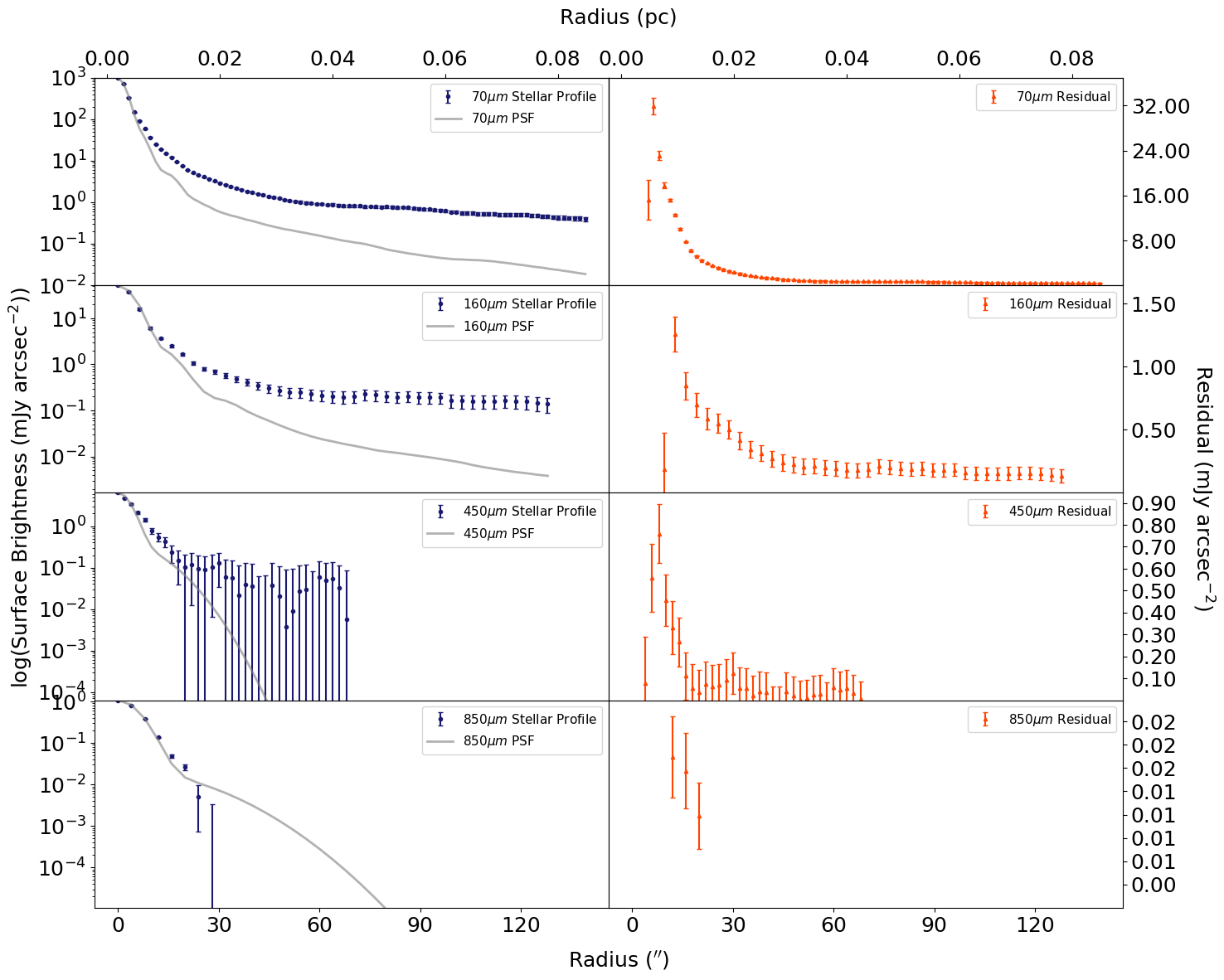}
  \subcaption{}
  \label{fig:rcas_RadialProfile}
  \end{subfigure}
  
\begin{subfigure}[b]{0.4\textwidth}
  \centering
  \includegraphics[width=\textwidth]{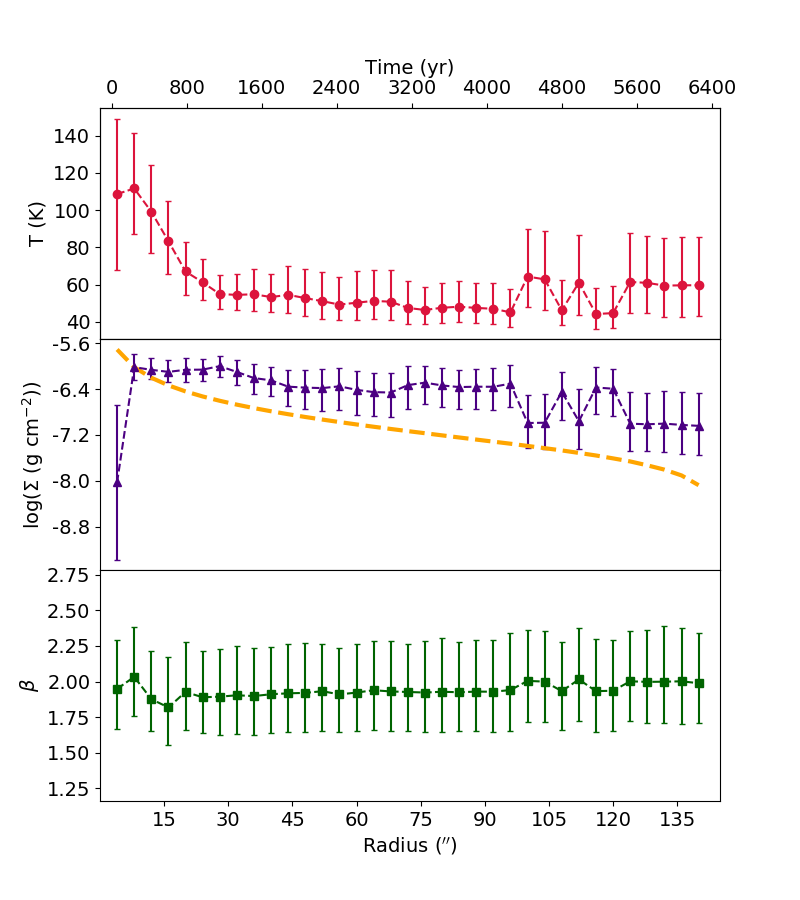}
  \subcaption{}
  \label{fig:rcas_TempDensBeta_Profile}
  \end{subfigure}
\begin{subfigure}[b]{0.4\textwidth}
  \centering
  \includegraphics[width=\textwidth]{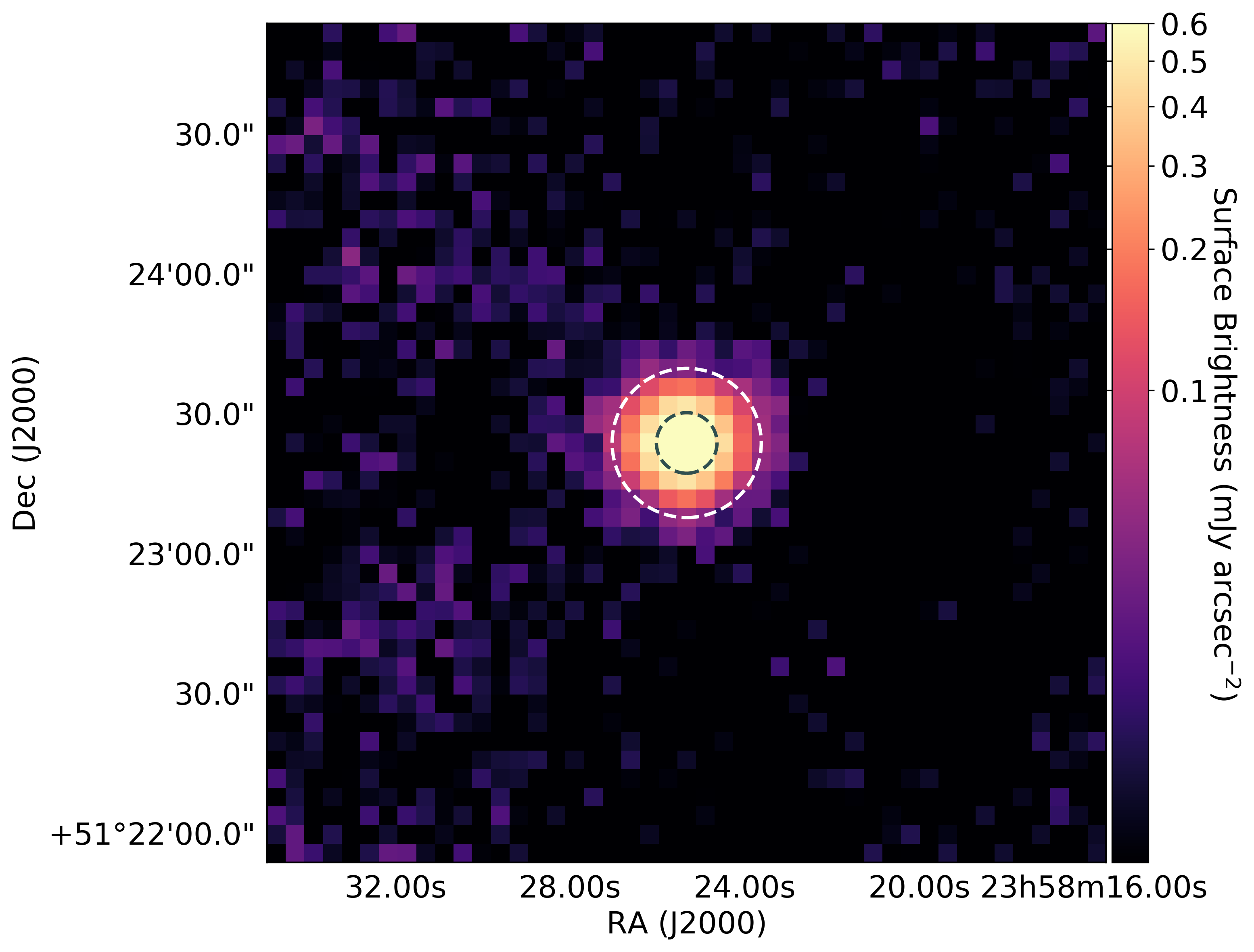}
  \subcaption{}
  \label{fig:rcas_ContourPlot_850}
  \end{subfigure}
  
  \caption{As Fig.~\ref{fig:cit6_All} for AGB star R Cas}
  \label{fig:rcas_All}
\end{figure*}

\begin{figure*}
\centering
\begin{subfigure}[b]{0.8\textwidth}
  \centering
  \includegraphics[width=\textwidth]{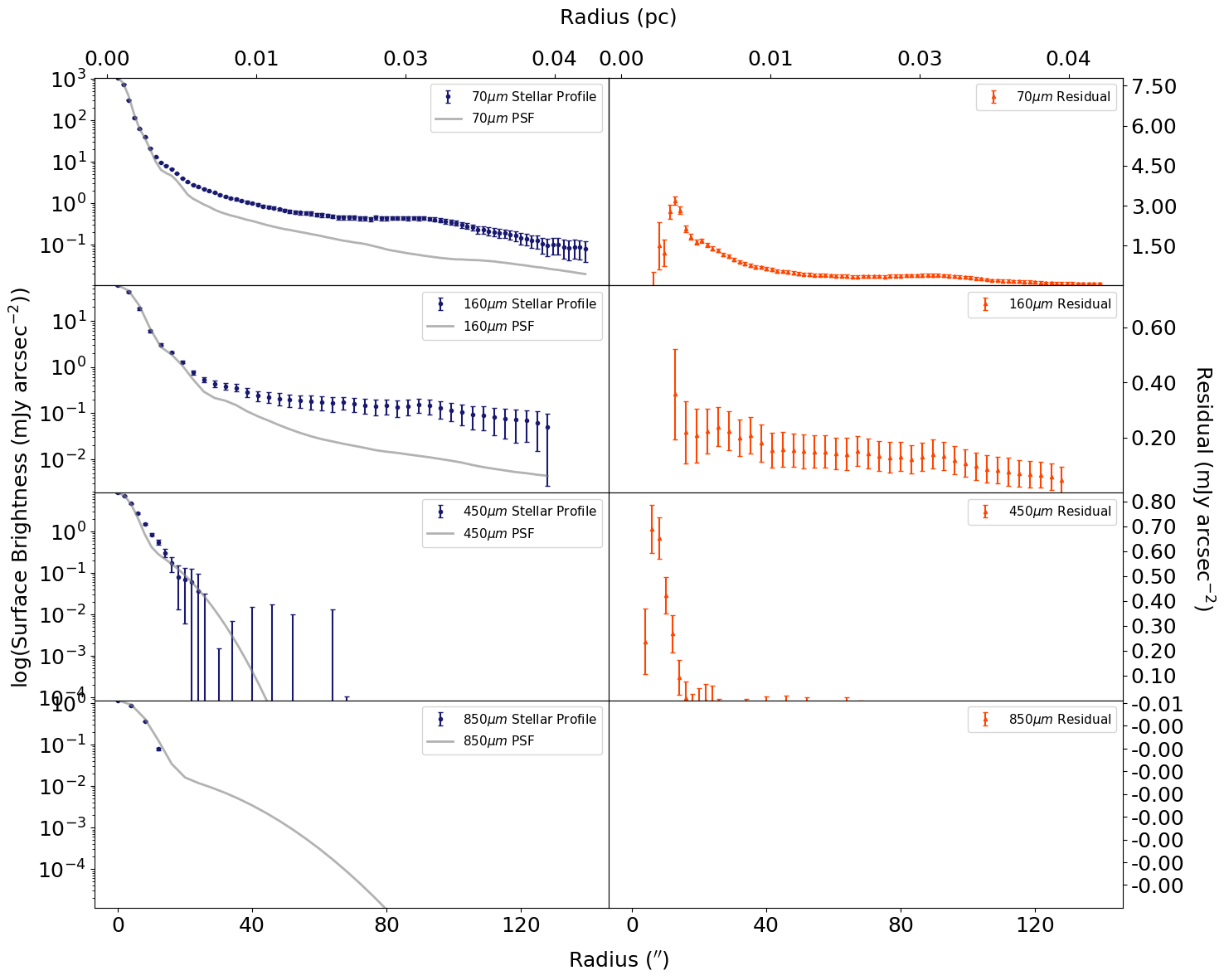}
  \subcaption{}
  \label{fig:rleo_RadialProfile}
  \end{subfigure}
  
\begin{subfigure}[b]{0.4\textwidth}
  \centering
  \includegraphics[width=\textwidth]{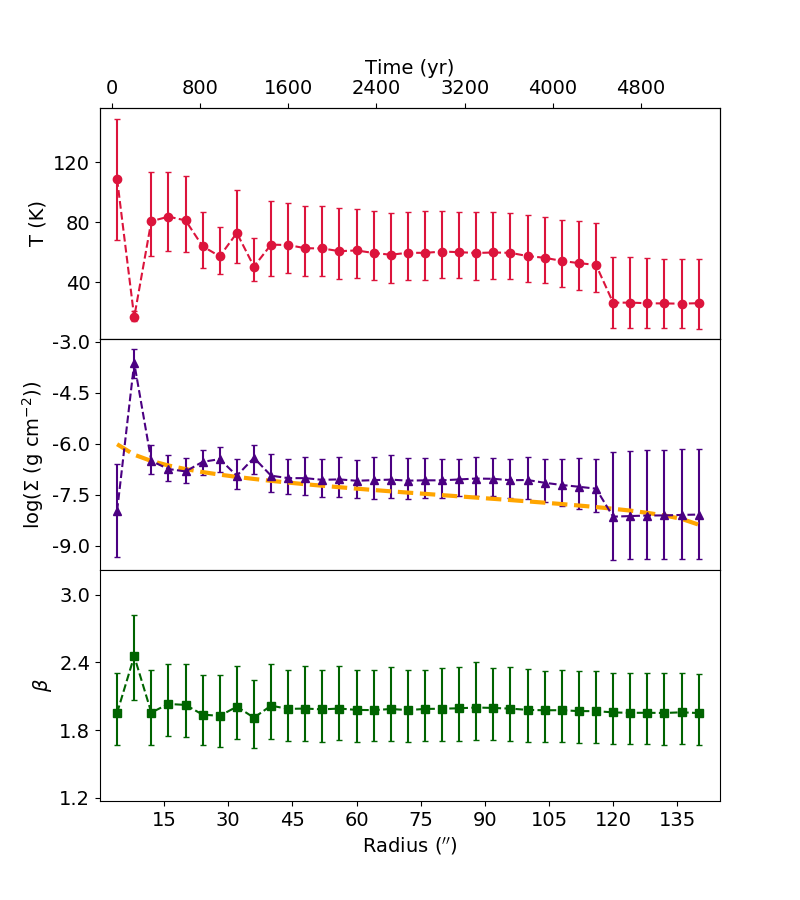}
  \subcaption{}
  \label{fig:rleo_TempDensBeta_Profile}
  \end{subfigure}
\begin{subfigure}[b]{0.4\textwidth}
  \centering
  \includegraphics[width=\textwidth]{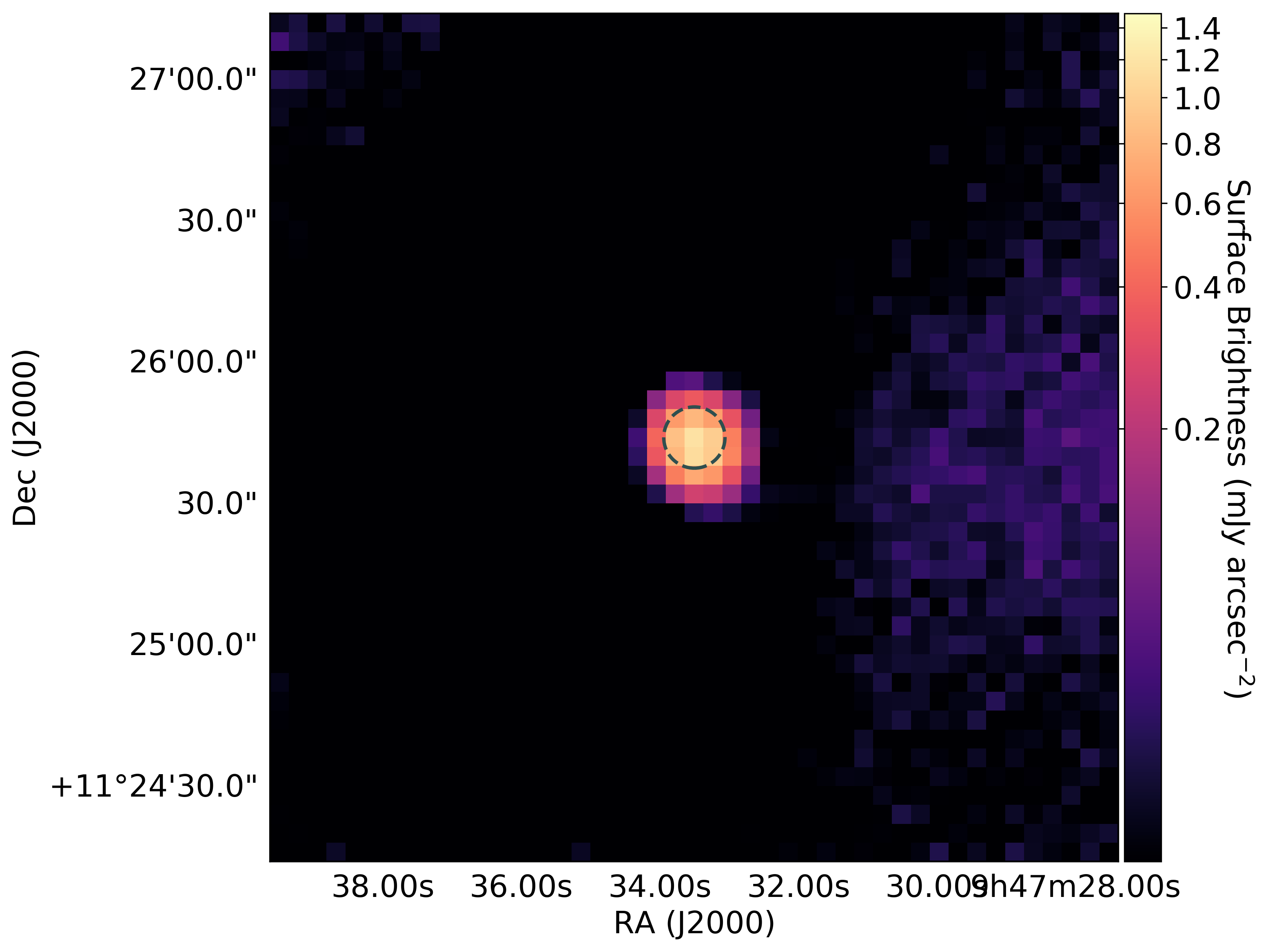}
  \subcaption{}
  \label{fig:rleo_ContourPlot_850}
  \end{subfigure}
  
  \caption{As Fig.~\ref{fig:cit6_All} for AGB star R Leo}
  \label{fig:rleo_All}
\end{figure*}

\begin{figure*}
\centering
\begin{subfigure}[b]{0.8\textwidth}
  \centering
  \includegraphics[width=\textwidth]{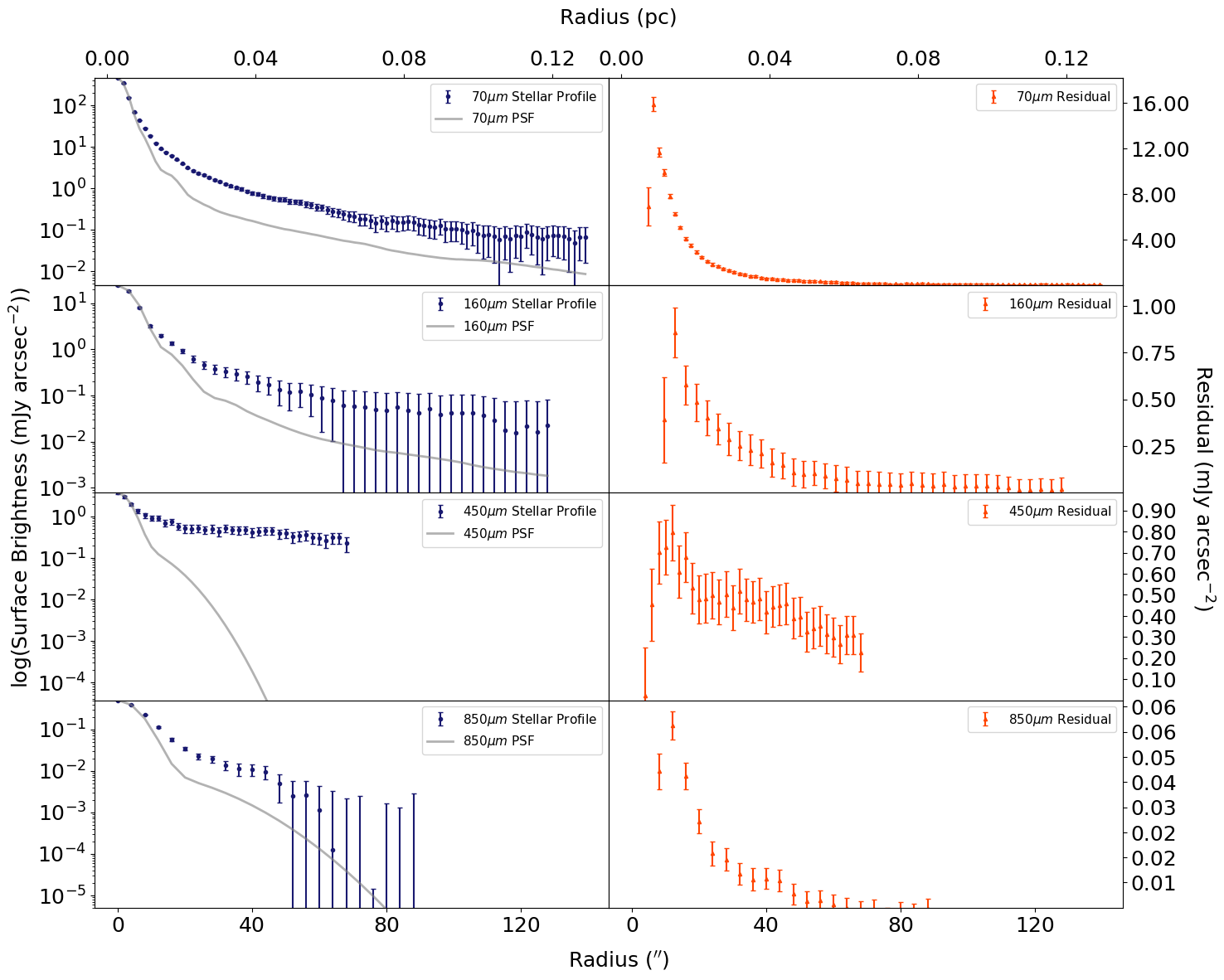}
  \subcaption{}
  \label{fig:rxboo_RadialProfile}
  \end{subfigure}
  
\begin{subfigure}[b]{0.4\textwidth}
  \centering
  \includegraphics[width=\textwidth]{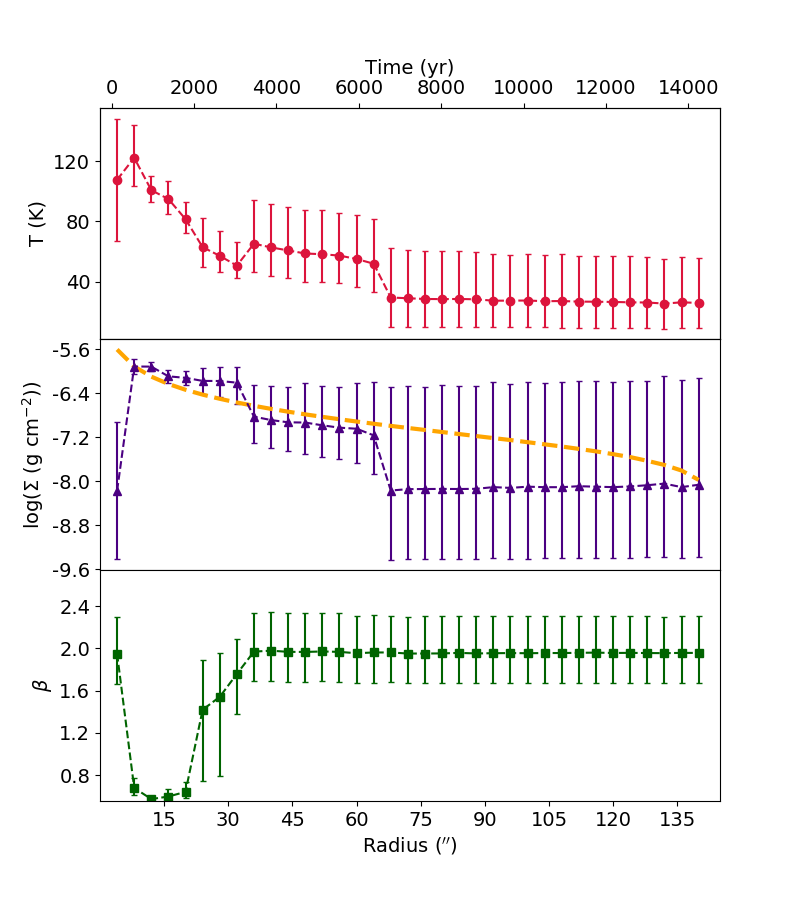}
  \subcaption{}
  \label{fig:rxboo_TempDensBeta_Profile}
  \end{subfigure}
\begin{subfigure}[b]{0.4\textwidth}
  \centering
  \includegraphics[width=\textwidth]{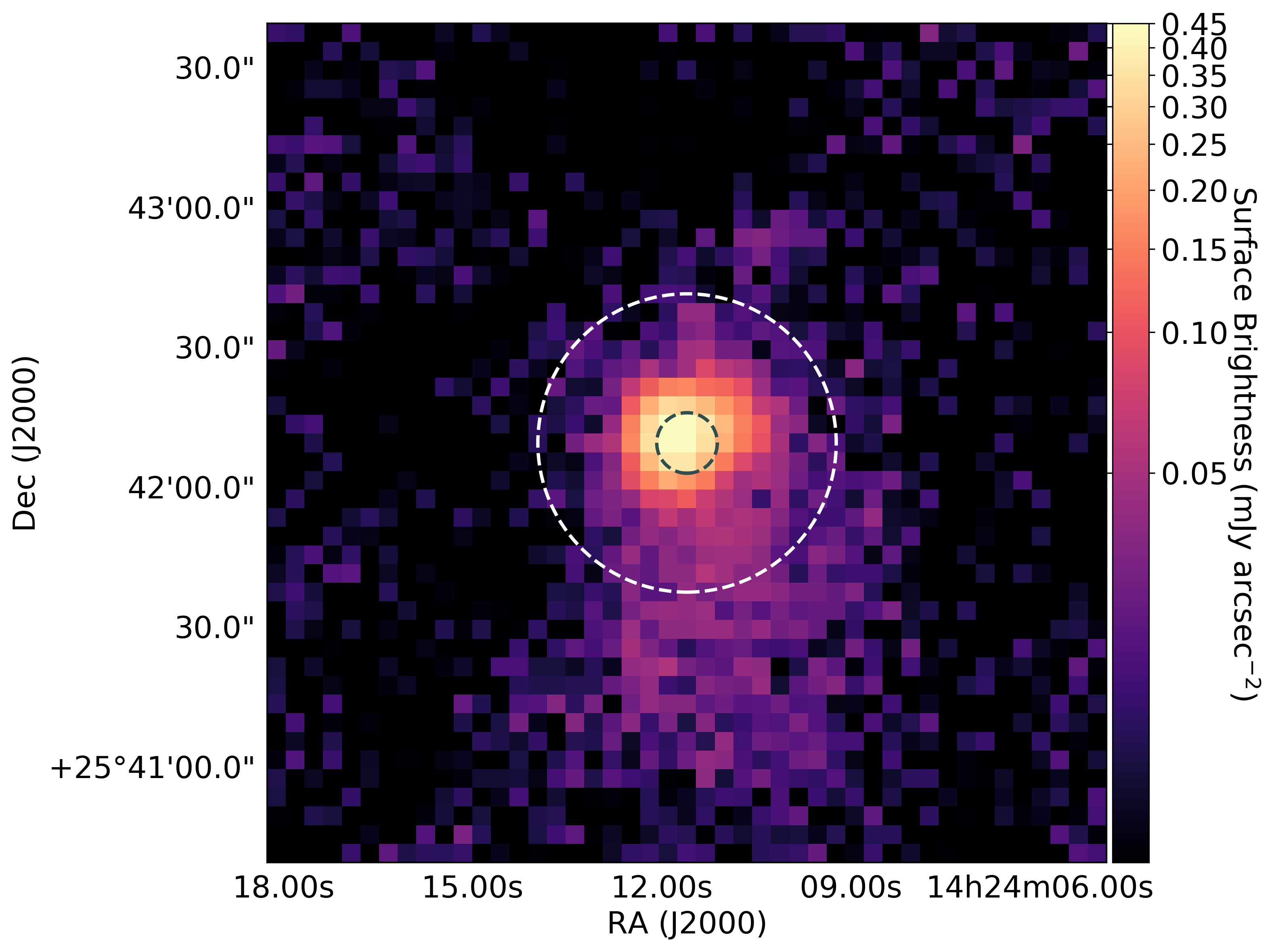}
  \subcaption{}
  \label{fig:rxboo_ContourPlot_850}
  \end{subfigure}
  
  \caption{As Fig.~\ref{fig:cit6_All} for AGB star RX Boo}
  \label{fig:rxboo_All}
\end{figure*}

\begin{figure*}
\centering
\begin{subfigure}[b]{0.8\textwidth}
  \centering
  \includegraphics[width=\textwidth]{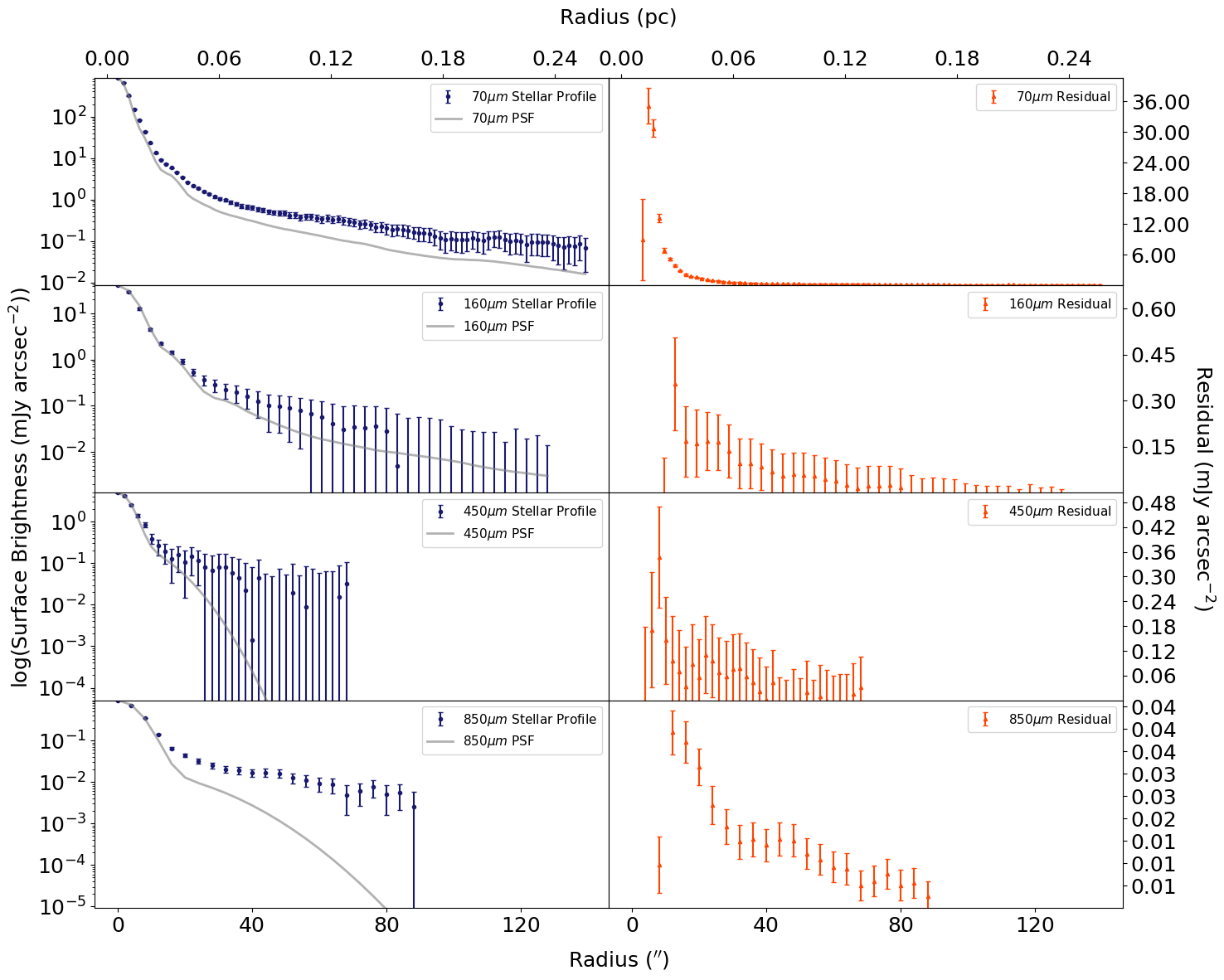}
  \subcaption{}
  \label{fig:txcam_RadialProfile}
  \end{subfigure}
  
\begin{subfigure}[b]{0.4\textwidth}
  \centering
  \includegraphics[width=\textwidth]{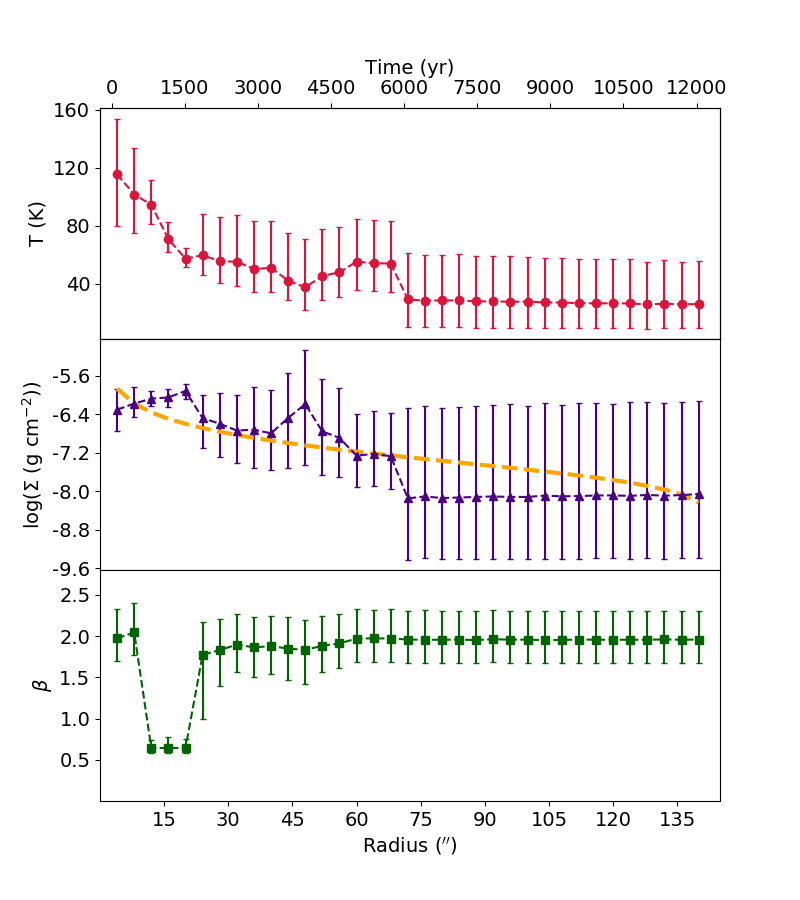}
  \subcaption{}
  \label{fig:txcam_TempDensBeta_Profile}
  \end{subfigure}
\begin{subfigure}[b]{0.4\textwidth}
  \centering
  \includegraphics[width=\textwidth]{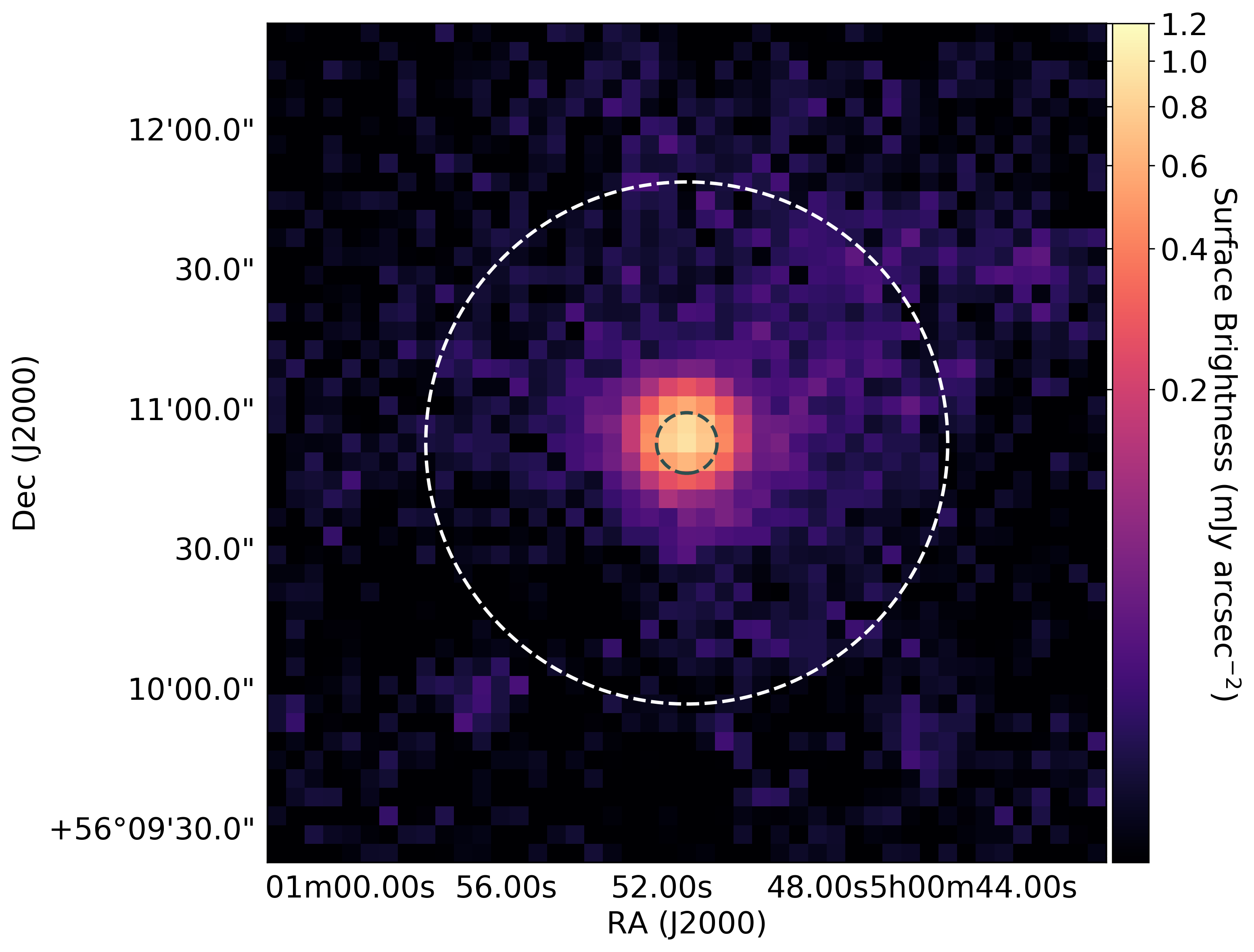}
  \subcaption{}
  \label{fig:txcam_ContourPlot_850}
  \end{subfigure}
  
  \caption{As Fig.~\ref{fig:cit6_All} for AGB star TX Cam}
  \label{fig:txcam_All}
\end{figure*}

\begin{figure*}
\centering
\begin{subfigure}[b]{0.8\textwidth}
  \centering
  \includegraphics[width=\textwidth]{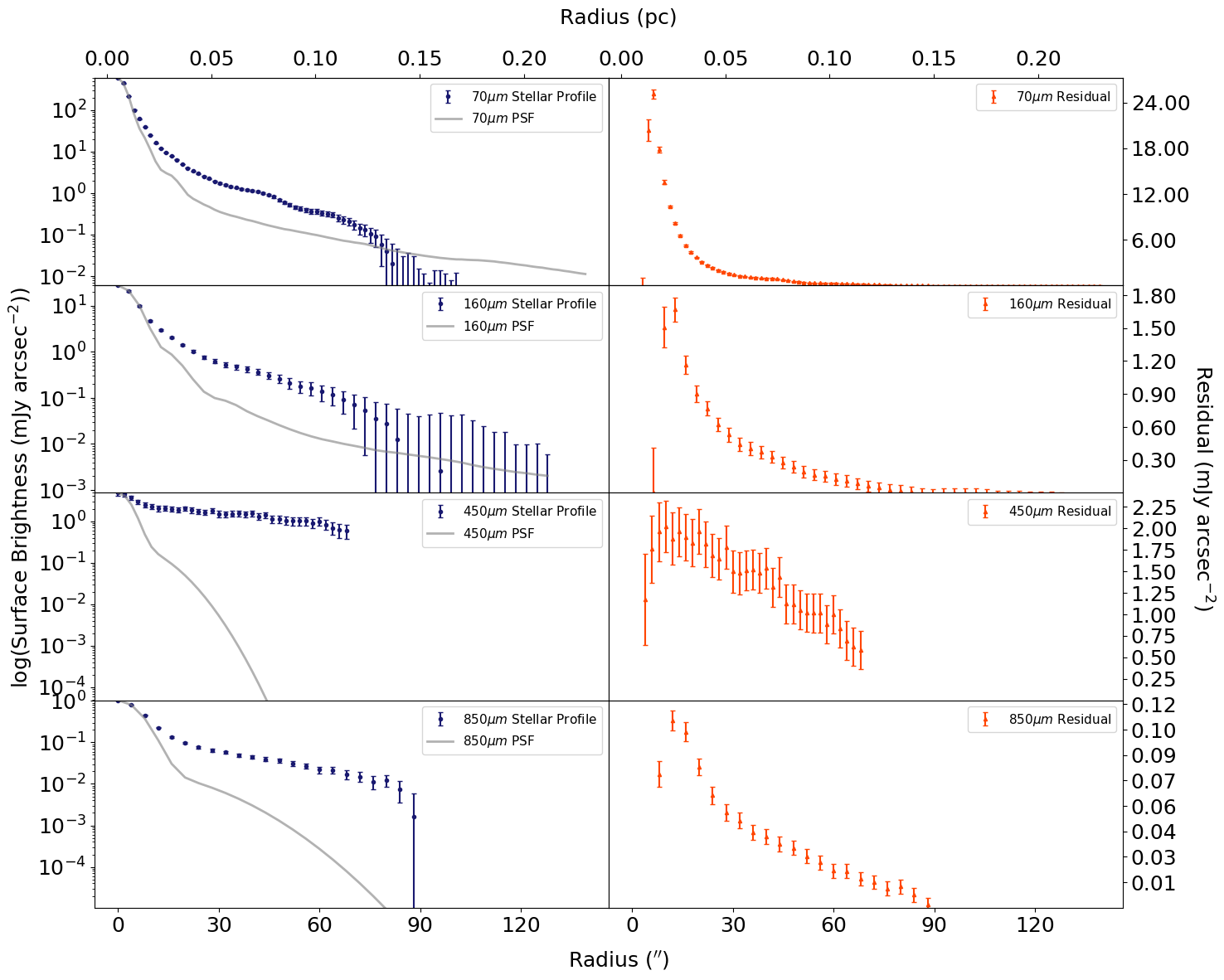}
  \subcaption{}
  \label{fig:waql_RadialProfile}
  \end{subfigure}
  
\begin{subfigure}[b]{0.4\textwidth}
  \centering
  \includegraphics[width=\textwidth]{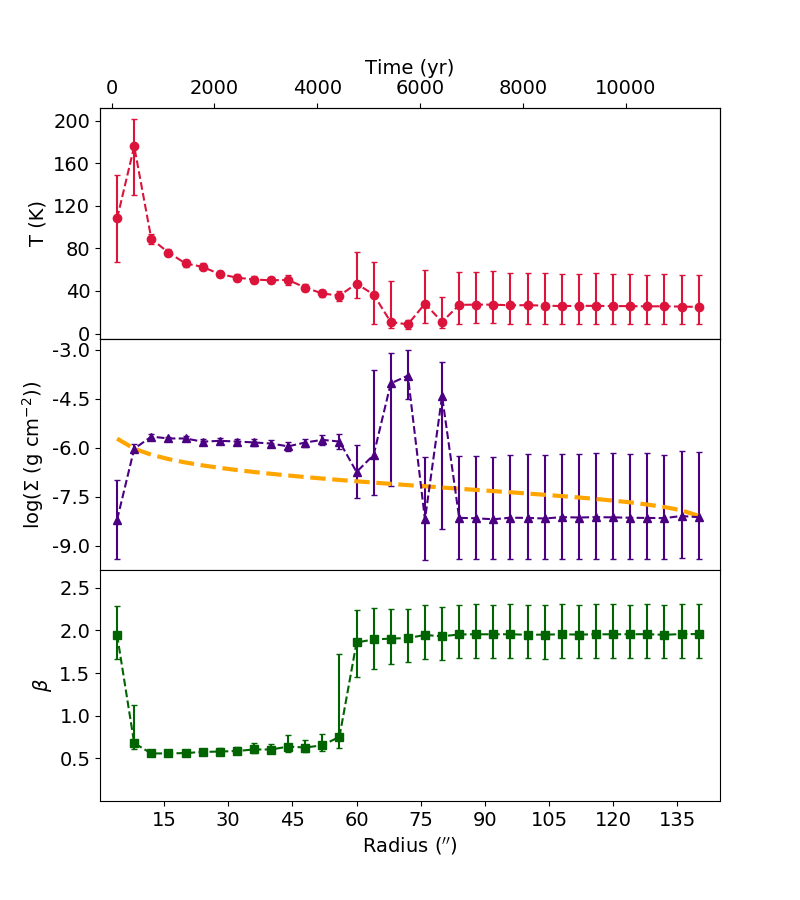}
  \subcaption{}
  \label{fig:waql_TempDensBeta_Profile}
  \end{subfigure}
\begin{subfigure}[b]{0.4\textwidth}
  \centering
  \includegraphics[width=\textwidth]{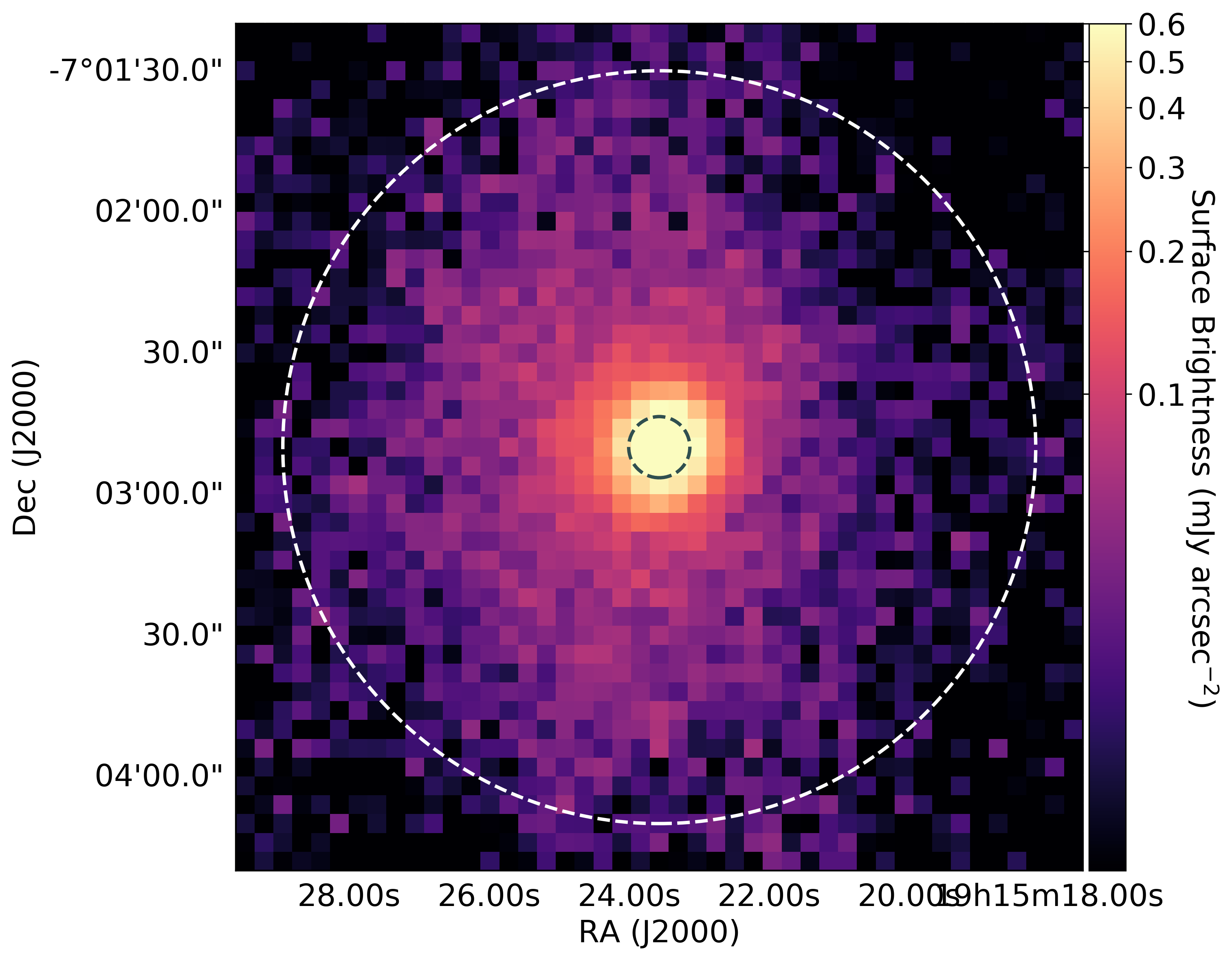}
  \subcaption{}
  \label{fig:waql_ContourPlot_850}
  \end{subfigure}
  
  \caption{As Fig.~\ref{fig:cit6_All} for AGB star W Aql}
  \label{fig:waql_All}
\end{figure*}

\begin{figure*}
\centering
\begin{subfigure}[b]{0.8\textwidth}
  \centering
  \includegraphics[width=\textwidth]{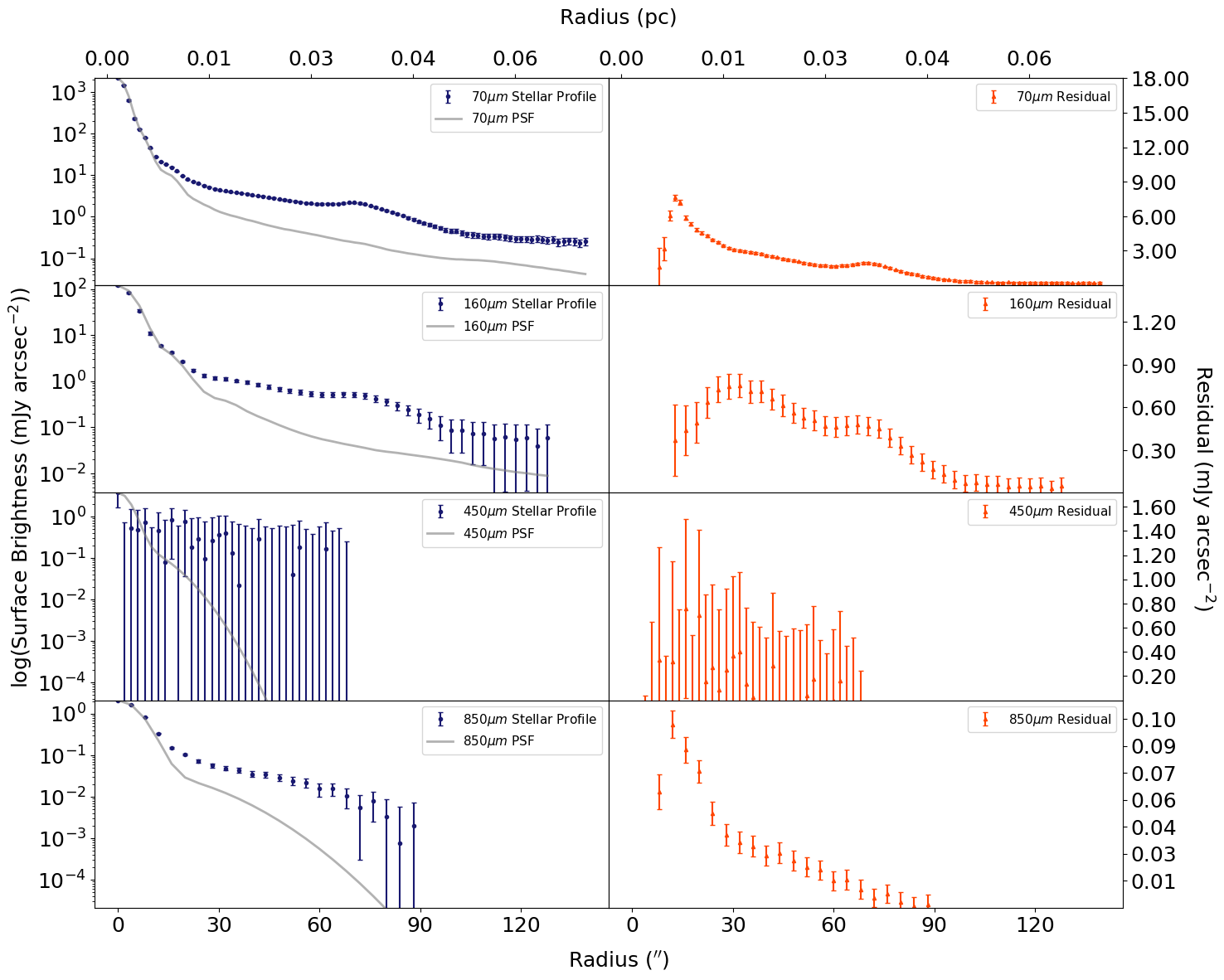}
  \subcaption{}
  \label{fig:whya_RadialProfile}
  \end{subfigure}
  
\begin{subfigure}[b]{0.4\textwidth}
  \centering
  \includegraphics[width=\textwidth]{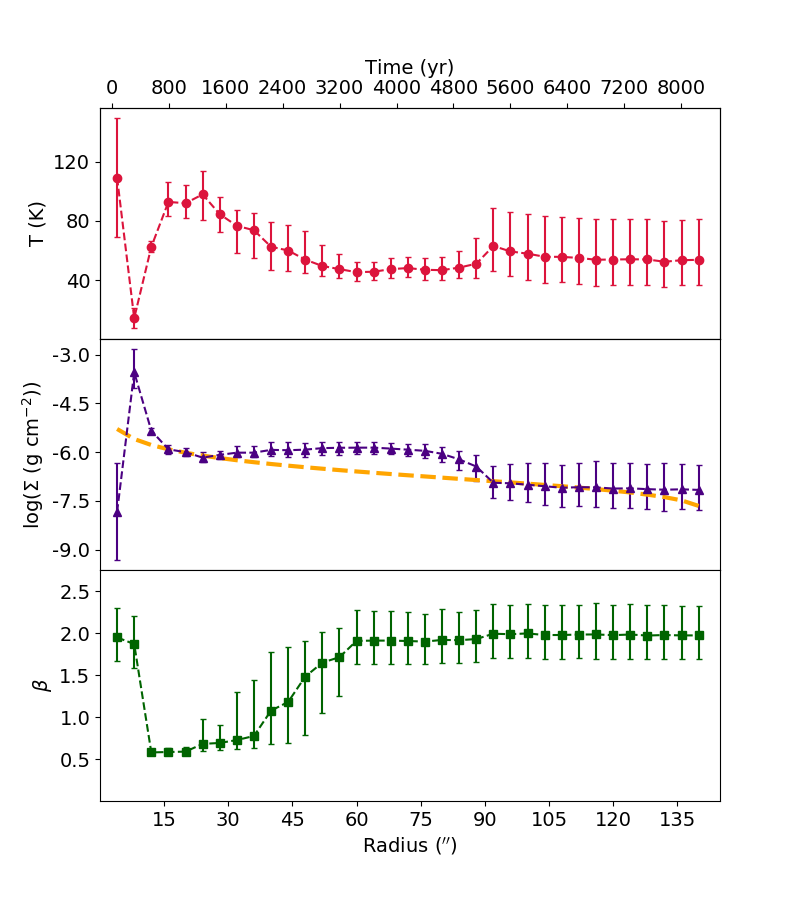}
  \subcaption{}
  \label{fig:whya_TempDensBeta_Profile}
  \end{subfigure}
\begin{subfigure}[b]{0.4\textwidth}
  \centering
  \includegraphics[width=\textwidth]{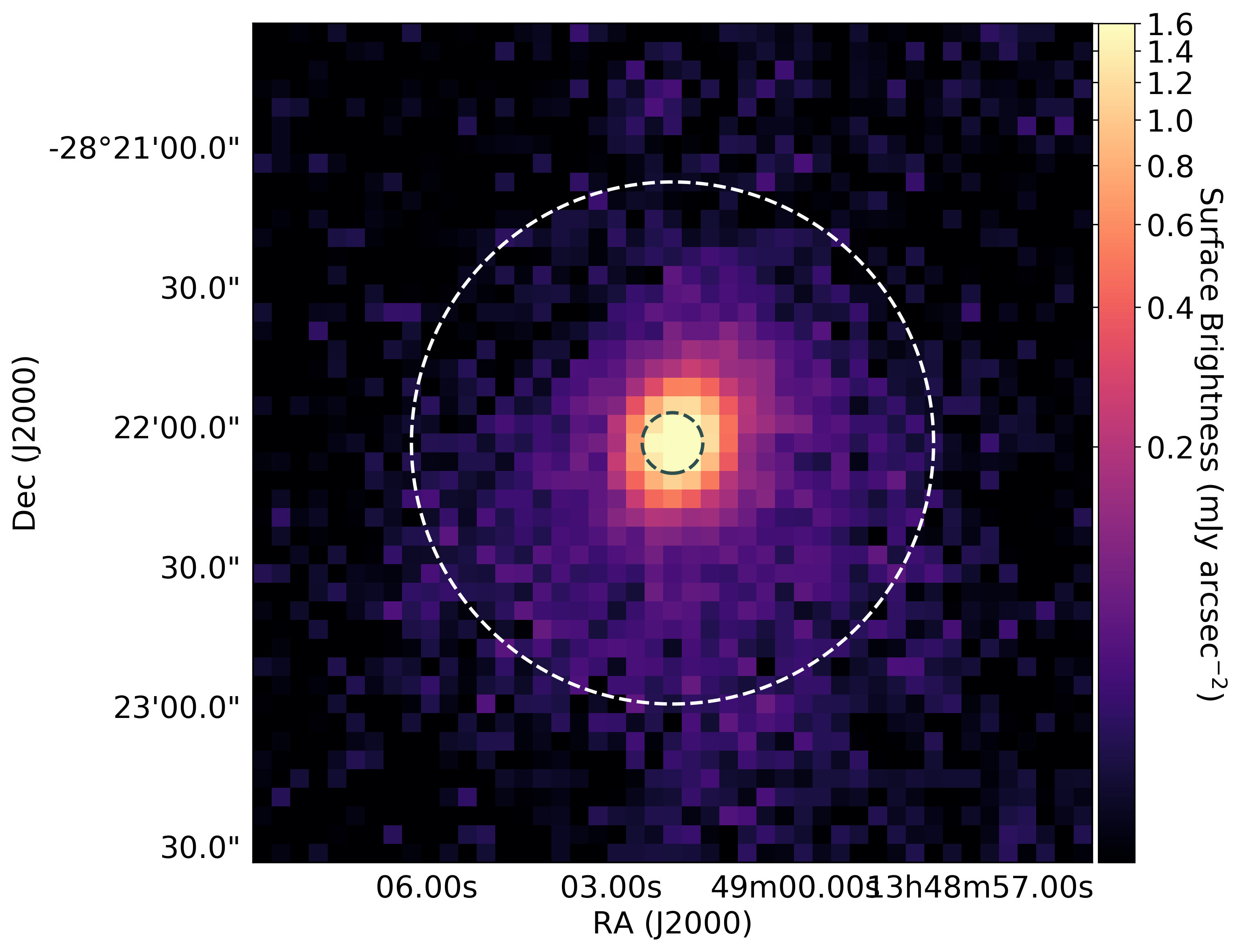}
  \subcaption{}
  \label{fig:whya_ContourPlot_850}
  \end{subfigure}
  
  \caption{As Fig.~\ref{fig:cit6_All} for AGB star W Hya}
  \label{fig:whya_All}
\end{figure*}

\end{document}